\documentclass{article}
\usepackage{amsmath,amssymb,dsfont,stfloats,color,url,bbm}
\usepackage{subfigure}
\usepackage{cite}
\usepackage{xcolor}
\usepackage{pgfplots, pgfplotstable}
\renewcommand\fbox{\fcolorbox{red}{white}}
\allowdisplaybreaks

\usepackage{mathtools}
\usepackage{cite}
\usepackage[margin=1in]{geometry}
\usepackage{amsmath}
\usepackage{amssymb}
\usepackage{amsthm}
\usepackage{latexsym}
\usepackage{verbatim}
\usepackage{subfigure}
\usepackage{psfrag}
\usepackage{color}
\usepackage{enumitem}

\usepackage[utf8]{inputenc} 
\usepackage{amsfonts} 
\usepackage[T1]{fontenc}
\usepackage{url}              

\usepackage{hyperref}
\hypersetup{
	colorlinks=true,
	linkcolor=blue,
	filecolor=blue,
	citecolor = blue,      
	urlcolor=cyan,}

\usepackage{esdiff}
\usepackage{pgf,tikz,psfrag}
\usepackage{yfonts}

\newtheorem{thm}{Theorem}
\newtheorem{prop}{Proposition}
\newtheorem{lem}{Lemma}
\newtheorem{asmp}{Assumption}
\newtheorem{cor}{Corollary}

\newtheorem{defi}{Definition}
\newtheorem{rem}{Remark}
\newtheorem{example}{Example}

\newtheorem{claim}{Claim}

\def\mathlette#1#2{{\mathchoice{\mbox{#1$\displaystyle #2$}}%
{\mbox{#1$\textstyle #2$}}%
{\mbox{#1$\scriptstyle #2$}}%
{\mbox{#1$\scriptscriptstyle #2$}}}}
\newcommand{\matr}[1]{\mathlette{\boldmath}{#1}}

\providecommand{\C}{\ensuremath{\mathbb{C}}}

\newcommand{\bm}{F} 

\providecommand{\GS}[2]{\mathcal{GS}(#1 \mid #2)}
\newcommand{\Op}[1]{\mathcal{O}(#1)}

\newfont{\bbb}{msbm10 scaled 700}

\newfont{\bb}{msbm10 scaled 1100}
\newcommand{\CC}{\mbox{\bb C}}
\newcommand{\PP}{\mbox{\bb P}}
\newcommand{\RR}{\mbox{\bb R}}

\newcommand{\EE}{\mbox{\bb E}}
\newcommand{\NN}{\mbox{\bb N}}

\newcommand{\av}{{\bf a}}
\newcommand{\bv}{{\bf b}}

\newcommand{\ev}{{\bf e}}

\newcommand{\gv}{{\bf g}}
\newcommand{\hv}{{\bf h}}

\newcommand{\mv}{{\bf m}}
\newcommand{\nv}{{\bf n}}

\newcommand{\rv}{{\bf r}}
\newcommand{\sv}{{\bf s}}

\newcommand{\uv}{{\bf u}}
\newcommand{\wv}{{\bf w}}
\newcommand{\vv}{{\bf v}}
\newcommand{\xv}{{\bf x}}
\newcommand{\yv}{{\bf y}}
\newcommand{\zv}{{\bf z}}
\newcommand{\zerov}{{\bf 0}}

\newcommand{\Am}{{\bf A}}
\newcommand{\Bm}{{\bf B}}
\newcommand{\Cm}{{\bf C}}
\newcommand{\Dm}{{\bf D}}
\newcommand{\Em}{{\bf E}}
\newcommand{\Fm}{{\bf F}}
\newcommand{\Gm}{{\bf G}}
\newcommand{\Hm}{{\bf H}}
\newcommand{\Id}{{\bf I}}

\newcommand{\Nm}{{\bf N}}

\newcommand{\Pm}{{\bf P}}
\newcommand{\Qm}{{\bf Q}}
\newcommand{\Rm}{{\bf R}}
\newcommand{\Sm}{{\bf S}}

\newcommand{\Um}{{\bf U}}

\newcommand{\Vm}{{\bf V}}
\newcommand{\Xm}{{\bf X}}
\newcommand{\Ym}{{\bf Y}}
\newcommand{\Zm}{{\bf Z}}

\newcommand{\Ac}{{\cal A}}

\newcommand{\Cc}{{\cal C}}
\newcommand{\Dc}{{\cal D}}

\newcommand{\Fc}{{\cal F}}

\newcommand{\Hc}{{\cal H}}

\newcommand{\Lc}{{\cal L}}
\newcommand{\Mc}{{\cal M}}
\newcommand{\Nc}{{\cal N}}

\newcommand{\Sc}{{\cal S}}

\newcommand{\Vc}{{\cal V}}

\newcommand{\Zc}{{\cal Z}}

\newcommand{\xiv}{\hbox{\boldmath$\xi$}}

\newcommand{\Sigmam}{\hbox{\boldmath$\Sigma$}}

\newcommand{\Xim}{\hbox{\boldmath$\Xi$}}

\newcommand{\diag}{{\hbox{diag}}}

\newcommand{\trace}{{\hbox{tr}}}

\renewcommand{\Re}{{\rm Re}}
\renewcommand{\Im}{{\rm Im}}
\newcommand{\eqdef}{\stackrel{\Delta}{=}}

\newcommand{\herm}{{\sf H}}

\newcommand{\transp}{{\sf T}}

\newcommand{\SNR}{{\sf SNR}}

\usepackage{stmaryrd} 

\title{Joint Message Detection and Channel Estimation for Unsourced Random Access in Cell-Free User-Centric Wireless Networks}

\pgfplotsset{compat=1.18}
\begin{document}
	\author{Burak \c{C}akmak, Eleni Gkiouzepi,  Manfred Opper, Giuseppe Caire
		\thanks{The work of Burak \c{C}akmak was supported by the Gottfried Wilhelm Leibniz-Preis 2021 of the German Science Foundation (DFG). The work of Giuseppe Caire was supported by BMBF Germany in the program of ``Souver\"an. Digital. Vernetzt.'' Joint Project 6G-RIC (Project IDs 16KISK030).}
        \thanks{ Burak \c{C}akmak, Eleni Gkiouzepi and Giuseppe Caire are with the Faculty of Electrical Engineering and Computer Science, Technical University of  Berlin, 10587 Berlin, Germany (emails: \{burak.cakmak,gkiouzepi,caire\}@tu-berlin.de).}
	\thanks{Manfred Opper is with the Faculty of Electrical Engineering and Computer Science, Technical University of Berlin, 10587 Berlin, Germany and the Centre for Systems Modelling and Quantitative Biomedicine, University of Birmingham, Birmingham B15 2TT, United Kingdom (email: manfred.opper@tu-berlin.de).}
	}
	\date{\today}
	\maketitle

	
	\maketitle
	\begin{abstract}
		We consider unsourced random access (uRA) in a cell-free (CF) user-centric wireless network, where a large number of potential users compete for a random access slot, while only a finite subset is active. The random access users transmit codewords of length $L$ symbols from a shared codebook, which are received by $B$ geographically distributed radio units (RUs), each equipped with $M$ antennas. Our goal is to devise and analyze a \emph{centralized} decoder to detect the transmitted messages (without prior knowledge of the active users) and estimate the corresponding channel state information. A specific challenge lies in the fact that, due to the geographically distributed nature of the CF network, there is no fixed correspondence between codewords and large-scale fading coefficients (LSFCs). This makes current activity detection approaches which make use of this fixed LSFC-codeword association not directly applicable. To overcome this problem, we propose a scheme where the access codebook is partitioned in location-based subcodes, such that users in a particular location make use of the corresponding subcode. The joint message detection and channel estimation is obtained via a novel {\em Approximated Message Passing} (AMP) algorithm for a linear superposition of matrix-valued sources corrupted by noise. 
		The statistical asymmetry in the fading profile and message activity leads to \emph{different statistics} for the matrix sources, which distinguishes the AMP formulation from previous cases. In the regime where the codebook size scales linearly with $L$, while $B$ and $M$ are fixed,  we present a rigorous high-dimensional (but finite-sample) analysis of the proposed AMP algorithm.  Exploiting this, we then present a precise (and rigorous) large-system analysis of the message missed-detection and false-alarm rates, as well as the channel estimation mean-square error. The resulting system allows the seamless formation of user-centric clusters and very low latency beamformed uplink-downlink communication without explicit user-RU association, 
		pilot allocation, and power control. This makes the proposed scheme highly appealing for low-latency random access communications in CF networks.	
	\end{abstract}

	
	\section{Introduction}  \label{intro}
{M}{ultiuser} multiple-input multiple-output (MU-MIMO) has been widely studied from an information theoretic point of view \cite{Caire-Shamai-TIT03,Viswanath-Tse-TIT03,Weingarten-Steinberg-Shamai-TIT06, Caire-Jindal-Kobayashi-Ravindran-TIT10} and has become an important component of the physical layer of cellular systems \cite{3gpp38211,Larsson-book} and wireless local area networks (see \cite{khorov2018tutorial,qu2019survey} and references therein). As a further extension of MU-MIMO, joint processing of spatially distributed remote radio units (RUs), which can be traced back to \cite{wyner1994shannon}, has been studied in different contexts and under different names such as {\em coordinate multipoint}, {\em cloud radio access network}, or {\em cell-free (CF) MIMO} \cite{ngo2017cell,8845768, 9064545,9336188,gottsch2022subspace}.  In this paper we focus on {\em user-centric} architectures as defined in \cite{9336188}, where each user is served by a localized cluster of nearby RUs, and each RU in turn serves a limited set of nearby users.
	
	A crucial aspect enabling spatial multiplexing is the availability of channel state information at the infrastructure side (i.e., base stations, RUs) in order to enable multiuser detection in the uplink (UL) and multiuser precoding in the downlink (DL). In particular, channel state information for the DL is obtained by exploiting the UL/DL channel reciprocity, which holds under mild conditions in Time Division Duplexing (TDD) systems \cite{marzetta2010noncooperative,Larsson-book}. Presently, most works on CF MIMO assumes that transmission resources (e.g., UL pilot sequences, time-frequency slots, and user-centric RU clusters) are permanently assigned to all users in the system.
	
	However, in highly dense scenarios,\footnote{Imagine a sports area with approximately 50,000 users (including those who are simultaneously inactive), all concentrated within a $300 \times 300 \, \text{m}^2$ area.} a static allocation of transmission resources would be impractical and highly inefficient due to the sporadic and intermittent activity exhibited by users.
	For such high-density systems, it is necessary to design a dynamic random access mechanism that allows users to access the system and request transmission resources only when they become active. 
	
\begin{figure}[t]
		\centering
			\includegraphics[width=0.7\linewidth]{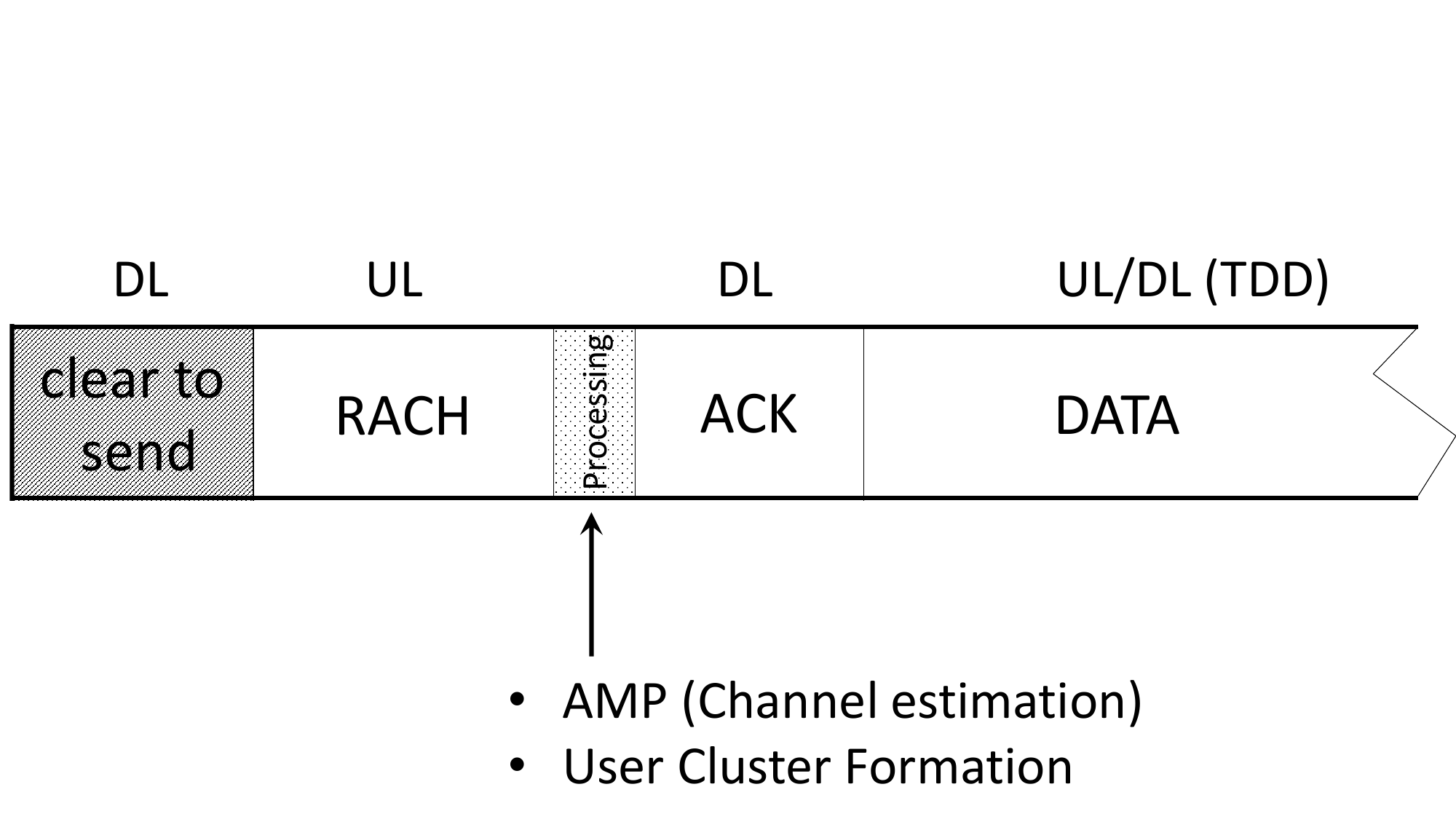}
		\caption{A schematic of the RACH slot, followed by a DL packet (referred to as ACK in this paper) 
			and subsequent UL/DL TDD slots for allocated traffic.} 
		\label{RACH}  
	\end{figure}

	We assume that all RUs periodically broadcast a clear-to-send signal followed by a {\em random access channel} (RACH) slot where users can transmit a random access message (see Fig.~\ref{RACH}).
	At the end of the RACH slot, the system must: 1) decode the random access messages and estimate the corresponding 
	channel vectors; 2) associate to each decoded message a cluster of RUs for further allocated communication;  
	3) send a DL message which may just contain the requested content by the random access users and/or an acknowledgement possibly containing resource allocation information (assigned UL and DL slots, pilot sequences, transmit power) for further communication, in a typical packet-reservation scheme  (e.g., see \cite{nanda1991performance}).
	Without going into unnecessary details, in this paper we are concerned with 
	the basic access protocol functions, namely: joint random access message detection and channel estimation, 
	seamless cluster formation, and ultra-low latency beamformed DL response, which for brevity will be referred to 
	as ``ACK message'' in the following.
	
	The high-level scheme	described above is conceptually similar to the so-called 2-step RACH scheme specified by 
	3GPP for fast random access (see \cite{liva2024unsourced} and references therein). This scheme makes use of 
	a codebook (i.e., a collection of codewords)~\footnote{For example, 
		in the 2-step RACH the codebook is formed by a collection of  Zadoff-Chu sequences \cite{liva2024unsourced}.} 
	that is fixed a priori and is common to all users. When a user wishes to access the channel, it picks a codeword
	and sends it. The system must generate a list of the transmitted codewords and, in our case, also
	an estimation of the corresponding channel coefficient vectors. Even though the system may not yet know the identity 
	of the random access user, it acknowledges that a user transmitting a certain $n$-th message (i.e., the $n$-th codeword) 
	in the detected list is requesting access and is associated with a channel estimate. Using this estimate, the system selects the set of $Q$ 
	{\em resource units} (RUs) with the largest channel magnitudes (where $Q$ is a suitable system parameter) and transmits an ACK message using some form of  DL precoding (e.g., {\em maximal-ratio transmission} (MRT), as in  \cite{ngo2017cell}). 
	
	In an information theoretic setting, the case where an arbitrary number of potential users (not all simultaneously active)
	make use of the {\em same} codebook to randomly access a shared channel, 
	and the system must decode the list of the transmitted messages, 
	is referred to as {\em unsourced random access} (uRA) \cite{Polyanskiy2017}.
	
	\subsection{Overview of Contributions}
	
	To overcome the challenge of the unknown association between {\em large scale fading coefficients} (LSFCs) and codewords,  we propose a location-based approach: the coverage area $\Dc$ is partitioned into $U$ regions $\{\Lc_u : u \in [U]\}$, called {\em locations}, where $U$ is a suitable { fixed integer}.  The locations are designed so that users within each location $\Lc_u$ experience a similar {  LSFC profile}, i.e., the collection of LSFCs  between a transmitter in location $\Lc_u$ and all the RUs is similar for all users in location $\Lc_u$. 
	The uRA codebook is partitioned in $U$ subcodes, respectively assigned to the $U$ locations.
	Users in a given location $\Lc_u$ are restricted to using the codewords of the associated $u$-th subcode. 
	Thus, the scheme allows general statistical asymmetry between users in different locations but enforces statistical symmetry between users within the same location. Note that conventional cellular systems also use { implicit 
		location information, enforcing a one-to-one correspondence between locations and cells}.  
	Thus, our approach can be seen as a generalization to the CF case of { current random access protocols  
		for  for cellular networks} (e.g., see also \cite{buzzi2022multi} for a location-based approach to beam alignment in mmWave CF networks).
	
	{ At a high level, we summarize our contributions as follows:
		\begin{enumerate}
			
			\item [(i)] 
			We introduce a (novel) multisource AMP (approximate message passing) framework for matrix-valued noisy linear observations that handles multiple independent signal sources, each of which can be described by an arbitrarily different distribution. We present a non-asymptotic (finite-sample) AMP analysis \cite{rush2016iterative,rush24} that provides explicit bounds on large-system approximation errors in terms of the $\mathcal L^p$ norm (Theorem 1). Our analysis yields the asymptotic consistency of the decoupling principles manifested in the AMP analysis and the replica-symmetric (RS) computation of the static problem. Consequently, the proposed AMP algorithm asymptotically achieves Bayesian optimality under the validity of the RS approach.			
			\item [(ii)] Utilizing the proposed AMP framework, we devise and analyze a centralized decoder to detect the transmitted messages and estimate the corresponding channel state information. In particular, we present  (rigorous) large-system analyses of the empirical missed-detection and false alarm-rates for the detection step (Theorem 2), and the empirical mean-square error for the subsequent channel estimation following the detection step (Theorem 3).
			
			\item [(iii)] Using the AMP large-dimensional outputs statistics, we provide expressions for the message
			missed-detection and false-alarm probabilities in near closed-form (a one-dimensional integral which is easily evaluated via Gauss-Chebyshev quadrature), as well as for the channel estimation mean-square error for the active detected users. 
			
			\item [(iv)] 
			We present insightful results on the achievable rate of beamformed ACK transmission in the DL. These results highlight the effectiveness of the proposed scheme for uRA, ensuring seamless connectivity and low latency in CF networks.
			
		\end{enumerate}
		
		We also compare with a previously proposed scheme for multicell networks, based on separated AMP at each RU and centralized message detection (see Section \ref{relatedworks}), which can be regarded as the present state-of-the-art, and show that our centralized approach yields superior performance. 
		
		Our novel (centralized) multisource AMP framework handles \emph{multiple} independent signal sources, each of which can be described by an arbitrarily different distribution. Interestingly, the multisource AMP is not significantly more complex than the previous multiple measurement vector approximated message passing (MMV-AMP) algorithm \cite{gerbelot2023graph}.  On the other hand, we note that our analysis is \emph{independent} of the previous studies. In particular, 
		the previous  MMV-AMP analysis \cite{Berthier20,gerbelot2023graph} refer to the convergence {\em in probability} of certain quantities  in the large-system limit. Our analysis yields convergence in the almost sure sense. Indeed, the latter is necessary to prove the almost sure convergence of the missed-detection and false-alarm rates, as well as the channel estimation error.  
		
	}

	\subsection{ Related Works}  \label{relatedworks}
	
	{ The uRA paradigm was proposed in  \cite{Polyanskiy2017} and motivated by ``Internet of Things'' applications,  
		where a huge population of users with very sporadic activity access at random a common channel. 
		In uRA, the receiver's task is to decode the list of {\em active} messages, i.e., the list of codewords present in the noisy superposition forming the received signal, without knowing a priori which user is transmitting. Since all users share the same codebook, the total number of users in the system can be arbitrarily large, as long as the number of active messages in each RACH slot is finite.
		As a matter of fact, it is immediate to recognize that any random access scheme in current wireless network standards is effectively an uRA scheme 
		(see  \cite{liva2024unsourced}).  Hence, uRA is the {\em appropriate} setting for the random access problem at hand.
		
		uRA has been intensively investigated in recent years from a coding/algorithmic view point (e.g., \cite{amalladinne2020coded,fengler2019sparcs,Fengler2021,Fengler2022,gao2022energy,liu2024coded,liu2024many,ebert2024multi,amalladinne2021unsourced,ebert2022coded}) and in terms of  information theoretic limits (e.g., \cite{Polyanskiy2017,gao2022energy,gao2024unsourced}). 
		While uRA  has not been extensively studied in the context of CF networks, it bears strong similarities with the {\em activity detection} (AD) problem \cite{liu2018massive1,liu2018massive2,Chen2018}. 
		In AD, each user is given a {\em unique} signature sequence, and the receiver aims to identify the list of active users. Simply put, {\em user} in activity detection plays the role of {\em message} in uRA. 
		However, it is important to note an important difference: in AD there is a fixed association between a user's signature sequence 
		and its set of large-scale fading coefficients (LSFCs), which describe the channel attenuation between the user and the RUs, since each signature sequence can only be transmitted from the location of the corresponding user.  In contrast, in uRA there is no such fixed association between codewords and LSFCs, since any message can be transmitted by any user at any location within the network coverage area. 	
		
		Various works have addressed the AD problem for multiantenna (MIMO) receivers. Notably, MMV-AMP algorithm \cite{javanmard2013state,Berthier20,pandit2020inference} is 
		used in \cite{liu2018massive1,liu2018massive2,Chen2018}, the covariance-based algorithm in \cite{Fengler2021}  (also analyzed in \cite{chen2021phase}), and the tensor-based modulation scheme in \cite{decurninge2020tensor} are well-known approaches. In particular, \cite{Fengler2021} shows that AMP performs well in the {\em linear regime}, where the signature block length $L$ scales linearly with the number of active users, and the number of antennas at the receiver $M$ is constant and typically small compared to $L$. In contrast, the covariance-based method achieves low activity detection error probability in the {\em quadratic regime}, where the number of active users scales as $L^2$, and the number of antennas $M$ scales slightly faster than $L^2$. Numerous works have followed \cite{Chen2018,Fengler2021}, and providing a comprehensive account here would be impractical.
		
		Some works have focused on the activity detection problem in CF systems. In particular,  \cite{Chen2019} proposes to run a separate AMP algorithm at each RU, followed by pooling the AMP outputs for centralized detection of the active users, while \cite{bai2022activity} considers a heuristic centralized AMP approach where observations from all RUs are jointly processed. 
		
		In this paper, we also explore a centralized AMP approach which is conceptually similar to that in \cite{bai2022activity}, but it can be rigorously analyzed in the finite (but large) dimensional regime.  More specifically, our location-based partitioned codebook system model reduces to the model in \cite{bai2022activity} by identifying each {\em user} in \cite{bai2022activity} with a {\em location} in our model, and restricting our location-based subcode to have a cardinality 1 (i.e., containing a single signature sequence).  However, as the system grows large this 
		yields a regime where the number of locations $U$ goes to infinity, while the size of each codebook remains finite (in fact, equal to 1).
		For this regime, a rigorous large-system analysis of the AMP is not known, and probably also does not hold unless under some restrictive conditions. Therefore, we refer to \cite{bai2022activity}  as {\em heuristic}. 
		
		Finally, in our numerical results, we compare our scheme to the separate AMP with centralized detection in \cite{Chen2019} 
		(shortly referred to as ``separate AMP'' scheme in the following),  suitably adapted to the uRA and CF settings. 
		This can be regarded as the de-facto state-of-the-art, since at present a random access scheme for CF user-centric networks
		is not specified in any existing standard, and current state-of-the-art schemes have addressed mostly the case of a single cell
		with a (concentrated antennas) MIMO receiver, and there is only a handful of works that have addressed 
		AD in the multicell case (e.g., \cite{Chen2019}).

	}

	\subsection{Organization}

	The paper is organized as follows: In Section~\ref{system}, we introduce the system model and the problem definition. Section~\ref{main_result} presents our main theoretical results. As applications of these results, Section~\ref{detection-and-estimation} delves into the analyses of the message detection algorithm and the subsequent channel estimation phase. In Section~\ref{cluster-formation-and-ACK}, we discuss cluster formation and beamformed ACH transmission.  Simulation results are presented in Section~\ref{Simresults}. We conclude the paper in Section~\ref{SecConc}. The proofs are given in the Appendix.

	\subsection{Notations} \label{notdef}
	We use the index set notation $[N] \eqdef \{1, \ldots, N\}$. Row vectors, column vectors, and matrices are denoted by boldfaced lower case letters, underlined bold-faced lower case letters, and boldfaced upper case letters, respectively (e.g., $\av$, $\underline{\av}$, and $\Am$).  The $i$th row and the $j$th column of $\Am$ are denoted by ${\av}_i$ and ${\underline\av}_j$, respectively. We use $\Am > \bf 0$ (resp. $\Am\geq  \bf 0$) to indicate that $\Am$ is positive definite (resp. positive semi-definite), 
	$(\cdot)^\transp$ and $(\cdot)^\herm$ to denote transpose and Hermitian transpose, and $\otimes$ to indicate the 
	Kronecker product. Also, $\Id_m$ indicates the $m \times m$ identity matrix. We denote the Frobenious norm of a matrix $\Am$ by $\| \Am \|_{\rm F} \eqdef \sqrt{ \trace( \Am^\herm \Am)}$.  We use the Matlab-like notation for which $\diag(\av)$ or $\diag(\underline{\av})$ is a diagonal matrix 
	with diagonal elements defined by the corresponding vector argument, and 
	${\rm diag}(\Am)$ returns a column vector of the diagonal elements of $\Am$. The multivariate circularly symmetric complex Gaussian distribution with mean $\mv$ and covariance $\Cm$ is denoted by $\mathcal{CN}(\mv,\Cm)$, while its density function is denoted by $\textswab{g}(\cdot \vert \mv, \Cm)$. The Bernoulli distribution with mean $\lambda$ is denoted by ${\rm Bern}(\lambda)$. We write $\av\sim P$  to indicate that the random vector $\av$ has a probability distribution $P$. We use the notation $\Am\sim_{\text{i.i.d.}} {\av}$ to indicate that the rows of the matrix $\Am$ are independently and identically distributed (i.i.d.) { realizations} of the random vector $\av$. Similarly, $\Am\sim_{\text{i.i.d.}}P$ indicates that the rows of $\Am$ are i.i.d. and have the probability distribution $P$. 
	$\mathbb E[\cdot]$ indicates the expectation with respect to the joint distribution of all the random variables appearing in the argument.

	\section{System Model and Problem Definition} \label{system}
	
\begin{figure}
		\centering	\includegraphics[width=0.6\linewidth]{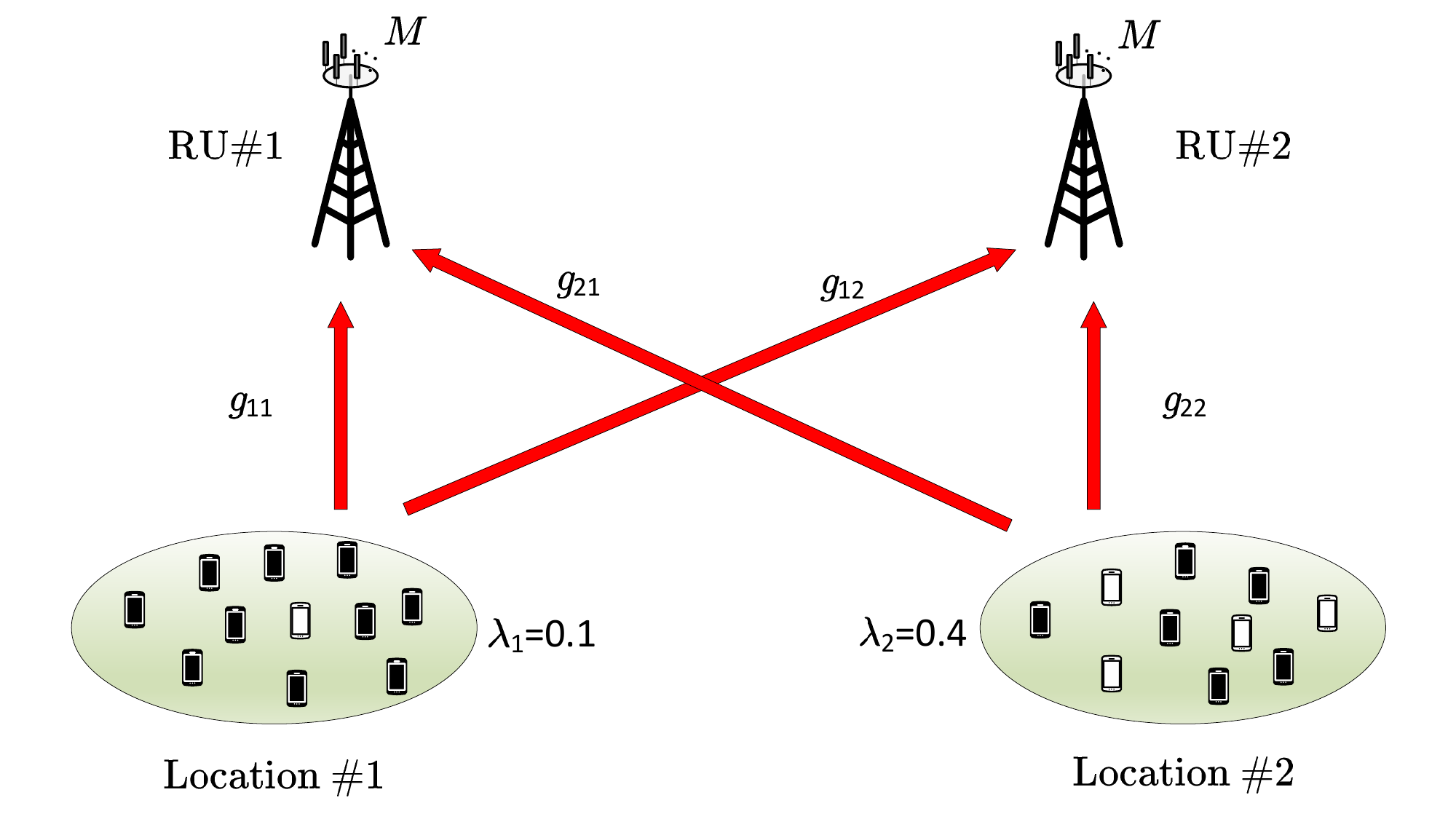}
		\caption{A simple $B = 2$, $U = 2$  toy model example of the proposed location-based approach.} 
		\label{fig1} 
	\end{figure}

	Consider a CF wireless network with $B$ radio units (RUs) with $M$ antennas each,  and a very large number of users geographically distributed in a certain coverage area $\Dc$.  In order to overcome the problem of the association between codewords and LSFCs, we propose a {\em location-based}  uRA approach, illustrated by the toy example of Fig.~\ref{fig1}.
	The coverage area $\Dc$ is partitioned into  $U$ zones, referred to as {\em locations} and denoted by $\Lc_1, \ldots, \Lc_U$ and we define the location-dependent LSFC profile $\{g_{u,b} : b \in [B]\}$ for location $\Lc_u$ such that the users in any given location have the { statistically} identical LSFC profile to all the RUs.
	
	Furthermore, following \cite{Polyanskiy2017,amalladinne2020coded,fengler2019sparcs,Fengler2021,Fengler2022,gao2022energy}, 
	an uRA codebook of block length $L$ and size $N$ is formed by a set of $N$ codewords arranged as the columns of 
	a matrix $\Sm \in \CC^{L \times N}$. With our location-based scheme, the codebook $\Sm$ is partitioned into
	$U$ subcodes, i.e., $\Sm = [\Sm_1, \ldots, \Sm_U]$ such that  $\Sm_u\in \CC^{L\times N_u}$ 
	and $\sum_u N_u = N$. 
	The random access users in location $\Lc_u$ are aware of their location and 
	make use of the corresponding codebook $\Sm_u$, i.e., any such user 
	in order to transmit message $n \in [N_u]$, sends the corresponding 
	codeword $\underline{\sv}_{u,n}$ (i.e., the $n$-th column of $\Sm_u$). We consider a random coding ensemble where each  $\underline{\sv}_{u,n}$ is generated with i.i.d. elements $\Nc(0,1/L)$.
	We also assume that each codeword of $\Sm_u$ may be chosen independently with probability $\lambda_u \in (0,1)$, where $\lambda_u$ reflects the random activity of users in location $\Lc_u$.\footnote{In practice, two users in the same location may choose the same codeword. In this case, the message is active and the corresponding channel vector is the sum of the independent identically distributed channel vectors. This presents no problem for the message detection and channel estimation algorithm presented in Section \ref{detection-and-estimation}, although the random access protocol will fail to reply with a individual ACK to the two users since it detects a single active message. This type of collisions are easily handled 
		by standard collision resolution mechanisms, and shall not be considered in this work.}
	
	The signal collectively received at the $F = BM$ RUs antennas over the $L$ symbols of the RACH slot is given by the $L \times F$ (time-space) matrix 
	\begin{eqnarray} 
		\Ym & = & \sum_{u=1}^U \Sm_u \Xm_u  + \Nm, \label{y_model}
	\end{eqnarray}
	where $\Nm\sim_{\text{i.i.d.}}\nv$ stands for the noise matrix for some RV $\nv\in\C^{\bm}$ and $\Xm_u \in \CC^{N_u \times F}$ is the matrix containing (arranged by rows) the aggregate channel vectors corresponding to 
	messages at location $\Lc_u$. If the $n$-th message is not active (i.e., not transmitted), then the $n$-th row $\xv_{u,n}$ of $\Xm_u$ is identically zero.  { If the $n$-th message is active (i.e., transmitted), then $\xv_{u,n}$ coincides with the 
		(unknown at the receiver) channel vector corresponding to the $n$-th active messages.} 
	Hence, we have $\Xm_u\sim_{\text{i.i.d.}}\xv_u$ with
	\begin{equation}
		\xv_u\sim {\rm a}_u\hv_u \; ,   \label{xu_distrib}
	\end{equation}
	{where $\hv_u\in \C^{F}$ is a RV distributed as the channel vector from any transmitter in location $\Lc_u$ and the $F$ receiving antennas at the RUs,  statistically independent of ${\rm a}_u\sim {\rm Bern}(\lambda_u)$. 
		In Section~\ref{fading_model}, we present a specific fading model for $\hv_u$.}
	
	For each $u\in [U]$, let $\widetilde N_u$ denote the random number (binomial $(\lambda_u, N_u)$ distributed) of active messages at location $u$ and let $\Pm_u\in \{0,1\}^{\widetilde N_u\times N_u}$ denotes a ``fat'' projection matrix that eliminates zero-rows from the 
	input signal $\Xm_u$. { Notice that $\Xm_u=\Pm_{u}^\top\Pm_u\Xm_u$.} Then, we introduce 
	\begin{align}
		\underline{\av}_u&\eqdef {\rm diag}(\Pm_u^\top \Pm_u) ~~~\text{(True Message Activity Vector)}\label{tsav}\\
		\Hm_u&\eqdef\Pm_u\Xm_u ~~~~~~~~~~~\text{(True Channel Matrix)}\label{tcm}\;.
	\end{align}
	The goal of the joint message detection and channel estimation algorithm is to recover the list of active messages (i.e., the binary vectors $\underline{\av}_u$) and estimate the corresponding channels $\Hm_u$  for each $u\in[U]$.

	{ \subsection{Channel Model Assumptions}\label{fading_model}
		
		While our analysis for joint message detection and channel estimation applies to any random vector $\hv_u \in \CC^{\bm}$ with the mild assumption that moments $\mathbb{E}[\Vert \hv_u\Vert^p]$ exist for all $p \in \NN$, it is important to develop an insightful fading model for $\hv_u$. 
		In the sequel, we make the following assumptions.
		
		\subsubsection{Nominal LSFC Assumption} 
		In general, the LSFCs are given by some distance-dependent 
		pathloss function indicating the signal power reduction along the propagation path \cite{toeltsch2002statistical,asplund2006cost}. In contrast, in several works treating the information theoretic analysis of wireless networks, the co-called 
		``Wyner model'' \cite{wyner1994shannon,shamai1997information,xu2011accuracy} 
		has been widely used. In the Wyner model, the LSFCs take only three values, 
		1 for the paths within the same cell (users within each cell to their base station), $\wp \in [0,1]$ for the path across adjacent cells, 
		and 0 for the paths across non-adjacent cells. We take an intermediate step and consider that users in the same location $\Lc_u$ have
		the same LSFC, indicated as $g_{u,b}$ to each $b$-th RU. The collection of LSFCs $\{g_{u,b} : b \in [B]\}$ is referred to as LSFC profile and it is common to all users in location $\Lc_u$ for $u \in [U]$. 
		
		This assumption, referred to as ``nominal'' LSFC assumption, is justified by the following rationale:
		\begin{itemize}
			\item We use the nominal LSFC assumption to devise a suitably simple {\em denoising function} in the proposed AMP algorithm. 
			Our analysis is valid even when the true statistics of the user channels does not obey the assumption. In this case, the denoising function
			derived from the nominal LSFC assumption is mismatched. Nevertheless, the analysis is still fully valid. 
			\item In our follow-up work \cite{elenisparcs} (published after the first submission of the present work), we considered the general case of users arbitrarily distributed  over the coverage area. In this case, the nominal LSFC assumption corresponds to the case of co-located users at given positions (e.g., centroids) of the locations. For uniformly distributed users, in \cite{elenisparcs} we have compared the matched case, where the AMP denoising function uses the true user distribution, and the mismatched case, where the AMP denoising function assumes the nominal LSFC values, 
			and we have verified that for typical pathloss functions and CF network layouts, with an appropriate choice of the locations, 
			the performance loss incurred by mismatch is virtually negligible. 
			\item Under the nominal LSFC assumption, the AMP output statistics are Gaussians under both the active message and the inactive message assumptions,  and the analysis of the missed-detection and false-alarm probabilities yields elegant and easy to evaluate near-closed form formulas, using the theory of quadratic forms of complex Gaussian RVs (see Appendix \ref{laplace-inversion}). 
			\item In this context, we find more appropriate to favor clean and (near-)closed form results based on a generalization of the Wyner model, that has a long tradition in information theory, rather than focusing on overly specific pathloss models that clutter the results witout bringing any funamental new insight into the problem. 
		\end{itemize}
		
		\subsubsection{Rayleigh Fading Assumption}
		We assume that the small-scale fading coefficients between any user and any RU antenna are i.i.d. $\sim \Cc\Nc(0,1)$ random variables (independent Rayleigh fading). Under this assumption, the channel vector between a user in $\Lc_u$ and the $M$-antenna array of RU $b$ follows $\Cc\Nc(\zerov, g_{u,b} \Id_M)$. While this is a common assumption, the model can be further generalized by considering Rician fading, antenna correlations, and other effects. 
		
		As a result of the above fading assumptions, the aggregated channel vector from a user in $\Lc_u$ to the $F \eqdef BM$ antennas of the RUs is a $1 \times F$ Gaussian vector distributed as
		\begin{equation} 
			\hv_u  =  [ \hv_{u,1}, \hv_{u,2}, \ldots, \hv_{u,B} ] \sim \Cc\Nc(\zerov, \Sigmam_u)\;,  \label{channel_u}
		\end{equation}
		where $\Sigmam_u$ is a $F \times F$ diagonal matrix given by
		\begin{equation}
			\Sigmam_u\eqdef {\rm diag}(g_{u,1},g_{u,2},\cdots,g_{u,B})\otimes \Id_M\;.\label{bdiag}
		\end{equation}
		
	}
	
	\section{The AMP Algorithm and Its High Dimensional Analysis} \label{main_result}
	
	We begin by formulating the AMP algorithm for the estimation of $
	\{\Xm_u : u \in [U]\}$, given $\{\Ym, \Sm_u: u \in [U]\}$. Subsequently, a comprehensive high-dimensional analysis of the AMP is presented. We conclude the section by showing the asymptotic consistency of the decoupling principles featured in the AMP analysis and the RS (replica-symmetric) calculation.
	
	For each $u \in[U]$, let $\Xm_u^{(t)}$ denote the AMP estimate of $\Xm_{u}$ at iteration step $t=1,2,\ldots$. We initiate the process with initial ``guesses'' that follow $\Xm_{u}^{(1)}\sim_{\text{i.i.d.}}\xv_u^{(1)}$ for some arbitrary RVs $\xv_u^{(1)} \in \C^{1\times \bm}$ (independent for each $u\in [U]$) with a bounded moment $\mathbb E[\Vert\xv_u^{(1)}\Vert^p]$ for each $p\in \NN$. For example, we can begin with $\Xm_{u}^{(1)}={\matr 0}$. For iteration steps $t=1,2,\ldots ,T$, the algorithm computes
	\begin{subequations}
		\label{AMP_alg}
		\begin{align}
			\matr \Gamma_u^{(t)}&=\Sm_u \Xm_u^{(t)}-
			{\alpha_u}\Zm^{(t-1)}\Qm_u^{(t)} \\
			\Zm^{(t)}&=\Ym-\sum_{u=1}^{U}\matr \Gamma_u^{(t)}\\
			\Rm_u^{(t)}&=\Sm_u^{\herm}\Zm^{(t)}+{\Xm}_u^{(t)}\\
			\Xm_u^{(t+1)}&=\eta_{u,t}(\Rm_u^{(t)})\;,  \label{ziopera}
		\end{align}	
	\end{subequations}
	with $\Zm^{(0)}=\matr 0$. Here, $\alpha_u\eqdef N_u/L$ and $\eta_{u,t}(\cdot):\C^{F}\to\C^{F}$ is an appropriately defined deterministic and $(u,t)$-dependent {\em denoiser} function with its application to a matrix argument, say $\Rm\in \CC^{N\times \bm}$, is performed row-by-row,~i.e., 
	\[\eta_{u,t}(\Rm)=
	[\eta_{u,t}({\rv}_1)^\top,
	\eta_{u,t}({\rv}_2)^\top,\cdots,
	\eta_{u,t}({\rv}_N)^\top]^\top\;.\]
	The recursive relation for updating the (deterministic) matrix $\Qm_u^{(t+1)}$ is given by
	\begin{equation}
		\Qm_u^{(t+1)} = \mathbb{E}[\eta_{u,t}'(\xv_{u}+\matr\phi^{(t)})] \quad \forall t\in[T], \label{onsager}
	\end{equation}
	where $\{\matr \phi^{(t)}\}_{t\in[T]}$ is a Gaussian process (see Definition~\ref{SEdef}) independent of the random vector~$\xv_{u}$ and $\eta_{u,t}'(\rv)$ denotes the Jacobian matrix of $\eta_{u,t}(\rv)$, i.e.\footnote{For a complex number $r = x + {\rm i}y$, the complex (Wirtinger) derivative is defined as $\frac{\partial }{\partial r}=\frac 1 2( \frac{\partial }{\partial x}-{\rm i}\frac{\partial }{\partial y})$.}, 
	\begin{equation} 	
		[\eta_{u,t}'({\rv})]_{ij}=\frac{\partial [\eta_{u,t}({\rv})]_{j}}{\partial r_i}\quad \forall i,j\in[F].  \label{jacobian}
	\end{equation}
	

	
	\begin{defi}[State Evolution]\label{SEdef}
		Let { \( \{ \matr{\psi}_u^{(t)} \in \mathbb{C}^{1\times F}\}_{t \in [T]}\) and} \(\{\matr{\phi}^{(t)} \in \mathbb{C}^{1\times F}\}_{t \in [T]}\) be independent and arbitrary zero-mean (discrete-time) Gaussian processes and $\{\matr{\psi}_u^{(t)}\}$ are independent for each $u\in[U]$. Their two-time covariances $\Cm_{u}^{(t, s)} \eqdef \mathbb E[(\matr\psi_u^{(t)})^\herm\matr\psi_u^{(s)}]$ and $\Cm^{(t,s)} \eqdef\mathbb E[(\matr\phi^{(t)})^\herm\matr\phi^{(s)}]$  for all $t,s\in[T]$ are constructed recursively according to 
		\begin{align}
			\Cm_{u}^{(t, s)}&=\alpha_u \mathbb E[(\xv_{u}-\xv_u^{(t)})^\herm ( \xv_{u}-\xv_u^{(s)})]\\
			\Cm^{(t,s)}&=\mathbb E[\nv^\herm\nv]+\sum_{u=1}^{U}\Cm_{u}^{(t, s)} \label{SE}\;,
		\end{align}
		where $\alpha_u= N_u/L$ and for $(u,t)\in[U]\times [T]$  we define the RVs $\xv_u^{(t+1)} =  \eta_{u,t}(\xv_{u}+\matr \phi^{(t)})$ (independent of $\xv_u^{(1)}$). 
	\end{defi}
	For the special case, $\bm=1$ and $U=1$, \eqref{SE} coincides with the classical SE formula~\cite{Bayati}. Note also that \eqref{SE} refers to the general two-time (i.e. $(t,s)$) characterization.  
	{ \subsection{The High-Dimensional Equivalence}
		We use the notion of concentration inequalities in terms of $\mathcal L^p$ norm:
		we write
		\begin{equation}
			a= \mathcal{O}(1) \label{opnotation}  
		\end{equation}	
		to imply that for each $p\in \NN$ there is a constant $C_p$ such that $\Vert {a} \Vert_{\mathcal L^p}\leq C_p$ with $\Vert {a}\Vert_{\mathcal L^p}\eqdef (\mathbb E\vert {a}\vert^p)^\frac {1}{p}$. In general, for any deterministic $\kappa>0$ (e.g., $\kappa=\sqrt{L},1/L$, etc.) we write, with some 
		abuse of notation,
		\begin{equation}
			\Am= \Op{\kappa} \quad \text{if} \quad \frac{1}{\kappa}\Vert\Am\Vert_{\rm F}=\Op{1}.\label{gopnotation}
		\end{equation}
		Moreover, we say ${\Am}\in \CC^{N\times F}$ is a high-dimensional equivalent of $\widehat\Am$, if $\widehat{
			\Am} -\Am= \Op{1}$ and denote this
		\begin{equation}\label{simeq}
			\widehat{\Am}\simeq \Am\;.
		\end{equation}
		In particular, if $\widehat{\Am}\simeq \Am$, then it follows from the Borel-Cantelli lemma that for any \emph{small} constant $\epsilon>0$ we have almost sure (a.s.) convergence as $N\to\infty$ 
		\begin{equation}
			\frac{1}{N^{\epsilon}}\Vert \widehat{\Am}-\Am \Vert_{\texttt{F}} \overset{a.s.}{\rightarrow} 0. \label{ascon}
		\end{equation}
		\begin{asmp} [Model Assumption]\label{as1}
			Consider the observation model \eqref{y_model}. Let the matrices $\{\{\Sm_u,\Xm_u\}_{u\in[U]},\Nm\}$ be mutually independent. For all $u\in[U]$, let $\Sm_u\sim_{\text{i.i.d.}}\Nc(0,\Id/L)$, $\Xm_{u}\sim_{\text{i.i.d.}}\xv_u$ and $\Nm\sim_{\text{i.i.d.}}\nv$ for some arbitrary RVs $\xv_u=\mathcal O(1)$ and $\nv=\mathcal O(1)$. Let $F,U$ and the aspect ratios $\alpha_u=N_u/L$ are all fixed w.r.t $L$.
		\end{asmp}

		It is important to note that the family of random variables $\Op{1}$ encompasses a wide range of distributions with heavy exponential tails. Indeed, $a^D = \Op{1}$ for a sub-Gaussian RV $a$ \cite{vershynin2018high} for any \emph{large} (constant) $D$.}
	
	\begin{thm}\label{Th1}
		Let the denoiser functions $\eta_{u,t}$ be differentiable and Lipschitz-continuous for all $(u, t) \in [U] \times [T]$. Suppose Assumption~\ref{as1} holds. Let $\matr \Theta_u\eqdef \Sm_u\Xm_u$. Then, we have for any $(u,t)\in[U]\times [T]$ 
		\begin{align}
			{  \matr \Gamma_u^{(t)}}&\simeq{ \matr \Theta_u+ \matr \Psi_{u}^{(t)}}\\
			\Rm_{u}^{(t)}&\simeq  \Xm_{u}+\matr \Phi_{u}^{(t)}\label{th1result}\;,
		\end{align}where $\matr\Psi_u^{(t)}\sim_{\text{i.i.d.}}\matr \psi_u^{(t)}$  and $\matr\Phi_u^{(t)}\sim_{\text{i.i.d.}}\matr \phi^{(t)}$ with the Gaussian processes $\matr \psi_u^{(t)}$
		and  $\matr \phi^{(t)}$ as in Definition \ref{SEdef} and { the elements in the sets $\{\matr\Psi_u^{(t)},\matr\Theta_u\}_{u\in [U]}$ and $\{\matr\Phi_u^{(t)},\Xm_u\}_{u\in [U]}$ are all mutually independent.}
		\begin{proof}
			See Appendix~\ref{Sketchofproof}   
		\end{proof}
	\end{thm}
	{ While our proof heavily relies on the assumption that the entries of $\Sm_u$ are i.i.d. zero-mean Gaussian, we expect that the typical universality phenomena of the AMP algorithms, which apply to random matrices with i.i.d. entries from a general sub-Gaussian family \cite{Bayati2015,wang2024universality} could apply to Theorem~\ref{Th1}. E.g., we expect that Theorem~1 could still hold (perhaps under a weaker notion than the finite sample notion ``$\simeq$'') when the entries of $\Sm_u$ are i.i.d. and $(\Sm_{u})_{ij} \in \frac{1}{\sqrt{2L}} \{\pm 1 +\pm {\rm i}\}$.
		
		From Theorem~\ref{Th1} we write the high-dimensional representations 
		\begin{align}
			\Zm^{(t)}&\simeq \Nm-\sum_{u\in[U]}\matr\Psi_u^{(t)} \label{pippa} \\
			\Xm_{u}^{(t+1)}&\simeq \eta_{u,t} (\Xm_{u}+\matr \Phi_{u}^{(t)})\;,\label{etause}
		\end{align}
		where  \eqref{pippa} follows from  $\Ym=\sum_{u\in [U]}\matr\Theta_u+\Nm$ and \eqref{etause} follows from the Lipschitz assumption of $\eta_{u,t}$. Thus, we have a complete high-dimensional representation of the AMP dynamics \eqref{AMP_alg}. While for the aforementioned application it is sufficient to have the high-dimensional representation of $\Rm_u^{(t)}$, we refer the reader to applications of the general complete high-dimensional representation of AMP dynamics in \cite{Bayati11,Berthier20,loureiro2021learning, gerbelot2022asymptotic, Cakmakisit24}.}

	From \eqref{etause} we can analyze the mean-square error (MSE) matrix of AMP at a given iteration step:
	\begin{cor}\label{Cormse}
		Let the premises of Theorem~\ref{Th1} hold.
		For any $(u,t)\in[U]\times[T+1]$  we have 
		\begin{align}
			&\frac{1}{N_u}(\Xm_u-\Xm_u^{(t)})^\herm (\Xm_u-\Xm_u^{(t)})\nonumber\\&=\mathbb E[(\xv_u-\xv_u^{(t)})^\herm (\xv_u-\xv_u^{(t)})]+\Op{L^{-\frac 1 2}}\;,
		\end{align}
		where the random vectors $\xv_u^{(t)}$ as in Definition \ref{SEdef}.
		\begin{proof}
			See Appendix~\ref{Proof_Cormse}.
		\end{proof}
	\end{cor}
	From Corollary~\ref{Cormse}, we have the convergence as $L\to \infty$
	\begin{align}
		&\frac{1}{N_u}(\Xm_u-\Xm_u^{(t)})^\herm (\Xm_u-\Xm_u^{(t)})\nonumber \\
		&\overset{a.s.}{\rightarrow}
		\mathbb E[(\xv_u-\xv_u^{(t)})^\herm (\xv_u-\xv_u^{(t)})]\;.
	\end{align}
	We can devise the denoiser function $\eta_{u,t}(\cdot)$ to minimize $\mathbb E[(\xv_u-\xv_u^{(t)})^\herm (\xv_u-\xv_u^{(t)})]$. This yields the minimum MSE (MMSE) estimator for $\xv_u$ from the observation $\rv = \xv_u + \matr \phi^{(t)}$, i.e., we let
	\begin{equation}
		\eta_{u,t}(\rv) = \mathbb E[\xv_u\mid \rv= \xv_u + \matr \phi^{(t)}]\;. \label{opt-denoiser}
	\end{equation}
	For the application at hand, from  \eqref{xu_distrib} we have   $\xv_u={\rm a}_u\hv_u$ with ${\rm a}_u\sim {\rm Bern}(\lambda_u)$. This yields the following useful decomposition of the posterior mean.
	\begin{rem}\label{MAPMSE}
		Consider the conditional mean $\eta(\rv)\eqdef\mathbb E[{\rm a}\hv\vert\rv]$  with ${\rm a}\sim {\rm Bern}(\lambda)$, $\hv$, and observation $\rv$ defined as a random vector jointly distributed with ${\rm a}$ and $\hv$. Then,  
		\begin{align}
			\eta(\rv)&=\frac{\lambda\int {\hv}~p(\rv,\hv\vert{\rm a}=1)\;{\rm d}\hv}{p(\rv)}
			\nonumber \\
			&=\frac{\lambda\int {\hv}~p(\rv,\hv\vert{\rm a}=1)\;{\rm d}\hv}{\lambda p(\rv\vert {\rm a}=1)+(1-\lambda)p(\rv\vert{\rm a}=0)} \nonumber \\
			&=\frac{\mathbb E[\hv\mid\rv,{\rm a}=1]}{1+\Lambda_{\rm map}(\rv)}\;,\label{etamap}
		\end{align}
		where we have defined $
		\Lambda_{\rm map}(\rv)\eqdef\frac{1-\lambda}{\lambda}\frac{p(\rv\vert{\rm a}=0)}{p(\rv\vert{\rm a}=1)}\;.$
		Note that $\Lambda_{\rm map}(\rv) \underset{{\rm a=1}}{\overset{{\rm a=0}}{\gtreqless}}1$ is
		the maximum a posteriori probability (MAP) decision test of ${a}$. \hfill $\lozenge$
	\end{rem}
	For example, let $\rv= {\rm a}{\hv} + \matr \phi$ where ${\rm a}\sim {\rm Bern}(\lambda)$,
	$\hv~\sim\mathcal{CN}(\matr 0,\Sigmam)$, and $\matr \phi\sim \mathcal {CN}(\matr 0,\Cm)$ are mutually independent. Then, $p(\rv | {\rm a}) = \textswab{g}(\rv\vert \zerov, {\rm a}\Sigmam + \Cm)$ and thus \eqref{etamap} is obtained explicitly using
	\begin{align}
		\mathbb E[\hv\mid\rv,{\rm a}=1]&=\rv(\Sigmam+\Cm)^{-1}\Sigmam\\
		\Lambda_{\rm map}(\rv)&=\frac{1-\lambda}{\lambda}\frac{\vert \Sigmam +\Cm\vert}{\vert \Cm \vert }{\rm e}^{-\rv(\Cm^{-1}-(\matr\Sigma+\Cm)^{-1})\rv^\herm}\;.
	\end{align}
	Moreover, after some simple algebra (omitted for the sake of brevity), 
	the Jacobian matrix $\eta'(\rv)$ (see \eqref{jacobian}) takes on the appealing compact form
	\begin{equation}	\eta'(\rv)=\frac{(\Sigmam+\Cm)^{-1}\Sigmam}{1+\Lambda_{\rm map}(\rv)}
		+\Lambda_{\rm map}(\rv)\Cm^{-1}\eta(\rv)^\herm \eta(\rv)\;. \label{etaprime}
	\end{equation}
	Also, from the fact that  $x{\rm e}^{-x}\leq 1,x\in \RR$, we have $\Vert \eta'(\rv) \Vert_{\rm F}\leq C$ for some constant $C$. 
	Hence, in this case $\eta(\rv)$ is Lipschitz continuous.  Note that the above derivation is general and must be the particularized for indices $u$ and $t$ to be used in  \eqref{AMP_alg}. 
	
	\subsection{Asymptotic Consistency with the Replica-Symmetric Calculation}\label{Replicaresults}
	Given that the denoiser functions $\eta_{u,t}$ correspond to the posterior mean estimators as described in  \eqref{opt-denoiser}, we can express the concentration of the MSE matrix at a given iteration step as
	\begin{align}
		&\frac{1}{N_u}(\Xm_u-\Xm_u^{(t+1)})^\herm (\Xm_u-\Xm_u^{(t+1)})\nonumber \\
		&\quad\overset{a.s.}{\rightarrow}{\rm mmse}\left(\xv_u\mid\xv_u+\zv(\Cm^{(t,t)})^{\frac 1 2}\right)\;.
	\end{align}
	Here, $\xv_u$ and $\zv\sim \mathcal {CN}(\matr 0;\Id_\bm)$ are independent and  we define the mmse covariance matrix as
	\begin{equation}
		{\rm mmse}(\xv\vert \yv)\eqdef\mathbb E[(\xv-\mathbb E[\xv\vert\yv])^\herm (\xv-\mathbb E[\xv\vert\yv])]\;.
	\end{equation}
	Notice that by particularizing the denoiser functions $\eta_{u,t}$ to the posterior mean estimators as in  \eqref{opt-denoiser} we have from the state-evolution equation \eqref{SE} that  the fixed point of $\Cm^{(t,t)}$, denoted as $\Cm^\star$, is the solution to the (matrix-valued) fixed-point equation 
	\begin{equation}
		\Cm^\star =\mathbb E[\nv^\herm\nv]+ \sum_{u=1}^{U} \alpha_u \, \text{mmse}\left(\xv_u \mid \xv_u + \zv(\Cm^\star)^{\frac{1}{2}}\right)\;.\label{cstar}
	\end{equation}
	
	\begin{example}\label{conmmse}
		Initialize the AMP algorithm \eqref{AMP_alg} with the choice $\Xm_u^{(0)}=\matr 0$ (for each $u\in[U]$). Let $\xv_u$ have a Bernoulli-Gaussian distribution (see \eqref{channel_u}) with a \emph{diagonal} covariance $\Sigmam_u$. Then, we have
		\begin{align}
			&\lim_{t\to \infty}{\rm mmse}\left(\xv_u\mid\xv_u+\zv(\Cm^{(t,t)})^{\frac 1 2}\right)\nonumber  \\&\quad ={\rm mmse}\left(\xv_u \mid \xv_u + \zv(\Cm^\star)^{\frac{1}{2}}\right)\;,
		\end{align}
		where $\Cm^\star$ satisfies
		the solution to the fixed-point equation \eqref{cstar}. Proving the uniqueness of the solution $\Cm^\star$ is relatively non-trivial we leave it as an open problem. A self-contained proof of the above statement is provided in Appendix~\ref{Prop1_proof}.  \hfill $\lozenge$
	\end{example}
	
	Subsequently, we derive through the  RS (Replica Symmetry) ansatz that the high-dimensional limit of the (exact) minimum MSE (MMSE)  coincides with ${\rm mmse}\left(\xv_u \mid \xv_u + \zv(\Cm^\star)^{\frac{1}{2}}\right)$.
	
	In general, we are interested  in the high dimensional limits of the normalized input-output mutual information and the MMSE which are defined as 
	\begin{align}
		\mathcal I&\eqdef \lim_{L\to\infty}\frac 1 L \mathcal I(\Xm;\Ym\vert \Sm)
		\\
		\matr \chi_u&\eqdef \lim_{L\to\infty}\frac{1}{N_u}{\rm mmse}(\Xm_{u}\vert\Ym,\Sm)\;,
	\end{align}
	where for short we write $\Xm \equiv[\Xm_1^\top,\Xm_2^\top,\ldots,\Xm_U^\top ]^\top$ and, as defined before, $\Sm\equiv[\Sm_1,\Sm_2,\cdots,\Sm_U]$, and the output matrix $\Ym$ is given in \eqref{y_model}. 
	Here, $\mathcal I(\cdot;\cdot)$ stands for the input-output mutual information in nats and we define the $\bm\times \bm$ MMSE covariance matrices (for each $u\in [U]$) 
	\begin{align}
		&{\rm mmse}(\Xm_{u}\vert\Ym,\Sm)\nonumber\\
		&=\mathbb E_{\Ym,\Xm}[(\Xm_{u}-\mathbb E[\Xm_{u}\vert \Ym,\Sm])^\herm(\Xm_{u}-\mathbb E[\Xm_{u}\vert \Ym,\Sm])]\;, 
	\end{align}
	where expectation is with respect to $\Xm$ and $\Ym$ for given $\Sm$ (here $\Sm$ plays the role of
	the so-called quenched disorder parameters in statistical physics \cite{tanaka2002statistical,guo2003multiuser,bereyhi2019statistical}). 
	We have computed these limiting expressions by means of the {\em replica-symmetric} (RS) ansatz \cite{tanaka2002statistical,guo2003multiuser,bereyhi2019statistical} and have the following claim, 
	dependent on the validity of the RS ansatz.
	\begin{claim}  \label{RS-claim}
		Suppose Assumption ~\ref{as1} holds. { Also let $\Nm\sim_{\text{i.i.d.}}\Nc(\matr 0,\Cm_{\rm n})$ with $\Cm_{\rm n}>\matr 0$.} Then, we have
		\begin{align}
			\mathcal I&=
			\sum_{u=1}^{U}{\alpha_u}\mathcal I \left(\xv_u;\xv_u+\zv(\Cm^\star)^{\frac 1 2}\right)\nonumber \\
			&~+{\rm tr}(\Cm_{\rm n}(\Cm^\star)^{-1})+\ln\frac{\vert\Cm^\star\vert}{\vert{\rm e}\Cm_{\rm n}\vert}\label{mutualinfo}\\
			\matr\chi_u&={\rm mmse}\left(\xv_u\mid\xv_u+\zv(\Cm^\star)^{\frac 1 2}\right)\;,\label{mmse}
		\end{align}
		where $\zv\sim \mathcal {CN}(\matr 0;\Id)$ is independent of $\xv_u$ and $\Cm^\star$ satisfies the solution of the fixed-point equation \eqref{cstar}. If the solution $\Cm^\star$ is not unique, we choose the solution minimizing \eqref{mutualinfo}. 
		\hfill $\square$
	\end{claim}
	
	For the special case $F=1$ and $U=1$ the formula coincides with \cite{guo2003multiuser}. In Appendix~\ref{free_energy} we derive the RS prediction of the mutual information formula \eqref{mutualinfo} which coincides with the ``free energy'' of statistical physics up to an additive constant. Similarly, one obtains the MMSE predictions in \eqref{mmse}. Specifically, by following the arguments in \cite[Appendix~E]{Tulino13}, one can introduce an auxiliary ``external field'' to the prior $p_{u}(\xv_u)\to p_{u}(\xv_u)\exp(\xv_u\Hm\xv_u^\herm)$ and compute the \emph{modified} free energy to obtain the desired input-output \emph{decoupling principle}, i.e., $(\xv_u;\xv_u+\matr \phi)$, in the MMSE (see \cite{guo2003multiuser} for the notion of decoupling principle). 
	
	Under the validity of the (non-rigorous) RS ansatz and assuming that $\Cm^\star$ in \eqref{cstar} has a unique solution, the AMP algorithm \eqref{AMP_alg}  with the denoisers \eqref{opt-denoiser} yields (upon the convergence as $t\to \infty$, see Remark~\ref{conmmse}) in the large system limit the Bayesian optimal estimate in Claim \ref{RS-claim}.
	
	{ Note also that if $\Cm^\star$ is not unique, we chose the solution of $\Cm^\star$ that minimizes the mutual information $\mathcal I$. In this case, $\Cm^{(t,t)}$ (from the state-evolution) might converge to a solution such as $\widetilde \Cm$, where $\widetilde \Cm\geq \Cm^\star$. 
		
		It is also worth noting that the RS result for the special case $F = 1$ and $U = 1$ has been rigorously proved in \cite{reeves2019replica,barbier2020mutual} and we expect that by applying the proof strategy of \cite{barbier2020mutual} to our general model the rigorous proof of the RS result can be obtained.}
	\section{Message Detection And Channel Estimation for uRA in Cell-Free Systems} \label{detection-and-estimation}
	We next apply the AMP algorithm in \eqref{AMP_alg} and its high-dimensional analysis to the uRA problem described in Section \ref{system}.  
	From Theorem~\ref{Th1}, we write the for the last iteration step $t=T$
	\begin{equation}
		\Rm\simeq \Xm_u+\matr\Phi_u\;,
	\end{equation}
	where, throughout this section, we set $\Rm_{u}\equiv\Rm_{u}^{(T)}$, $\matr\Phi_{u}\equiv\matr\Phi_{u}^{(T)}$ and $\Cm\equiv\Cm^{(T,T)}$ for the sake of notational compactness. Furthermore, we restrict our attention to the signal model for the rows of $\Xm_u$ (for each $u\in[U]$) as 
	\begin{equation}
		\Xm_{u}\sim_{\text{i.i.d.}}{\rm a}_{u}\hv_{u}\label{DCP}\;,
	\end{equation}
	with ${\rm a}_{u}\sim {\rm Bern}(\lambda_u)$ being independent of { the RV $\hv_u\in \C^{\bm}$ and $\hv_u=\Op{1}$. We assume that the model parameters describing the distribution of $\hv_u$ are all known at the receiver. For example, in the case of the fading model in \eqref{channel_u}, i.e., $\hv_u\sim \Nc(\matr 0,\Sigmam_u)$, the covariances $\{\matr \Sigma_u\}$ are all known at the receiver. } 
	
	\subsection{Message Detection}

	In uRA, the first task of the centralized receiver is to produce an estimate of the list of the active messages, i.e., 
	the list of columns of $\Sm_u$ that have been transmitted for each location $\Lc_u$. {
		Equivalently, this consists of estimating vector of binary message activities $\underline\av_{u}\in \{0,1\}^{N_u}$ 
		defined in \eqref{tsav}.}
	
	In random access problems, missed-detection  (an active message detected as inactive) and false-alarm
	(an inactive message detected as active) events play generally different roles. 
	In fact, in the presence of a {\em false-alarm} (FA) event, the system assumes that some user requests access, while in reality such user does not exist.  
	In this case, according to the uRA protocol sketched in Section \ref{intro},
	some transmission resource in the DL is wasted since the system sends an ACK message to an inexistent user, 
	and eventually will be released after some timeout.
	In the presence of a {\em missed-detection} (MD) event, some user requesting access is not recognized by the system. 
	In this case, it will receive no ACK in the DL will try again in a subsequence RACH slot after some timeout. 
	Depending on whether resource waste or delayed access is the most stringent system constraint, the roles of FA and MD are clearly asymmetric.  Therefore, minimizing the {\em average} error probability is not operationally meaningful. 
	
	With this motivation in mind, we consider a Neyman-Pearson detection approach \cite{poor1998introduction},  operating at some desired point of the MD--FD tradeoff curve. { 
		Firstly, from the high-dimensional representation \eqref{DCP}, we have the convergence of each row of $\Rm_u$ as (see Appendix~\ref{proof_cordecop})
		\begin{equation}
			\rv_{u,n}\overset{{\mathcal L}^p}{\rightarrow} \underbrace{\xv_{u,n}+\matr\phi_{u,n}}_{\eqdef \widetilde\rv_{u,n}} \label{dPT}
		\end{equation}
		as $L\to \infty$ for any fixed $p\in\NN$. Here $\overset{{\mathcal L}^p}{\rightarrow}$ stands for the convergence RV in the $\mathcal L^p$ norm. We then devise the decision test function according to the statistics of $\widetilde\rv_{u,n}$. Specifically, we have the conditional densities
		\begin{align}
			p(\widetilde\rv_{u,n}\vert {\rm a}_{u,n}=0)&= \textswab{g}(\widetilde\rv_{u,n}\vert \zerov, \Cm)\\
			p(\widetilde\rv_{u,n}\vert {\rm a}_{u,n}=1)&=\int p_u(\hv)\textswab{g}(\widetilde\rv_{u,n}\vert\hv,\Cm){\rm d}{\hv}\;,
		\end{align}
		where $p_u(\hv)$ is the density $\hv_u$. 
		Hence, the resulting likelihood ratio for the detection problem is given by
		\begin{equation}
			\Lambda_u(\rv)
			\eqdef \frac{\textswab{g}(\rv\vert \zerov, \Cm)}{\int p_u(\hv)\textswab{g}(\rv\vert\hv,\Cm){\rm d}{\hv}}\label{LR_u}\;.
		\end{equation}
			For example, particularizing $\hv_u$ to the fading model \eqref{channel_u}, the likelihood ratio in \eqref{LR_u} reads 
			\begin{equation}
				\Lambda_u(\rv)=\frac{\vert \Sigmam_u + \Cm\vert}{\vert\Cm\vert}{\rm e}^{-\rv\left(\Cm^{-1}-(\matr \Sigma_u + \Cm)^{-1}\right)\rv^\herm}\;.   \label{Qform}
			\end{equation}
	\begin{defi}[Message detection]\label{detectest}
		We define the recovery of the binary message activity ${\rm a}_{u,n}$ as
		\begin{equation}
			{\widehat{\rm a}}_{u,n}\eqdef{\rm u}(\nu_u-\Lambda_u(\rv_{u,n}))\qquad \forall (u,n)\in [U]\times [N_u]\;,  \label{LLR-Test} 
		\end{equation}
		where ${\rm u}(\cdot)$ is the unit-step function, $\Lambda_u(\rv)$ is as in \eqref{LR_u} and $\nu_u\in(0,\infty)$ is the decision threshold to achieve a desired trade-off between the MD and FA proabilities.  \hfill $\lozenge$
	\end{defi}
	{ Let us denote the actual set of active messages and the estimated set of active messages, respectively, as
		\begin{align}
			\mathcal A&\eqdef  \{(u,n)\in[U]\times[N_u]: {\rm a}_{u,n} = 1\}\label{Ac}\\
			\widehat{\mathcal A}&\eqdef \{(u,n)\in[U]\times[N_u]: \widehat {\rm a}_{u,n} = 1\}\label{achat}\;.
		\end{align}
		We then analyze the message detection in terms of the MD  and FA rates which are denoted respectively,
		\begin{align}
			\beta_{\rm e}^{\rm md} &\eqdef \frac{\big\vert {\widehat\Ac}^{\rm c}\cap \Ac \big\vert }{\big\vert  \Ac \big\vert}  \\
			\beta_{\rm e}^{\rm fa} &\eqdef \frac{\big\vert \widehat{\Ac}\cap \Ac^{\rm c} \big\vert }{\big\vert  \Ac^{\rm c}\big\vert} \label{md}\;,
		\end{align}
		where $\mathcal B^{\rm c}$ denotes the complement of the set $\mathcal B$. Here, and throughout the sequel, we use the subscripts $(\cdot)_{\rm e}$ and $(\cdot)_{\infty}$ in the notation to emphasize that the quantities are empirical and deterministic, respectively.
		
		\begin{thm}\label{the2}
			Let premises of Theorem~1  hold such that $\Xm_{u} \sim_{\text{i.i.d.}} {\rm a}_{u} \hv_{u}$ as in \eqref{xu_distrib}. Let either the decision test function $\Lambda_u(\rv)$ or its reciprocal, $1/\Lambda_u(\rv)$, be Lipschitz continuous. Then, as $L \to \infty$, we have  
			\[ \beta_{\rm e}^{\rm md}\overset{(a.s.)}{\rightarrow}\beta_{\infty}^{\rm md}~~\text{and}~~\beta_{\rm e}^{\rm fa}\overset{(a.s.)}{\rightarrow}\beta_{\infty}^{\rm fa}\;,\]
			where we define
			\begin{align}
				\beta_{\infty}^{\rm md} &\eqdef\frac{1}{Z}\sum_{u\in [U]}\alpha_u\lambda_u\mathbb{P}(\Lambda_u(\hv_u+\matr\phi) \geq \nu_u) \label{cmd}\\
				\beta^{\rm fa}_{\infty} &\eqdef\frac{1}{\widetilde Z}\sum_{u\in [U]}\alpha_u(1-\lambda_u)\mathbb{P}(\Lambda_u(\matr\phi) < \nu_u)\;\label{cfa}\;.
			\end{align}
			Here, $\matr\phi \sim \mathcal{CN}(\matr 0, \Cm)$ an arbitrary RV and for short we define the constants 
			\[Z\eqdef \sum_{u\in[U]}\alpha_u\lambda_u ~~\text{and}~~\widetilde Z\eqdef \sum_{u\in[U]}\alpha_u(1-\lambda_u)\;.\] 
			\begin{proof}
				See Appendix~\ref{pthe2}.\end{proof}
	\end{thm}}
	{ Note that e.g. $\beta_{\infty}^{\rm md}$ refers to a convex combination of the probabilities $\mathbb{P}(\Lambda_u(\hv_u+\matr\phi) \geq \nu_u)$. In fact, $\mathbb{P}(\Lambda_u(\hv_u+\matr\phi) \geq \nu_u)$ coincides with the asymptotic (for large $L$) 
		value of the MD rate at location $\mathcal L_u$ (see Appendix~\ref{pthe2}). Similarly, $\mathbb{P}(\Lambda_u(\matr\phi) < \nu_u)$ coincides with the asymptotic value of the FA rate at location $\mathcal L_u$.
		
		In passing, we note from \eqref{Qform} that for the fading model in \eqref{channel_u}, i.e., $\hv_u\sim{\Nc}(\matr 0,\matr\Sigma_u)$, $\Lambda_u(\rv)$ is Lipschitz continuous. Furthermore,  $\ln\Lambda_u(\rv)$ is a Hermitian quadratic form of $\rv$.  Hence, the probabilities $\mathbb{P}(\Lambda_u(\hv_u+\matr\phi) > \nu_u)$ and $\mathbb{P}(\Lambda_u(\matr\phi) < \nu_u)$} can be computed in closed form or tightly numerically approximated using the method of Laplace inversion and Gauss-Chebyshev quadrature as in \cite{ventura1997impact}. 
	As this aspect may be of some interest (e.g. it was not noticed in \cite{Chen2019,bai2022activity}), it might be useful for the reader to present this calculation explicitly in Appendix~\ref{laplace-inversion}. 
}

\subsection{Channel Estimation}  \label{chest-sect}	

We define the AMP channel estimation at the output of the $T$-th iteration (see \eqref{ziopera}) as
\begin{equation}
	\widehat\hv_{u,n} = \eta_{u}(\rv_{u,n})\quad \forall  (u,n) \in [U]\times[N_u]  \;, \label{AMP-ch-est}
\end{equation}
where we recall that $\rv_{u,n}\equiv \rv_{u,n}^{(T)}$ and similarly we set $\eta_{u}(\cdot)\equiv\eta_{u,T}(\cdot)$. Recall also the actual set of active messages $\Ac $ defined in \eqref{Ac}  and the {\em estimated} set of active messages resulting from the message detection test $\widehat\Ac $ defined in \eqref{achat}. Further, $\widehat{\Ac}$ is partitioned in two disjoint sets: the set of messages 
that are genuinely active and those that are inactive (FA events):
\begin{align}
	\mathcal A_{\rm d}&\eqdef \widehat{\Ac}\cap \Ac=  \{(u,n)\in [U]\times [N_u]:{\rm a}_{u,n}\widehat{\rm a}_{u,n}=1\}\\
	\mathcal A_{\rm fa}&\eqdef \widehat{\Ac} \cap \Ac^c \nonumber \\
	&= \{(u,n)\in [U]\times [N_u]:(1-{\rm a}_{u,n})\widehat{\rm a}_{u,n}=1\}\;.
	\label{setA_hatA}
\end{align}
From an operational viewpoint, it is important to notice that the channel estimation error 
is relevant only for  the messages in set $\Ac_{\rm d}$.  In fact, these messages correspond to users which effectively  wish to access the network. Then, the quality of the corresponding channel estimates directly impacts the achievable 
data rates to/from these users (see Section \ref{cluster-formation-and-ACK}).  
In contrast, messages in $\Ac_{\rm fa}$ do not correspond to any real user.  Hence, the quality of the  corresponding channel estimates is irrelevant for the subsequent data communication 
phase. Nevertheless, FA events (i.e., messages in $\Ac_{\rm fa}$) 
trigger the transmission of some ACK message in the DL. 
Hence, they cost some RU transmit power 
and cause some multiuser interference to the legitimate users (i.e., the users transmitting the messages in $\Ac_{\rm d}$).
In Section \ref{cluster-formation-and-ACK} we shall take these effects into account 
when considering the ACK beamformed transmission.

{Thus, we analyze the channel estimator in terms of the empirical mean-square-error for channels in $\mathcal A^{\rm d}$ and the powers consumed for the FA rates:
	\begin{align}
		{\rm mse}_{\rm e}^{\rm d}&\eqdef \frac{1}{\vert\mathcal A^{\rm d} \vert}\sum_{(u,n)\in\mathcal A^{\rm d}} \Vert \hv_{u,n}-\widehat\hv_{u,n} \Vert^2\\
		{\rm pow}_{\rm e}^{\rm fa}&\eqdef 
		\frac{1}{\vert\mathcal A^{\rm fa} \vert}\sum_{(u,n)\in\mathcal A^{\rm fa}} \Vert\widehat\hv_{u,n} \Vert^2\;.
	\end{align}
	
	\begin{thm}\label{the3}
		Let premises of Theorem~1  hold such that $\Xm_{u} \sim_{\text{i.i.d.}} {\rm a}_{u} \hv_{u}$ as in \eqref{xu_distrib}. Let either the decision test function $\Lambda_u(\rv)$ or its reciprocal, $1/\Lambda_u(\rv)$, be Lipschitz continuous. Then, as $L\to\infty$, \[{\rm mse}_{\rm e}^{\rm d}\overset{(a.s.)}{\rightarrow}{\rm mse}_{\infty}^{\rm d}~~\text{and}~~{\rm pow}_{\rm e}^{\rm fa}\overset{(a.s.)}{\rightarrow}{\rm pow}_{\infty}^{\rm fa}\;,\] where we define 	
		\begin{align}
			{\rm mse}_{\infty}^{\rm d}&\eqdef\frac {1}{{Z\beta_{\infty}^{\rm d}}}\sum_{u\in [U]}
			{\alpha_u\lambda_u\mathbb P(\mathcal D_u)}\nonumber \\
			&\qquad  \qquad\times\mathbb E\left[\Vert \hv_u-\eta_{u}(\hv_{u}+\matr\phi)\Vert^2\mid \mathcal D_{u}\right] \label{errorAMP1}\\\
			{\rm pow}^{\rm fa}_{\infty}&\eqdef \frac{1}{\widetilde Z\beta_{\infty}^{\rm fa}}\sum_{u\in [U]}{\alpha_u(1-\lambda_u)\mathbb P(\mathcal F_u)}
			\nonumber \\
			&\qquad \qquad\qquad  
			\times\mathbb E\left[\Vert\eta_{u}(\matr \phi)\Vert^2\mid \mathcal F_{u}\right]\;.  \label{errorAMP2}
		\end{align}
		Here,  we let $\beta_\infty^{\rm d}\eqdef 1-\beta_{\infty}^{\rm md}$ and $\beta_{\infty}^{\rm md}$, $Z$ and $\widetilde Z$ are as in Theorem~\ref{the2}. Moreover,  $\matr\phi\sim \mathcal {CN}(\matr 0,\Cm)$ is independent of $\hv_u$ and we define the events
		\begin{align}
			\Dc_u &\eqdef \{ \matr\phi,\hv_u \in \CC^{F} : \Lambda_u(\hv_u+\matr\phi) <\nu_u\}\label{Du}\\
			\Fc_u &\eqdef \{\matr\phi \in \CC^{F} : \Lambda_u(\matr\phi) < \nu_u\}\;.\label{Fu}
		\end{align}
		\begin{proof}
			See Appendix~\ref{pthe3}.
		\end{proof}
	\end{thm}
Note, for example, that ${\rm mse}_{\infty}^{\rm d}$ refers to a convex combination of the terms $\mathbb E\left[\Vert \hv_u-\eta_{u}(\hv_{u}+\matr\phi)\Vert^2\mid \mathcal D_{u}\right]$ over locations, which is
is the asymptotic value (for large $L$) of the channel estimation error at location $u$.}

\subsection{Comparison with Genie-Aided MMSE Channel Estimation}
{ In a low MD rate regime, i.e. $\beta^{\rm md}_{\infty}\ll 1$, it is interesting to compare the asymptotic channel estimation error \eqref{errorAMP1} with the error performance of  the \emph{genie-aided}  MMSE (i.e., the  exact MMSE conditioned on the knowledge of the true active messages $\Ac$),
	\begin{equation}
		{\rm mmse}^{\rm genie}_{\rm e}\eqdef \frac{1}{\vert \mathcal A\vert} {\left\Vert \Hm-\mathbb E[\Hm\vert \Ym,\{\Sm_u\},\Ac] \right\Vert_{\rm F}^2 }\;,
	\end{equation}
	where for short we write $\Hm \equiv[\Hm_1^\top, \Hm_2^\top,\ldots,\Hm_U^\top ]^\top$. Recall \eqref{tcm} where $\Pm_u\in \{0,1\}^{\tilde N_u\times N_u}$ stands for a projection matrix that eliminates the zero-rows of $\Xm_u$. Note that the knowledge of $\Ac$ is equivalent to the knowledge of $\{\Pm_u\}$ (see definition above \eqref{tsav}). Then, we write 
	\begin{eqnarray} 
		\Ym & = & \sum_{u=1}^U \Sm_u \Pm_u^\top\Pm_u \Xm_u  + \Nm \\
		& = & \sum_{u=1}^U \tilde{\Sm}_u \Hm_u  + \Nm \label{y_model_reduced}\;,
	\end{eqnarray}
	where for convenience we define  $\widetilde \Sm_u \eqdef  \Sm_u\Pm_u^\top$. 
	Thus, the \emph{genie-aided} MMSE reads as 
	\begin{align}
		{\rm mmse}_{\rm e}^{\rm genie}=\frac{1}{\vert \mathcal A\vert} {\left\Vert \Hm-\mathbb E[\Hm\mid \Ym,\{\widetilde\Sm_u\}] \right\Vert_{\rm F}^2 }\;.
	\end{align}
	In particular, for the fading model in \eqref{channel_u}, we can use standard random-matrix theory analysis 
	to derive the large $L$ limit of ${\rm mmse}_{\rm e}^{\rm genie}$.
	\begin{prop}\label{geniemmse}
		Let Assumption~\ref{as1} hold. Furthermore, let
		$\Nm\sim_{\text{i.i.d}}\Nc(\zerov, \sigma^2\Id_\bm)$ and 
		$\Hm_u\sim_{\text{i.i.d}}\Nc(\zerov, \matr\Sigma_u)$ with $\matr \Sigma_u={\rm diag}(\tau_{u1}, \tau_{u2},\cdots, \tau_{uF}),\forall u\in[U]$. Then, as $L\to \infty$ we have 
		\[{\rm mmse}_{\rm e}^{\rm genie}\overset{(a.s.)}{\rightarrow}{\rm mmse}^{\rm genie}_{\infty}\;, \]
		where we define
		\begin{equation}
			{\rm mmse}^{\rm genie}_{\infty}\eqdef\frac{1}{Z}\sum_{(u,f)\in [U]\times [F]}\alpha_u\lambda_u \frac{\tau_{uf}c_f^\star }{\tau_{uf}+c_f^\star }\;.
		\end{equation}
		Here, each $c^{\star}_{f}$ is the unique solution of
		(w.r.t. $c_f^\star$)
		\begin{equation} 
			c_f^\star =\sigma^2+\sum_{u\in [U]}\lambda_u\alpha_u
			\frac{\tau_{uf}c_f^\star }{\tau_{uf}+c_f^\star }  \label{fixed-point}	\;.
		\end{equation}
		\begin{proof}
			See Appendix~\ref{pthe4}.
		\end{proof}
	\end{prop}
}

\section{Cluster formation and beamformed ACK transmission} \label{cluster-formation-and-ACK}

{ Finally, we would like to share some insights related to beamformed ACK transmission in the
	in the DL. Throughout the section, we restrict our attention to the fading model in \eqref{channel_u}, i.e., $\hv_u\sim\Nc(\matr 0,\matr\Sigma_u)$ and $\Sigmam_u= {\rm diag}(g_{u,1},g_{u,2},\cdots,g_{u,B})\otimes \Id_M$. We also consider Gaussian noise as $\Nm\sim_{\text{i.i.d.}}\Nc(\matr 0,\sigma^2\Id_F)$.
	Defining the UL signal-to-noise ratio $\SNR$ as the average transmitted signal energy per symbol (for any active message) over the noise variance per component, the relation between $\SNR$ and $\sigma^2$ is $\sigma^2 = 1/ (L \; \SNR)$ (see also Remark \ref{zioSNR} 
	later in this section on SNR normalization).}

In the ACK slot (see the protocol frame structure of Fig.~\ref{RACH}), the system sends ACK messages\footnote{As already explained, this may correspond to a requested DL data packet and/or some PRMA resource allocation. The specific protocol details go beyond the scope of this paper and shall not be discussed further.} 
to all users whose RACH message is detected as active. 
For each detected active message $(u,n) \in \widehat{\Ac}$, 
a user-centric cluster $\Cc_{u,n}$ of $Q$ RUs with {\em strong} channel state is selected, where the cluster size 
$Q \geq 1$ is a system parameter. 
We consider simple cluster-based MRT transmission, which does not require computationally expensive matrix inversion as other types of DL precoders \cite{ngo2017cell,9336188}.\footnote{Notice that we advocate MRT only for the fast ACK transmission to the random access users, which is a small fraction of all the users in the system. Of course, for the TDD data slots, the plethora of precoding techniques widely studied in the rich CF literature can be used  \cite{ngo2017cell,7827017, 8845768, 9064545,9336188,gottsch2022subspace}.}
Here and throughout the sequel we consider the per-RU block partition of the channel estimates in \eqref{AMP-ch-est} as
\begin{equation}		
	\widehat{\hv}_{u,n} = [\widehat\hv_{u,n,1},\widehat\hv_{u,n,2},\ldots,\widehat\hv_{u,n,B}] \; .
\end{equation}	
The user-centric cluster can be established on the basis of the nominal LSFCs, i.e., 
for each $(u,n) \in \widehat{\Ac}$ a cluster is formed by the $Q \geq 1$ RUs with largest
coefficients $\{ g_{u,b}: b \in [B]\}$.  
This selection is {\em static} as it depends only on the network geometric layout.\footnote{ 
	As an alternative, a {\em dynamic} selection based on the channel estimates is possible. 
	Namely,  for each $(u,n) \in \widehat{\Ac}$ a cluster is formed by the $Q$ RUs 
	with largest estimated channel strengths $\| \widehat{\hv}_{u,n,b}\|^2$.
	While finite dimensional simulations shows a slight advantage for dynamic selection, practicality (ease of routing DL message bits to the RUs forming a cluster) and analytical tractability motivates us to focus on static cluster formation.} 

In particular, for a given network geometry, the static cluster formation determines a bipartite graph with $[U]$ location nodes and $[B]$ RU nodes, such that each location $u$ is assigned to a cluster $\Cc_u \subset [B]$ formed by the 
$Q$ RUs with largest  LSFCs relative to that location, and each RU $b$ is assigned a set of served locations
$\Sc_b \subseteq [U]$. 
Let $N_{\rm ack}$ denote the block length of the ACK slot expressed in channel uses. 
The $M \times N_{\rm ack}$ space-time signal matrix transmitted by RU $b$ in the ACK slot is given by 
\begin{equation}
	\Xim_b =\sqrt{\rho_{\rm DL}} \sum_{(u,n) \in \widehat{\Ac} : u \in \Sc_b} \widehat{\hv}_{u,n,b}^\herm \xiv_{u,n}, \label{MRT}
\end{equation} 
where $\rho_{\rm DL}$ is the DL transmit power normalization factor and $\xiv_{u,n} \in \CC^{1 \times N_{\rm ack}}$ is the ACK codeword, containing DL data and/or protocol information as described in Section \ref{intro}. 
Notice that all RUs $b \in \Cc_u$ send the same codeword $\xiv_{u,n}$ in response to the detected message $(u,n) \in \widehat{\Ac}$. Also, notice that (\ref{MRT}) corresponds to the standard MRT approach (e.g., see \cite{Larsson-book,7827017}) where the precoding vector for $\xiv_{u,n}$ 
is given by the Hermitian transpose of the estimated channel vector
$\widehat{\hv}_{u,n,b}$, i.e., the estimate of the $1 \times M$ section of the channel vector 
$\hv_{u,n}$ corresponding to the $b$-th RU. 
The system has no means to distinguish between actually active messages $(u,n) \in \Ac_{\rm d}$ and 
FA events $(u,n) \in \Ac_{\rm fa}$. Therefore, it sends an ACK codeword for all messages detected 
in $\widehat{\Ac} = \Ac_{\rm d} \cup \Ac_{\rm fa}$. 

The ACK codebooks are constructed such that 
codewords are mutually uncorrelated and have the same unit energy per symbol\footnote{In fact, 
	we may assume that codewords $\xiv_{u,n}$ belong to distinct codebooks, one for each message $(u,n)$, 
	and that these codebooks are independently generated from some particular random coding ensemble. Since any given active user
	is aware of its own transmitted uRA message, then it can use the corresponding codebook knowledge 
	for decoding the ACK message.}
\begin{equation}
	\frac{1}{N_{\rm ack}} \EE[ \xiv_{u,n} \xiv_{u',n'}^\herm ] = \left \{ \begin{array}{ll}
		1 & (u,n) = (u',n') \\
		0 & (u,n) \neq (u',n')  \end{array} \right .
	\label{ACK-codebook}\;.
\end{equation}
For a given channel state and AMP estimation realization, the 
average transmit power of RU $b$ (expressed as energy per channel use) is given by 
\begin{eqnarray}
	P_{{\rm tx},b} & = & \frac{1}{N_{\rm ack}} \trace \left ( \EE[ \Xim_b \Xim_b^\herm ] \right ) \nonumber \\
	& = & \rho_{\rm DL} \sum_{(u,n) \in \widehat{\Ac} : u \in \Sc_b} \|  \widehat{\hv}_{u,n,b} \|^2 \label{Ptxb}\;.
\end{eqnarray}
Separating the messages in $\Ac_{\rm d}$ and in $\Ac_{\rm fa}$ in the sum in \eqref{Ptxb} and taking expectation
over all the system variables (channel state, AMP estimation outcome, message activity), we find the average transmit power
\begin{equation}
	\overline{P_{{\rm tx},b}} = \rho_{\rm DL} \sum_{u \in \Sc_b} \lambda_u N_u \Zc_{u,b} \label{Ptxb_av}\;,
\end{equation}
where we define
\begin{align}
	\Zc_{u,b} & \eqdef \PP(\mathcal D_u) \EE\left[\Vert \eta_{u}(\hv_{u,b} + \zv_b \Cm_b^{\frac 1 2})\Vert^2 \; | \; \zv, \hv_u  \in \Dc_u \right]   \nonumber \\
	&  +~ \PP(\mathcal F_u)(\lambda_u^{-1} - 1)\EE\left[\Vert \eta_{u}(\zv_b \Cm_b^{\frac 1 2})\Vert^2 \; | \; \zv \in \Fc_u \right]  \label{ZZ}\;,
\end{align}
where $\hv_{u} \sim \Cc\Nc(\zerov, \Sigmam_u)$, $\zv \sim_{\rm i.i.d.}  \Cc\Nc(0,1)$ are mutually independent, 
$\hv_{u,b}$ and $\zv_b$ denote the $b$'th segment of size $1 \times M$ of $\uv_u$ and $\zv$, respectively, and the events $\Dc_u$ and $\Fc_u$ are defined in \eqref{Du} and \eqref{Fu}, respectively. 
Without going into details, \eqref{ZZ} can be explained as follows: each location $u$ contributes on average with
$\lambda_u N_u$ genuinely active messages, of which a fraction $\PP(\mathcal D_u)$ is detected as active, such that the
average size of the set $\{ (u,n) \in \Ac_{\rm d} : u \in \Sc_b \}$ is $\PP(\mathcal D_u) \lambda_u N_u$. Similarly, the average size of the set  $\{ (u,n) \in \Ac_{\rm fa} : u \in \Sc_b \}$ is $\PP(\mathcal F_u) (1 - \lambda_u)  N_u$. 

Before moving on, the following two remarks are in order: 

\begin{rem}   \label{finiteL-remark}
	Up to this point, the treatment of the uRA mechanism in the RACH slot has been rigorous: the AMP, message detection, and channel estimation, are all defined for finite system parameters
	$L, \{N_u\}, B, M$, and the asymptotic statistics and corresponding analysis holds in the limit of large $L$, fixed ratios 
	$U_u/L = \alpha_u$ and fixed (not growing with $L$) number of RUs $B$ and antennas per RU $M$. 
	
	However, in the data transmission slot (the ACK slot treated in this section), the DL serves $O(L)$ users
	with $F = BM$ antennas. Even under perfect channel state knowlegde, the DL rate per user vanishes as $O(1/L)$ since the 
	channel (a vector Gaussian broadcast channel \cite{Caire-Shamai-TIT03,Weingarten-Steinberg-Shamai-TIT06}) has only at most $\min\{ F, L \sum_{u\in [U]} \lambda_u \alpha_u\}$  degrees of freedom.   
	
	Hence, when considering the DL performance in terms of ACK message rate, we shall informally
	use the asymptotic statistics for large $L$, but provide expressions (in this section) and results (in Section \ref{Simresults}) 
	for finite $L$: we fix some large but finite $L$ and $\{\alpha_u\}$ and let 
	$N_u$ be the integer rounding of  $\alpha_u L$ for $u \in [U]$.
	\hfill $\lozenge$
\end{rem}

\begin{rem}   \label{zioSNR}
	It is important to clarify the SNR and power normalization adopted in this work. 
	In the random CDMA literature (e.g., see \cite{tulino2004random} and references therein) it is customary to 
	define a symbol as formed by $L$ ``chips'' where $L$ is the so-called {\em spreading factor}, 
	such that unit average energy per symbol yields
	average energy per chip equal to $1/L$.  Letting $W$ denote the (spread-spectrum) signal bandwidth in Hz, 
	one symbol interval corresponds to $L/W$ seconds.  SNR (transmit power over noise power in Watts) 
	is given by $\SNR = \frac{1/(L/W)}{N_0 W} = 1/(L N_0)$, which is the same normalization used in this paper (with $\sigma^2$ 
	{\em in lieu} of $N_0$, with the same meaning of the variance per component of the additive complex circularly symmetric Gaussian noise). Hence, this paper is consistent with the standard random CDMA literature by identifying a channel use (or symbol) 
	of the channel models in \eqref{y_model} (for the RACH slot) and in \eqref{received_ACK} (for the ACK slot) with 
	a ``chip'' of the random CDMA standard model. For some (large) fixed value of $L$, we fix a 
	reasonable value of the parameter $\SNR$ which corresponds to the ratio of the received signal power
	over noise power over the system bandwidth, for a small distance between transmitter and receiver, and let 
	$\sigma^2 = 1/(L \SNR)$. 
	\hfill $\lozenge$
\end{rem}


As a general system design rule of thumb, we consider a balanced total power expenditure in the UL and DL. With our normalizations, 
the power (average energy per channel use) spent by the uRA users is
\[ \overline{P_{\rm uRA}} = \sum_{u=1}^U \lambda_u N_u/L = \sum_{u=1}^U \lambda_u \alpha_u. \]
The total DL power is obtained by summing $\overline{P_{{\rm tx},b}}$ over $b \in [B]$. Imposing equality in the UL RACH and ACK slots powers we find the normalization factor
\begin{equation} 
	\rho_{\rm DL} = \frac{1}{L} \; \frac{\sum_{u=1}^U \lambda_u \alpha_u}{\sum_{b=1}^B \sum_{u \in \Sc_b} \lambda_u \alpha_u \Zc_{u,b}} \label{DLnormfactor}\;.
\end{equation}
Next, we evaluate the performance of the ACK slot by considering a lower bound to the achievable {\em ergodic rate}
obtained by Gaussian random coding and treating interference as noise at the receiver of the uRA 
active user sending message $(u,n) \in \Ac_{\rm d}$, referred to for brevity as ``user $(u,n)$''. 
Notice that we are not interested in users in $\Ac_{\rm fa}$ since these users actually do not exist. 

As said, each RU $b$ sends the corresponding space-time codeword $\Xim_b$ (see \eqref{MRT}) simultaneously. 
The block of DL received signal samples (dimension $1 \times N_{\rm ack}$) at the receiver of user $(u,n)$ is given by 
\begin{align} 
	&\yv_{u,n}^{\rm ack}  = + \wv_{u,n} +   \sum_{b=1}^B \hv_{u,n,b} \Xim_b  \nonumber \\
	&  =  +  \wv_{u,n} +\sqrt{\rho_{\rm DL}} \left ( \sum_{b \in \Cc_{u}} \hv_{u,n,b} \widehat{\hv}^\herm_{u,n,b} \right ) \xiv_{u,n}   \label{useful} \\
	& +  \sqrt{\rho_{\rm DL}} \sum_{(u',n') \in \widehat{\Ac} \setminus (u,n)} \left ( \sum_{b \in \Cc_{u'}} \hv_{u,n,b} 
	\widehat{\hv}^\herm_{u',n',b} \right ) \xiv_{u',n'} \label{received_ACK}\;,
\end{align}
where $\wv_{u,n} \sim_{\rm i.i.d.}  \Cc\Nc(0, \sigma^2)$. A lower bound on the ergodic rate is obtained 
by the so-called ``Use and then Forget'' (UatF) bound, widely used in 
the CF/massive MIMO literature 
(e.g., see \cite{Larsson-book,9064545,9336188,9069486,gottsch2022subspace,miretti2022team,miretti2023ul}). 
We identify the useful signal term in (\ref{useful}) and the interference plus noise term in (\ref{received_ACK}).
Adding and subtracting the useful signal with mean coefficient
$\sqrt{\rho_{\rm DL}} \left (   \sum_{b \in \Cc_{u}} \EE[ \hv_{u,n,b} \widehat{\hv}^\herm_{u,n,b} | (u,n) \in \Ac_{\rm d}] \right ) \xiv_{u,n}$ 
and using the classical worst-case uncorrelated additive noise result (e.g., see \cite{Larsson-book}), the UatF lower bound yields
	\begin{align}
		R_{u,n}^{\rm UatF} = \log \left ( 1 + \frac{\left | \sum_{b \in \Cc_u} \Mc_{u,b} \right |^2}
		{\sigma^2/\rho_{\rm DL} + \sum_{b \in \Cc_{u}} \Vc_{u,b} 
			+ L \sum_{u' \in [U]} \sum_{b \in \Cc_{u'}}  \lambda_{u'} \alpha_{u'} g_{u,b} \Zc_{u',b} } \right ) \label{UatF}
	\end{align}
where, for all $(u,b) \in [U] \times [B]$, we define the mean and variance of the useful signal term as
\begin{align}
	\Mc_{u,b} & \eqdef   \EE\left [ \hv_{u,b} \;  \eta_{u}(\hv_{u,b} + \zv_b \Cm_b^{\frac 1 2})^\herm | \hv_u, \zv \in \Dc_u \right ] \label{mean} \\
	\Vc_{u,b} & \eqdef   {\rm Var}\left ( \hv_{u,b} \;  \eta_{u}(\hv_{u,b} + \zv_b \Cm_b^{\frac 1 2})^\herm | \hv_u, \zv \in \Dc_u \right ) \label{variance},
\end{align}
and where the coefficients $\{\Zc_{u,b}\}$ and the definition of $\hv_u, \zv, \hv_{u,b}, \zv_b$,  
are as in \eqref{ZZ}.
In particular, the multiuser interference terms in the denominator of the fraction inside the ``log'' in (\ref{UatF}) follows from 
noticing that for $(u,n) \neq (u',n')$ the channel $\hv_{u,n,b}$ and the estimate channel $\widehat{\hv}_{u',n',b}$ are statistically independent, Hence, for $(u'n') \in \widehat{\Ac} \setminus (u,n)$ we can write
\begin{eqnarray}
	\EE[|\hv_{u,n,b}  \widehat{\hv}^\herm_{u',n',b}|^2] 
	& = & \EE[ \widehat{\hv}_{u',n',b} \hv^\herm_{u,n,b} \hv_{u,n,b} \widehat{\hv}^\herm_{u',n',b} ]  \nonumber \\
	& = & g_{u,b} \EE[ \| \widehat{\hv}_{u',n',b} \|^2 ]. 
\end{eqnarray}
Then, by separating the contributions of $(u',n') \in \Ac_{\rm d}$ and $(u',n') \in \Ac_{\rm fa}$ and following a similar argument leading to
(\ref{Ptxb_av}) we arrive at the final expression of the multiuser interference term.

As a term of comparison, the UatF bound in the ``genie-aided'' conditions 	$\widehat{\Ac} = \Ac$ (perfect message detection)
and $\widehat{\hv}_{u,n,b}  = \hv_{u,n,b}$ (perfect CSI estimation) yields 
	\begin{align}
		R^{\rm UatF, genie}_{u,n} &= \log \left ( 1 + \frac{M^2 \left ( \sum_{b\in \Cc_u} g_{u,b} \right )^2}{\sigma^2/\rho_{\rm DL} + 
			M \sum_{b\in \Cc_u} g_{u,b} + L M \sum_{u' \in [U]} \sum_{b \in \Cc_{u'}} \lambda_{u'} \alpha_{u'} g_{u,b} g_{u',b}} \right ) \label{UatF_genie}
	\end{align}

Comparing the cumulative distribution function (CDF) of $R^{\rm UatF}_{u,n}$ 
$R^{\rm UatF, genie}_{u,n}$ over the user population for a given network topology, we can appreciate how close 
the effective uRA message detection and channel estimation performs with respect to an ideal system using the same MRT 
DL precoding.

Notice also that the ratio $\sigma^2/\rho_{\rm DL}$ in \eqref{UatF} and \eqref{UatF_genie}, 
with our normalizations, is a constant that depends on $\SNR$ and on the system geometry, but it is independent of $L$, consistent with the fact that the total UL and DL power is balanced.  In contrast, the multiuser interference term grows linearly with  $L$. As observed in Remark \ref{finiteL-remark}, the per-user rate in the DL will vanish as $L \rightarrow \infty$, consistent with the fact that the system degrees of freedom is bounded by $F = BM$ (total number of antennas), while the number of users grows unbounded. As a matter of fact, our analysis yields practically relevant results when fixing $L$ to some large but finite value, such that the asymptotic statistical analysis of the AMP algorithm is meaningful, and then determine 
the other parameters as a function of $L$ (in particular, $N_u = \alpha_u L$ for some proportionality factors 
$\{\alpha_u\}$, and $B, M$ significantly smaller than $L$).


\section{Numerical Examples}\label{Simresults}

\begin{figure}[t]
	\centering
	\begin{tabular}{p{0.001\linewidth}c}
		\rotatebox{90}{\hspace{13mm}\small{ }} & 
		\includegraphics[width=0.7\linewidth]{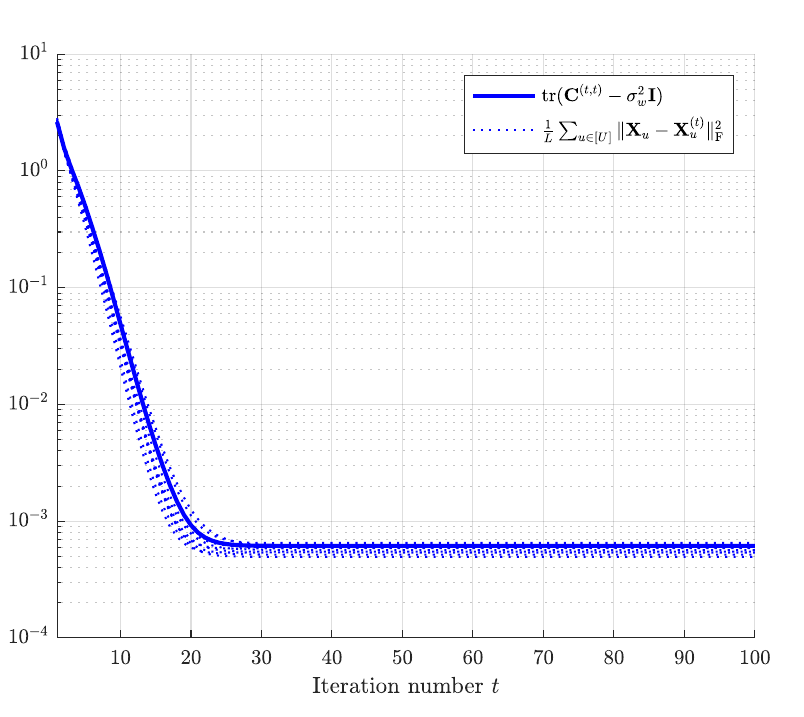}
		\\[-3mm]
		& \small{}\\[-1mm]
	\end{tabular}
	\caption{The { empirical} MSE of AMP for { $10$ instances of  AMP dynamics} and its theoretical deterministic 
		prediction for the toy model example (Fig.~\ref{fig1}) with $N_u= 2048$, $L = 1024$, $\wp = 1/2$, and UL $\SNR = 10$ dB .} 
	\label{fig.MMSE}    
\end{figure}

{ In simulations, we restrict the fading model in \eqref{channel_u}, 
	Gaussian noise as $\Nm\sim_{\text{i.i.d.}}\Nc(\matr 0,\sigma^2\Id_F)$ with $\sigma^2=1/(L{\rm SNR})$.}	
As a warm-up, let us consider the linear Wyner model \cite{wyner1994shannon} with only $U = B = 2$ 
locations and RUs, as depicted in Fig.~\ref{fig1}.  We set the $2\times 2$ LSFC matrix as
$$
[ g_{ub} ] = \begin{bmatrix}
	1 &\wp \\
	\wp &1
\end{bmatrix}\;,
$$
where $\wp \in [0,1]$ is the crosstalk coefficient in location $u$ and RU $b \neq u$. 
The message activity probabilities are set to $\lambda_1 = 0.1$ and $\lambda_2 = 0.2$. 
The number of antennas per RU is $M = 2$ and the SNR parameter is set to $\SNR=10$dB. 
We consider, $N_u = 2048$ codewords per location with block length $L=1024$. 
To visualize the AMP convergence and the agreement of the asymptotic analysis with the empirical finite-dimensional 
behavior, we track the total empirical squared error as a function of the AMP iteration index $t$. 
In particular,  combining Corollary~\ref{Cormse} and the SE in (\ref{SE}) for $s = t$ we have that 
\begin{equation}
	\frac{1}{L} \sum_{u\in U}\Vert \Xm_u-\Xm_u^{(t)}\Vert_{\rm F}^2\overset{a.s.}{\rightarrow}{\rm tr}(\Cm^{(t,t)}-\sigma^2\Id)\label{cmse}.
\end{equation}
In  Fig.~\ref{fig.MMSE} we show the left- and righthand side of 
\eqref {cmse} (referred to as total normalized MSE)  
as a function of the iteration index $t$ 
for the simple model of Fig.~\ref{fig1} with the parameters given above.
Notice that for the actual simulated finite-dimensional system the total normalized MSE is a random trajectory
function of $t$.   In the figure, we show 10 independent realizations and the deterministic trajectory 
predicted by the SE. 

\begin{figure}[t]
	\centering
	\begin{tabular}{p{0.001\linewidth}c}
		\rotatebox{90}{\hspace{13mm}\small{ }} & 
		\includegraphics[width=0.70\linewidth]{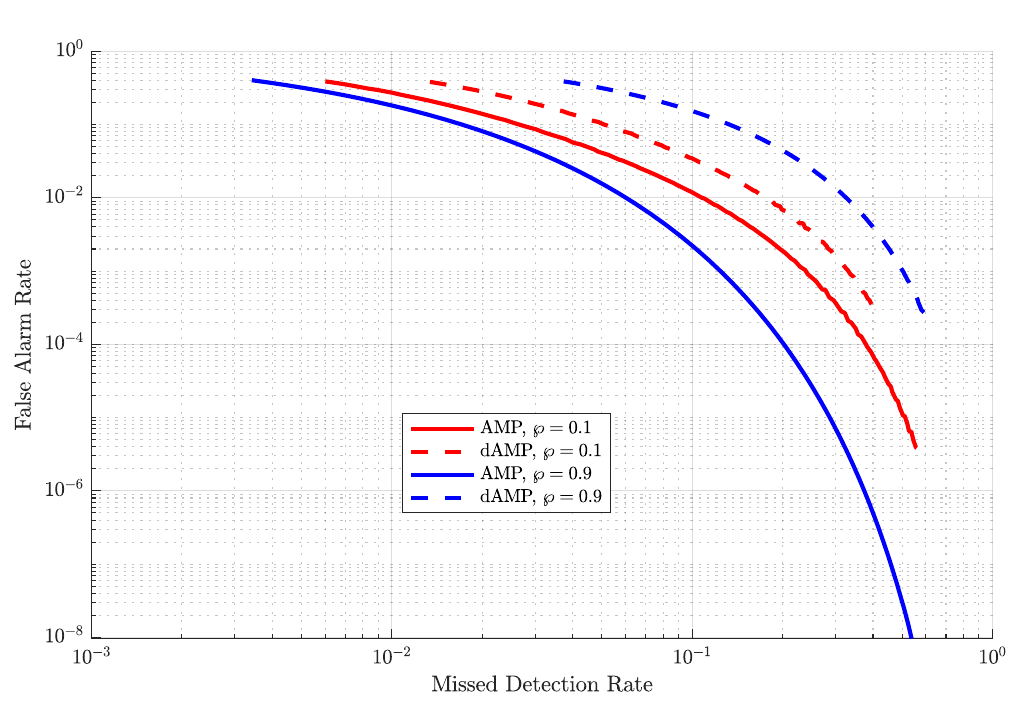}
		\\[-3mm]
		& \small{}\\[-1mm]
	\end{tabular}
	
	\caption{The performance comparison of the AMP and sAMP algorithms for the toy model example 
		(Fig.~\ref{fig1}) with  $N_u= 2048$, $L = 1024$, and $\SNR=10$~dB.}
	\label{fig.SimpleBetas}
\end{figure}

Next, we compare the detection performances of the {\em separated AMP} (sAMP) of \cite{Chen2019}
and the proposed multisource AMP algorithm in Fig.~\ref{fig.SimpleBetas}.
sAMP applies a conventional AMP separately for each RU, and then combines the LLRs for the message activity 
detection.  We observe that the proposed centralized AMP approach 
generally outperforms sAMP and its performance gain increases with the 
crosstalk parameter $\wp$. 

\begin{figure}[t]
	\centering
	\includegraphics[width=0.7\linewidth]{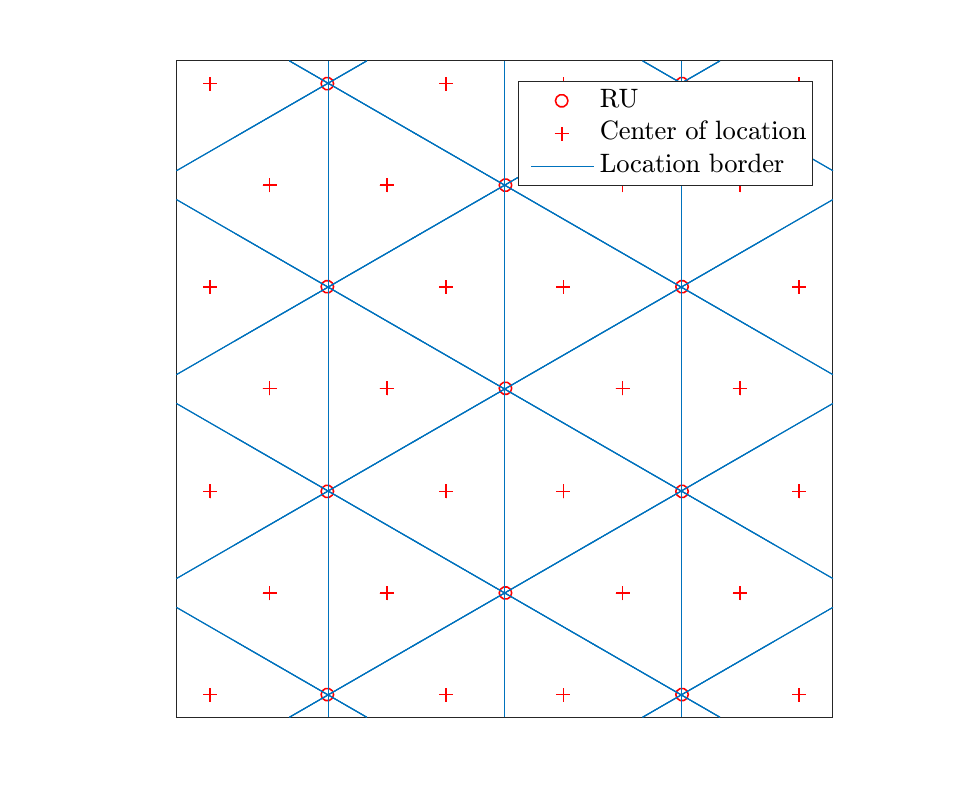}
	\caption{Hexagonal network topology with equilateral triangle tiles. The crosses denote the user locations 
		and the circles denote the RUs.} 
	\label{RUs} 
\end{figure}

\begin{figure}[t]
	\centering
	\includegraphics[width=0.7\linewidth]{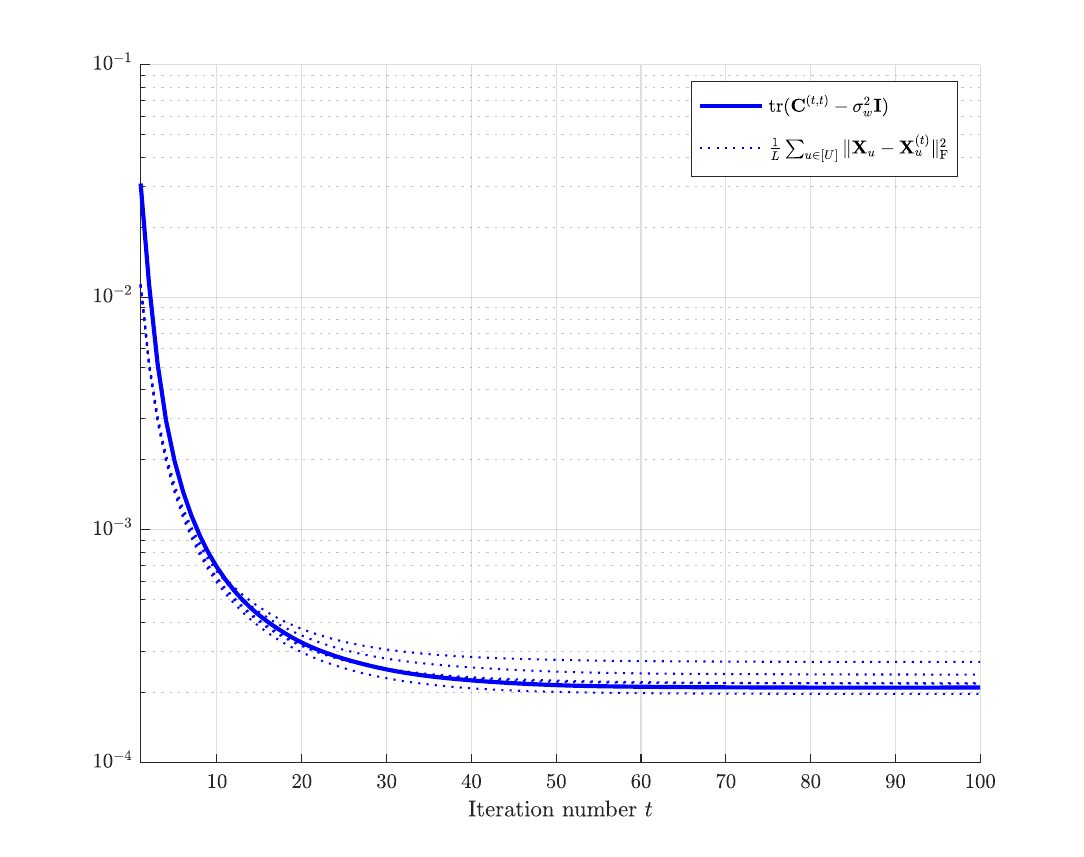}
	\caption{The { empirical} MSE of the AMP iterations (\ref{AMP_alg}) for { $10$ independent instances of AMP dynamics} its theoretical prediction for the cell-free network of Fig.~\ref{RUs}) with $N_u= 2048$, $L = 1024$, $M = 2$, and  $\SNR_{\rm rx} = 10$ dB.} 
	\label{SEconverge_10dB} 
\end{figure}

\begin{figure}
	\centering
	\includegraphics[width=0.7\linewidth]{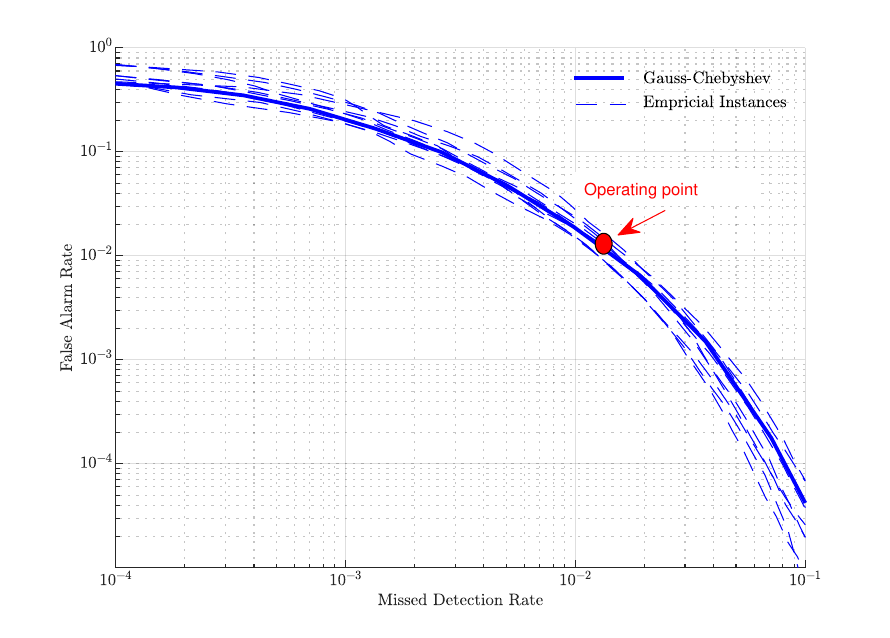}
	\caption{Operational tradeoff curve of the message detector with { $10$ independent instances from AMP} for the CF network of Fig.~\ref{RUs} with 
		$N_u= 2048$, $L = 1024$, $M = 2$, and $\SNR_{\rm rx}=10$~dB. } 
	\label{Pfa_Pmd_realnet} 
\end{figure}

\begin{figure}
	\centering
	\includegraphics[width=0.7\linewidth]{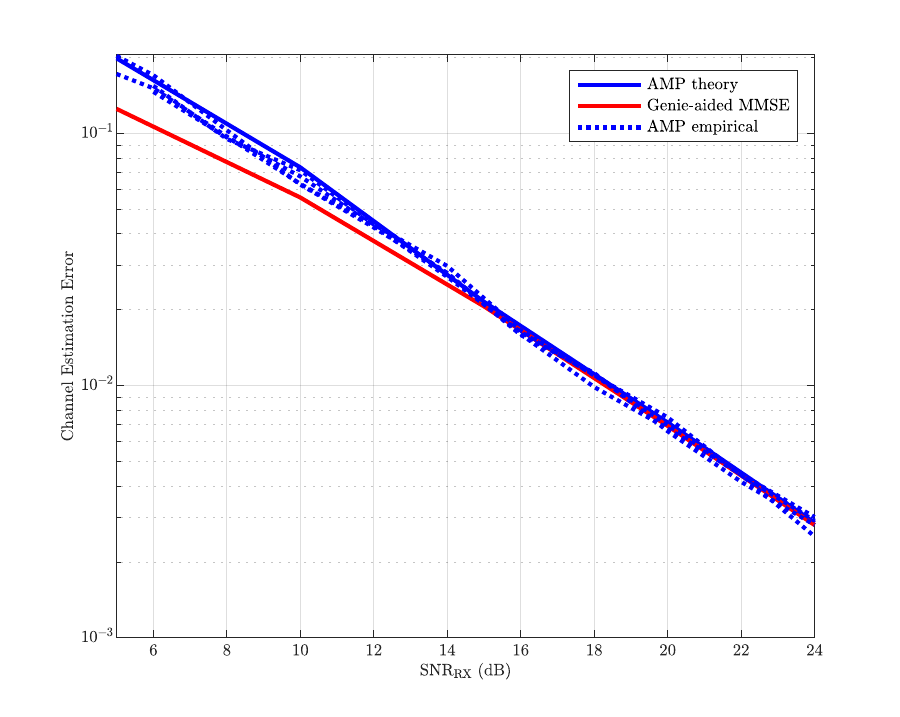}
	\caption{Channel estimation error for the genie-aided MMSE, the asymptotic theory (AMP), and $4$ independent instances of AMP.} 
	\label{fig.genie_aided}
\end{figure}

Then, we  consider a more realistic CF user-centric network. 
The coverage area is divided into $U=16$ location tiles with equilateral triangular shape of side $100$m, arranged
in a hexagonal pattern as depicted in Fig.~\ref{RUs}.
We position $B=12$ RUs at the corners of the triangular location tiles. 
The message probabilities are set as $\lambda_u \in \{0.003, 0.002\}$, repeated in a fixed alternating pattern, 
such that $\lambda_1=0.003, \lambda_2=0.002$, and so on. 
Each location-based uRA codebook $\Sm_u$ consists of $N_u= 2048$ codewords with block length $L = 1024$. 
We use  torus-topology to calculate distances in order to approximate an infinitely large network without border effects. The distance-dependent pathloss (PL) function is given by
\begin{equation}
	\text{PL}=\frac{1}{1 + \left(\frac{d}{d_0}\right)^\gamma}\;,
\end{equation}
with pathloss exponent $\gamma=3.67$, and  3dB cutoff distance 
$d_0=13.57 \text{m}$ (this is the distance at which the received signal power from a given RU is attenuated by 3dB due to the pathloss). 
In order to set the UL transmit power (i.e.,  the parameter $\SNR$), we consider the pathloss between a location and its nearest RU. Letting $\varsigma$ denote such minimum distance and 
$\text{PL}_{\varsigma} = 1/(1 + \left(\frac{\varsigma}{d_0}\right)^\gamma)$ the corresponding pathloss, 
we set the received SNR (at any nearest RU antenna) to be some reasonable value $\SNR_{\rm rx}$ and, as a consequence, we let  $\SNR  = \SNR_{\rm rx} / \text{PL}_{\varsigma}$. 
For the system described above, we illustrate the theory-simulation consistency for the 
MSE of the AMP iterations, i.e. \eqref{cmse},  in Fig.~\ref{SEconverge_10dB} 
for $M = 2$ and $\SNR_{\rm rx} = 10$dB (again, we show 10 independently generated trajectories of the finite-dimensional system and the deterministic trajectory predicted by the SE). 
For the same system parameters, Fig.~\ref{Pfa_Pmd_realnet} shows the
theoretical predictions and the finite dimensional simulation of the MD and FA  probabilities. 
The theoretical prediction is obtained using the Gauss-Chebyshev Laplace inversion method in Appendix~\ref{laplace-inversion}. 
For each location $u$ the detection threshold in the Neyman-Pearson test
\eqref{LLR-Test} is varied in order to obtain the tradeoff curve. 
Fig.~\ref{Pfa_Pmd_realnet} shows the average tradeoff curve over all locations. 
In particular, the red dot corresponds to the point at which, for each location $u$, the threshold is set such that 
{ $\PP(\mathcal D_u)+\PP(\mathcal F_u) = 1$ where the events $\Dc_u$ and $\Fc_u$ are defined in \eqref{Du} and \eqref{Fu}, respectively and the probabilities  $\PP(\mathcal D_u)$ and $\PP(\mathcal F_u)$ coincides with the large-system limits of the $\mathcal L_u$-dependent correct detection and false alarm rates, respectively.}  
All subsequent results are given for this choice, i.e.,  the system operates such that in each location the FA and MD  
probabilities are equal. 
Next, we consider the channel estimation error for the messages in $\Ac_{\rm d}$ (active messages detected as active) for the system described above. With reference to Theorem~\ref{the3} and Proposition~\ref{geniemmse}, we can express the asymptotic total channel estimation 
error calculated by summing the errors (normalized by the total number of RU antennas) 
over all locations. This comparison is shown in Fig.~\ref{fig.genie_aided}. In general, we observe that when operating with sufficiently low { $\beta_{\infty}^{\rm md}$ and $\beta_{\infty}^{\rm fa}$} (in our case, $\approx 10^{-2}$, as seen from Fig.~\ref{Pfa_Pmd_realnet}), the channel estimation performance of the joint AMP-based message detection and channel estimation is very close to the performance of the genie-aided MMSE estimation and no further reprocessing is needed.\footnote{In some works, a reprocessing approach is advocated. Using AMP for
	activity detection, and then applying linear MMSE estimation conditioned on the set of (detected) active messages. 
	See e.g., \cite{Fengler2022}.} 

\begin{figure}
	\centering
	\includegraphics[width=1\linewidth]{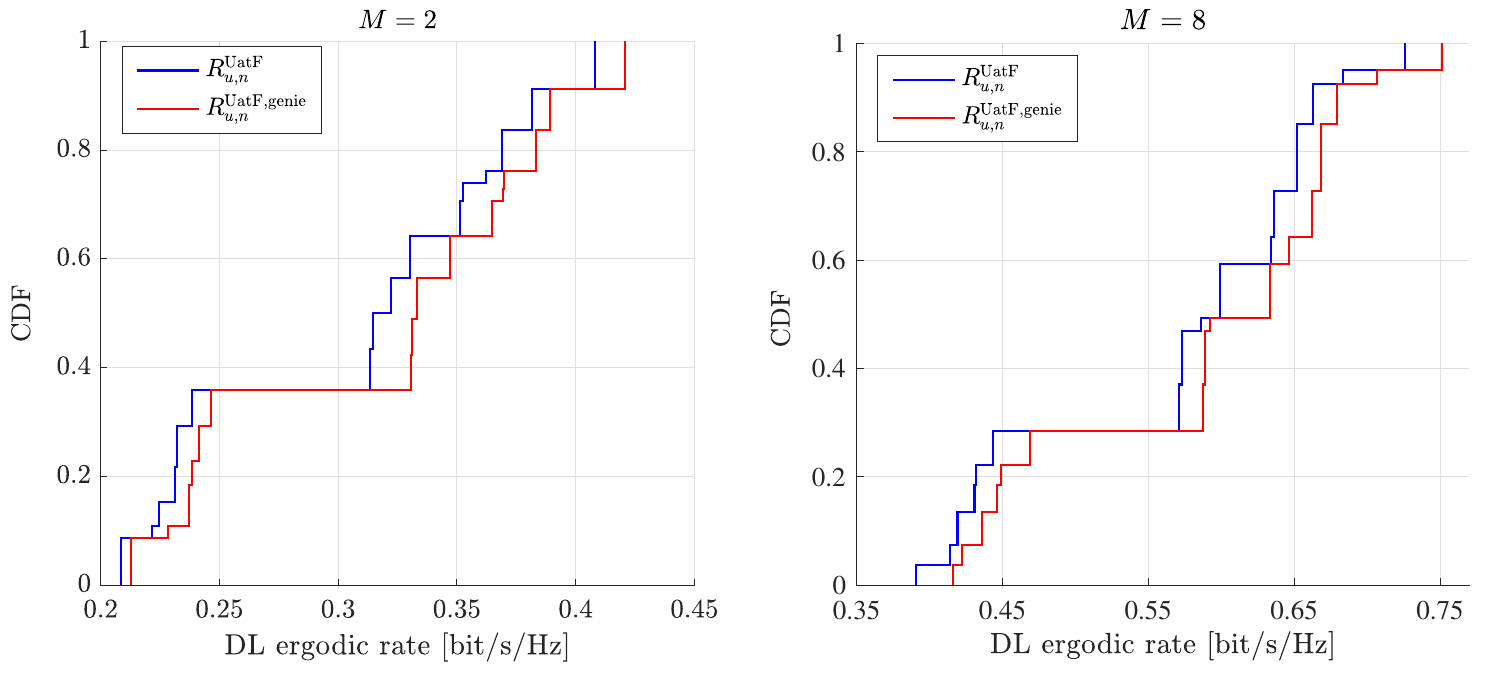}
	\caption{Achievable per-user ergodic rate CDF of the DL ACK message under MRT precoding for the CF network  of Fig.~\ref{RUs} with  $N_u= 2048$, $L = 1024$, and $\SNR_{\rm rx}=10$~dB.} 
	\label{fig.ergodic}
\end{figure}

According to the scheme described in Section \ref{cluster-formation-and-ACK}, 
the RACH slot is followed by a DL transmission (referred to as ACK slot) to the users corresponding to the detected 
uRA messages $\widehat{\Ac}$. In this example, we consider that each location is served by the $Q = 3$ 
nearest RUs using MRT based on the estimated channel vectors. 
Fig.~\ref{fig.ergodic} shows the CDF of the per-user rate (computed via the UatF achievable lower bound \eqref{UatF} ) 
and, for comparison,  the corresponding CDF of the genie-aided UatF bound given in  \eqref{UatF_genie}. 
Since these rate expressions depend only on the location index $u$ and not on the message index $n$, 
these CDFs have a staircase shape with jumps corresponding to the rates at different locations $u$. 
For example, if all locations have distinct rate values, we have $U$ jumps and the $u$-th jump has amplitude (assuming 
very small { $\beta_{\infty}^{\rm md}$ and $\beta_{\infty}^{\rm fa}$}, i.e., $\Ac = \widehat{\Ac}$) 
equal to  $\lambda_u N_u / (\sum_{u'} \lambda_{u'} N_{u'})$, i.e., the relative fraction of active users in location $u$.  
Fig.~\ref{fig.ergodic} shows that when the system is sized to operate with low message error probability and low channel estimation error, as in this case, the achievable per-user rate is very close to the ideal case (perfect uRA message detection and CSI). We also illustrate the effect of the number of antennas per uRA on the user rates for $M=\{2, 8\}$. Indeed, the median user rate for $M = 2$ is $\approx 0.32$ bit/symbol, while for $M = 8$ it increases to $\approx0.6$ bit/symbol.



\section{Conclusion}\label{SecConc}

We considered the problem of unsourced random access (uRA) in cell-free (CF) wireless networks. The model is an extension of what already considered in several works for the case of uRA with a massive MIMO receiver. The difference is that, in the CF case, 
the total number of antennas $BM$ is partitioned in $B$ groups of $M$ antennas each, and these are allocated to geographically separated RUs. While this may seem a minor modification, the problem is significantly 
more complicated by the fact that the statistics of the channel vectors describing the propagation between 
users and RUs dependents on the user (transmitter) location, via the set of distance-dependent
large-scale fading coefficients between the user and the RUs. In the uRA setting, users can transmit any codeword 
from a common codebook. It is therefore not possible to associate in a fixed manner a given access codeword to a given set of large-scale propagation coefficients (unlike the problem of activity detection with known user locations and fixed association between activity signature sequences and users). We circumvent this problem by proposing a partitioned access codebook associating sub-codebooks to locations, such that users in a given location select codewords from the corresponding  codebook. For this system, we propose a centralized AMP algorithm for joint uRA message detection and 
CSI estimation. The rigorous analysis of this  AMP scheme is novel and does not reduces immediately to 
a known case for which the asymptotic SE analysis and concentration result has been rigorously proved. 
In fact, our results yield immediately other known
concentration results the classical AMP and the MMV-AMP, obtained as 
special cases of our general setting for $U = F = 1$ and for $U = 1, F > 1$, respectively.
In addition, we also show the consistency of the replica symmetry (non-rigorous) results with the rigorous SE analysis. 

Based on our analysis, a rigorous (asymptotic) characterization of the statistics at the output of the AMP 
detection/estimation scheme
are given. Using these statistics, we can provide very accurate and almost closed-form expressions for the FA and MD probabilities of the uRA messages. Also, using the asymptotic statistics, simple and almost closed-form expressions for the 
achievable user rate in a subsequent transmission phase are obtained. 
For simplicity, we focused here on the downlink transmission based on MRT, which requires only per-RU local combining CSI. However, in a similar manner, expressions for other forms of combining/precoding can be obtained for both downlink and 
uplink transmission (in the latter case, after the RACH slot the user would append a data communication slot
as in the scheme of \cite{Fengler2022}). 

An interesting extension is the case where the number of antennas per RU scales with the number of uRA messages (i.e., with $L$). In this case, we let the number of antennas per RU be equal to $m M$, where $M$ is fixed (constant with $L$) and $m/L = \beta$. Then, each block of $M$ antennas forms a system identical to the one presented in this paper, for which the concentration holds. However, for the sake of message detection and data transmission, the $m$ group of antennas at each RU can be ``pooled'' together. This yields vanishing message error probability (both FA and MD) and non-vanishing per-user rate as $L \rightarrow \infty$. Again, these results will be presented in a follow-up paper.

We notice that the symbol synchronous frequency-flat channel model used in most existing uRA. MIMO papers (starting from  \cite{Polyanskiy2017}, all related papers referenced here including the present paper), is unrealistic. Propagation goes through frequency-selective multipath channels, and transmission in the RACH slot cannot be perfectly chip-level synchronous at each RU receiver, since propagation distances are different for different users and 
RUs. Hence, the extension to a wideband frequency-selective asynchronous model, 
where the unknown channels are vectors of
multipath impulse responses is called for. 

We emphasize the broader applicability of our theoretical results. They go beyond the context of wireless networks and are relevant to the general problem of ``multiple measurement vector'' compressed sensing involving 
the superposition of multiple signal sources, each with different statistical characteristics. This suggests the relevance of our results to a wider range of applications.

{ Another intriguing direction is to develop the AMP framework using randomly permuted Fourier-type codebooks \cite{ma2017orthogonal} instead of codebooks with i.i.d. entries. From a computational perspective, randomly permuted Fourier-type codebooks and their corresponding AMP algorithms (which include matrix-vector products) can be computed very efficiently. Recently, we have introduced a new AMP framework in this direction. For further details, we refer to \cite{Cakmak25}. } 
\section*{Acknowledgement}
The authors would like to thank the associate editor and reviewers, whose comments have greatly improved the quality of the presentation. 
\appendix

\section{Preliminaries: Concentration Inequalities with \texorpdfstring{$\mathcal L^p$}{TEXT} norm} \label{preliminariesop}
{ We recall the notion of concentration in terms of $\Op{\kappa}$ notation introduced in \eqref{opnotation}-\eqref{gopnotation}}. 
In this section, we present its elementary properties which are frequently used in the following proofs.  	
\begin{lem}\label{lemma:op_properties}
	Consider the (scalar) random variables ${\rm a} = \Op{\kappa}$ and ${\rm b} = \Op{\tilde\kappa}$. Then, we have
	\begin{align}
		{\rm a + b} &= \Op{\max(\kappa, \tilde \kappa)}\label{eq:op_sum}\\
		{\rm ab}&= \Op{\kappa \tilde \kappa}\label{eq:op_prod}\\
		\sqrt{\rm a}&=\Op{\sqrt{\kappa}}.\label{eq:op_sqrt_pos}
	\end{align}
	\begin{proof}
		The results \eqref{eq:op_sum} and \eqref{eq:op_prod} follow from the Minkowski inequality  (i.e., $\Vert {\rm a}+{\rm b} \Vert_{\mathcal L^p}\leq\Vert {\rm a}\Vert_{\mathcal L^p}+\Vert{\rm b} \Vert_{\mathcal L^p}$) and H\"older’s inequality (i.e.,  $\Vert {\rm a}{\rm b} \Vert_{\mathcal L^p}\leq\Vert {a}\Vert_{\mathcal L^{2p}}\Vert{\rm b} \Vert_{\mathcal L^{2p}}$), respectively. The result \eqref{eq:op_sqrt_pos} follows from the inequality $\Vert {a}^{1/2} \Vert_{\mathcal L^{p}}\leq \Vert {\rm a}^{1/2} \Vert_{\mathcal L^{2p}}=\Vert {\rm a} \Vert_{\mathcal L^{p}}^{\frac 1 2}$\;.
	\end{proof}
\end{lem}
\begin{lem}\label{keycon}\cite[Lemma~7.8]{erdHos2017dynamical}
	Consider the random vector $\underline\av\in \CC^{N}$ where $\underline{\av} \sim _{\rm i.i.d.} {\rm a }$ 
	with ${\rm a}=\Op{1}$. Then,
	\[\frac 1{N} \sum_{n\in[N]}{\rm a}_n=\mathbb E[{\rm a}]+\Op{N^{-\frac1 2}}\;.\]
\end{lem}	

\begin{lem}\label{eq:ip_concentration}
	Consider the random vectors $\underline\av,\underline\bv\in \C^{N}$ where  $\underline \av\sim_\text{i.i.d.} {\rm a}$ and  $\underline \bv\sim_\text{i.i.d.} {\rm b}$  with ${\rm a}=\Op{1}$ and ${\rm b}=\Op{1}$. Then, for any $\widehat{\underline\av}\simeq\underline\av$ and $\underline{\widehat \bv}\simeq \underline\bv$, we have
	\begin{equation}
		\frac{1}{N} \widehat{\underline \av}^\herm\underline{\widehat \bv}= \mathbb E[{\rm a}^* {\rm b}] + \Op{N^{-\frac 1 2}}\;.
	\end{equation}
	\begin{proof}
		Let $\underline{\matr \delta_a}\eqdef\widehat{\underline\av}-{\underline\av} $ and $\underline{\matr \delta_b}\eqdef\widehat{\underline\bv}-{\underline\bv} $ with noting that e.g. $\underline{\matr \delta_{a}}=\Op{1}$. From Lemma~\ref{keycon} we have $ \underline{\av}=\Op{\sqrt{N}}$ and $ \underline{\bv}=\Op{\sqrt{N}}$. Hence, we get from the arithmetic rules  \eqref{eq:op_sum} and \eqref{eq:op_prod}  that
		\begin{equation}
			\underline{\widehat \av}^\herm\underline{\widehat \bv}-{\underline \av}^\herm\underline{ \bv}=\underline{ \av}^\herm\underline{\matr\delta_b}+\underline{ \bv}^\herm\underline{\matr\delta_a}+\underline{\matr \delta_a}^\herm \underline{\matr\delta_b}=\Op{\sqrt{N}}\;.
		\end{equation}
		From the product rule \eqref{eq:op_prod} we have
		${\rm a}^\star {\rm b}=\Op{1}$. Thus, from Lemma~\ref{keycon} we have the concentration
		$\frac{1}{N} {\underline \av}^\herm\underline{\bv}= \mathbb E[{\rm a}^\star {\rm b}] + \Op{N^{-1/2}}$. This completes the proof. 
	\end{proof}
	
\end{lem}

\section{Proof of Theorem~\ref{Th1}}\label{Sketchofproof}
In this section, we provide a nearly self-contained proof of Theorem~\ref{Th1}. Auxiliary lemmas are proved in Appendix~\ref{proof_auxilary} in order to maintain the flow of exposition.

It is useful to define the dynamics of residuals  for each $(u,t)\in[U]\times[T]$\
\begin{subequations}
	\begin{align}
		\widehat{\matr\Psi}_u^{(t)}&\eqdef \matr \Gamma_u^{(t)}-\matr\Theta_u\\\
		\widehat{\matr\Phi}_u^{(t)}&\eqdef \Rm_u^{(t)}-\Xm_u\;.
	\end{align} 
\end{subequations}
We recall that $\Ym=\sum_{u\in [U]}\matr\Theta_u+\Nm$. Then, from the original AMP dynamics \eqref{AMP_alg} we have the recursion of ``residual-AMP'' dynamics for $t=1,2,\cdots, T$  as
\begin{subequations}
	\label{AMP_algnew}
	\begin{align}
		\widehat{\matr\Psi}_u^{(t)}&=\Sm_u\Fm_u^{(t)}-\alpha_u\Qm_u^{(t)}\Zm^{(t-1)}\\
		\Zm^{(t)}&=\Nm-\sum_{u\in [U]}\widehat{\matr\Psi}_u^{(t)}\\
		\widehat{\matr\Phi}_u^{(t)}&=\Sm_u^\herm\Zm^{(t)}+\Fm_u^{(t)}\\
		\Fm_u^{(t+1)}&= \eta_{u,t}(\Xm_u+\widehat{\matr\Phi}_u^{(t)})-\Xm_u\;,
	\end{align}
\end{subequations}
with the initialization ${\Fm}_u^{(1)}= \Xm_u^{(1)}-\Xm_u$. 

Note that our goal is to verify the high-dimensional representations (for any $(u,t)\in[U]\times[T]$) 
\begin{subequations}
	\label{provethis}
	\begin{align}
		\widehat{\matr \Psi}_u^{(t)}&\simeq{\matr \Psi}_u^{(t)}\\
		\widehat{\matr \Phi}_u^{(t)}&\simeq \matr\Phi_u^{(t)}.
	\end{align}   
\end{subequations}
To this end, we use the idea of the ``Householder Dice'' representation of AMP dynamics introduced in \cite{lu2021householder}. Such a representation can be seen as an intermediate step in the commonly used ``conditioning'' AMP proof technique \cite{Bolthausen}. Namely, it clarifies how to represent the original AMP dynamics---that are coupled by a random matrix ($\Sm_u$, in our case)---as equivalent dynamics without a random matrix (i.e., $\Sm_u$-free). This step has no approximation and requires no assumption other than the underlying random matrix assumption (i.e., $\Sm_u\sim\mathcal {CN}(\matr 0,\Id_{N_u}/L)$). We also refer the reader to \cite{li2022non,Cakmakisit24} where a similar AMP proof strategy is used.

At a high level, we have two main steps in the proof. In the first step (Section~\ref{step1}), we use the block Gram-Schmidt process to express the dynamics of the residual-AMP \eqref{AMP_algnew} in terms of an equivalent $\Sm_u$-free dynamics, called the ``Householder Dice representation''. In the second step (Section~\ref{step2}), we use the properties of the notion of $\mathcal L^p$ concentration given in Appendix~\ref{preliminariesop} to establish the high-dimensional equivalence of the Householder Dice representation, and thereby it will give us the high-dimensional equivalent of the original residual-AMP dynamics \eqref{AMP_algnew}.

For convenience, we will use the following notation for matrices throughout the proof: Consider matrices $\Am$ and $\Bm$, both with the same number of rows, denoted by $N$. Then, we define
\begin{equation}\label{eq:nip}
	\langle \Am,\Bm\rangle\eqdef \frac{1}{N} \Am^{\herm}\Bm
\end{equation}
whenever the dependency on $N$ is clear from the context.

\subsection{Step-1: The Householder Dice equivalent}\label{step1}
We will use an adaptive representation of the Gaussian random matrices (i.e., $\Sm_u$ in the context) by using the following result (see also \cite[Eq.~(11)]{lu2021householder}).
\begin{lem}\label{G_const}
	Let the random matrices $\Vm\in \C^{L\times \bm}$, $\Um\in \C^{N\times \bm}$ and $\tilde{\Sm}\in \C^{L\times N}$ be mutually independent. Let  $\Pm_\Vm^\perp\eqdef \Id-\frac 1 L\Vm\Vm^\herm$ be a projection matrix, $\Um\sim_\text{i.i.d.} \mathcal{C N}(\matr 0,\Id_\bm)$ and $\tilde\Sm\sim_\text{i.i.d.} \mathcal{C N}(\matr 0,\Id_N/L)$. Let 
	\begin{equation}
		\Sm=\frac{\Vm \Um^\herm}{L}+\Pm_\Vm^\perp\tilde\Sm\;.
	\end{equation}
	Then, $\Sm\sim\tilde\Sm$ and $\Sm$ is (statistically) independent of $\Vm$. 
	\begin{proof}
		Let ${\rm vec}(\Sm)$ denote the column vector obtained by stacking the column vectors of $\Sm$ below one another, i.e. 
		${\rm vec}(\Sm)= 
		[\underline\sv_1^\top,\underline\sv_2^\top,\cdots \underline \sv_N]^\top$. Then, conditioned on $\Vm$, it is easy to verify that ${\rm vec}({\Sm})$ is a complex Gaussian vector with zero mean and covariance 
		\begin{align}
			&\Id_N\otimes \left(\frac{1}{L^2}\Vm\Vm^\herm+\frac 1 L(\Pm_\Vm^\perp)^2\right)\nonumber \\
			&=\Id_N\otimes \left(\frac{1}{L^2}\Vm\Vm^\herm+\frac 1 L\Pm_\Vm^\perp\right)\\
			&=\Id_N \otimes \frac 1 L\Id_L\;.
		\end{align}
		Thereby, conditioned on $\Vm$, we have $\Sm\sim_\text{i.i.d.} \mathcal{C N}(\matr 0,\Id_N/L)$. Furthermore, since the conditional distribution of $\Sm$ is invariant from $\Vm$, $\Sm$ and $\Vm$ are independent. 
	\end{proof}
\end{lem}
Note that we can also consider the conjugate representation of Lemma~\ref{G_const} as 
\begin{equation}
	\Sm^\herm= \frac{\Um\Vm^\herm}{L}+\tilde\Sm^\herm \Pm_\Vm^\perp\label{conjlemma}\;.
\end{equation}
For example, let us consider a matrix $\Zm \in \CC^{L\times \bm,}$ with $\text{span}(\Zm) = \text{span}(\Vm)$. So that $\Pm_{\Vm}^\perp\Zm=\matr 0$. Hence, we have the representation of the product $\Sm^\herm\Zm $ 
\begin{align}
	\Sm^\herm\Zm&= \Um\langle \Vm,\Zm \rangle
\end{align} 
which involves solely the lower-dimensional random element $\Um\sim_\text{i.i.d.} \mathcal{C N}(\matr 0,\Id_F)$. 

To extend this idea to e.g. the dynamics  $\Sm_u^\herm\Zm^{(t)}$ (and similarly to the dynamics $\Sm_u\matr\Fm_u^{(t)}$), we will employ the block Gram-Schmidt process to construct a set of orthogonal matrices, denoted as 
\[\Vm^{(1:t)} \equiv \{\Vm^{(1)}, \Vm^{(2)}, \ldots, \Vm^{(t)}\}\]
such that $\text{span}(\Vm^{(1: t)}) = \text{span}(\Zm^{(1: t)})$. We then recursively apply Lemma~\ref{G_const} to obtain an $\Sm_u$-free equivalent of the dynamics $\Sm_u^\herm\Zm^{(t)}$.

We first introduce the block Gram-Schmidt notation:
Let $\Vm^{(1:t-1)}=\{\Vm^{(s)}\in \CC^{N\times \bm}\}_{s\in[t-1]}$ such that $\langle \Vm^{(i)}, \Vm^{(j)}\rangle=\mathcal \delta_{ij}\Id_\bm$ for all $i,j$. Then, for any matrix $\Bm\in \CC^{N\times \bm}$, by using the block Gram-Schmidt (orthogonalization) process, 
we can always construct the new (semi-unitary) matrix
\[ \Vm^{(t)}\eqdef \GS{\Bm}{\Vm^{(1:t-1)}} 
\]
such that $\langle \Vm^{(t)}, \Vm^{(s)}\rangle=\delta_{st}\Id_\bm$  and we have the block Gram-Schmidt decomposition
\begin{equation}
	\Bm=\sum_{1\leq s\leq t}\Vm^{(s)}\langle\Vm^{(s)},\Bm \rangle\;.
\end{equation}
We refer to Appendix~\ref{BGS} for the explicit definition of the block Gram-Schmidt notation.
\subsubsection{Iteration Step \texorpdfstring{$t=1$}{Lg}} 
We begin with deriving the Householder Dice representation of the { residual}-AMP dynamics \eqref{AMP_algnew} for the iteration step $t=1$.  Let  
\begin{equation}
	\Vm_u^{(1)}=\GS{\Fm_u^{(1)}}{\emptyset}\;.
\end{equation}
We recall that $\alpha_u=N_u/L$. Then, we use Lemma~\ref{G_const} to represent $\Sm_u$ as
\begin{equation}
	\Sm_u={\sqrt{\alpha_u}}\frac{\Gm_{u}^{(1)}(\Vm_{u}^{(1)})^\herm}{N_u}+ \Sm_u^{(0,1)}\Pm_{\Fm_u^{(1)}}^\perp\label{iter1}
\end{equation}
where we introduce the arbitrary (lower-dimensional) random element $\Gm_u^{(1)}\sim\Nc(\matr 0,\Id_\bm)$ and arbitrary random matrix $\Sm_u^{(0,1)}\sim \Sm_u$. 
Note that  $\Pm_{\Fm_u^{(1)}}^\perp\Fm_u^{(1)}=\matr 0$. Then, from   \eqref{iter1} we have
\begin{align}
	\widehat{\matr \Psi}_u^{(1)}&=\Sm_u\Fm_u^{(1)}={\sqrt{\alpha_u}}\Gm_{u}^{(1)}\langle\Vm_{u}^{(1)},\Fm_{u}^{(1)} \rangle\;.
\end{align}
To derive the $\Sm_u$-free representation of $\widehat{\matr \Phi}_u^{(1)}$ we first construct the $L\times \bm$ unitary matrix 
\[\Vm^{(1)}=\GS{\Zm^{(1)}}{\emptyset}\]
and use Lemma~\ref{G_const} to represent $\Sm_u^{(0,1)}$ in \eqref{iter1} ~as
\begin{equation}
	\Sm_u^{(0,1)}= \frac{\Vm^{(1)}(\Um_u^{(1)})^\herm}{L}+ \Pm_{\Zm^{(1)}}^\perp\Sm_u^{(1,1)}\label{X01}\;,
\end{equation}
where we introduce a new (arbitrary) random element $\Um_u^{(1)}\sim\Nc(\matr 0,\Id_\bm)$ and a new random matrix $\Sm_u^{(1,1)}\sim \Sm_u$. Hence, \eqref{iter1} reads as
\begin{align}
	\Sm_u&=  {\sqrt{\alpha_u}}\frac{\Gm_{u}^{(1)}(\Vm_{u}^{(1)})^\herm}{N_u}+\frac{\Vm^{(1)}(\widehat\Um_u^{(1)})^\herm}{L}\nonumber\\&+\Pm_{\Zm^{(1)}}^\perp\Sm_u^{(1,1)}\Pm_{\Fm_u^{(1)}}^\perp\label{Sm11}\;,
\end{align}
where for short we define $\widehat{\Um}_u^{(1)}\eqdef \Pm_{\Fm_u^{(1)}}^\perp{\Um}_u^{(1)}$. Note that $\Pm_{\Zm^{(1)}}^\perp\Zm^{(1)}=\matr 0$. Then, from \eqref{Sm11}  we have 
\begin{align}\label{eq:rmf1st}\widehat{\matr\Phi}_u^{(1)}&=\Sm_u^\herm\Zm^{(1)}+\Fm_u^{(1)}\\
	&=\Vm_{u}^{(1)}\frac{\langle\Gm_{u}^{(1)},\Zm^{(1)}\rangle}{\sqrt{\alpha_u}}+\widehat\Um_u^{(1)}\langle\Vm^{(1)},\Zm^{(1)}\rangle+\Fm_u^{(1)}\\
	&=\widehat\Um_u^{(1)}\langle\Vm^{(1)},\Zm^{(1)}\rangle\nonumber\\
	&+\Vm_u^{(1)}(\frac{\langle\Gm_{u}^{(1)},\Zm^{(1)}\rangle}{\sqrt{\alpha_u}}+\langle\Vm_u^{(1)},\Xm_u^{(1)}\rangle)\label{eq:rmf2nd}\;.
\end{align}  
In step \eqref{eq:rmf2nd} we use the block Gram-Schmidt decomposition 
\begin{equation}	\Fm_u^{(1)}=\Vm_u^{(1)}\langle\Vm_u^{(1)},\Fm_u^{(1)}\rangle\;.
\end{equation}
This completes the Householder Dice representation of the dynamics \eqref{AMP_algnew} for $t=1$.
\subsubsection{Iteration Step \texorpdfstring{$t=2$}{Lg}}
Moving on to the second iteration step, we fix $t=2$ but mimic the notations in a manner that the arguments can be recalled for any $t>1$. We construct the matrices (for each $u\in[U]$)
\begin{align}
	\Vm_u^{(t)}&=\GS{\Fm_u^{(t)}}{\matr \Vm_u^{(1:t-1)}} . \label{wrfl}
\end{align}
Then, using Lemma~\ref{G_const}, we represent $\Sm_u^{(t-1,t-1)}$ (i.e. $\Sm_u^{(1,1)}$ in \eqref{Sm11}) as
\begin{equation}
	\Sm_u^{(t-1,t-1)}= {\sqrt{\alpha_u}}\frac{\Gm_{u}^{(t)}(\Vm_{u}^{(t)})^\herm}{N_u}+ \Sm_u^{(t-1,t)}\Pm_{\Vm_u^{(t)}}^\perp\;,
\end{equation}
where we introduce the new  random element $\Gm_u^{(t)}\sim \Gm_u^{(1)}$ and a new random matrix $\Sm_u^{(t-1,t)}\sim \Sm_u$.  Then, \eqref{Sm11} reads as
\begin{align}
	\Sm_u&={\sqrt{\alpha_u}}\sum_{s=1}^{t}\frac{\widehat\Gm_{u}^{(s)}(\Vm_{u}^{(s)})^\herm}{N_u}+\sum_{s=1}^{t-1}
	\frac{\Vm^{(s)}(\widehat\Um_u^{(s)})^\herm}{L}\nonumber\\&+
	\Pm_{\Zm^{(1:t-1)}}^\perp\Sm_u^{(t-1,t)}\Pm_{\Fm_u^{(1:t)}}^\perp\;\label{desired0}\;.
\end{align}
Here the matrices $\Pm_{\Fm_u^{(1:t)}}^\perp$ and $\Pm_{\Zm^{(1:t-1)}}^\perp$ denote the projections onto the orthogonal complement of the subspace spanned by $\Fm_u^{(1:t)}$ and $\Zm^{(1:t-1)}$, respectively. For example, 
\[  \Pm_{\Zm^{(1:t-1)}}^\perp=\Id-\frac 1 L \sum_{s=1}^{t-1}\Vm^{(s)} (\Vm^{(s)})^\herm\;.\]
Also, for convenience, we have defined 
\begin{align}
	\widehat{\Gm}_u^{(t)} &\eqdef \Pm_{\Zm^{(1:t-1)}}^\perp {\Gm}_u^{(t)}\\
	\widehat{\Um}_u^{(t)} &\eqdef \Pm_{\Fm^{(1:t)}}^\perp {\Um}_u^{(t)}\;.
\end{align}
By using the representation \eqref{desired0} we then obtain 
\begin{align}
	&\widehat{\matr \Psi}_u^{(t)}={\sqrt{\alpha_u}}\sum_{s=1}^{t}\widehat\Gm_{u}^{(s)}\langle\Vm_{u}^{(s)},\Fm_{u}^{(t)} \rangle\nonumber \\&+{\alpha_u}\sum_{s=1}^{t-1}\Vm^{(s)}\langle\widehat\Um_{u}^{(s)},\Fm_{u}^{(t)} \rangle-{\alpha_u}\Zm^{(t-1)}\Qm_u^{(t)}
	\\
	&={\sqrt{\alpha_u}}\sum_{s=1}^{t}\widehat\Gm_{u}^{(s)}\langle\Vm_{u}^{(s)},\Fm_{u}^{(t)} \rangle\nonumber \\&
	+{\alpha_u}\sum_{s=1}^{t-1}\Vm^{(s)}\left(\langle\widehat\Um_{u}^{(s)},\Fm_{u}^{(t)} \rangle-\langle\Vm^{(s)},\Zm^{(t-1)}\rangle\Qm_u^{(t)}\right)\;.
\end{align}
Here, the latter step uses  the block Gram-Schmidt decomposition 
\begin{equation}
	\Zm^{(t-1)}=\sum_{s=1}^{t-1}\Vm^{(s)}\langle\Vm^{(s)},\Zm^{(t-1)}\rangle\;.\label{bgsZm}
\end{equation}
We now construct $\Vm^{(t)}=\GS{\Zm^{(t)}}{\Vm^{(1:t-1)}}$ and
\begin{equation}
	\Sm_u^{(t-1,t)}= \frac{\Vm^{(t)}(\Um_{u}^{(t)})^\herm}{L}+ \Pm_{\Vm^{(t)}}^\perp\Sm_u^{(t,t)}\;,
\end{equation}
with the arbitrary random matrix $\Sm_u^{(t,t)}\sim \Sm_u$ and an arbitrary random element ${\Um}_u^{(t)}\sim {\Um}_u^{(1)}$.Plugging this results in \eqref{desired0} leads to 
\begin{align}
	\Sm_u&= {\sqrt{\alpha_u}}\sum_{s=1}^{t}\frac{\widehat\Gm_{u}^{(s)}(\Vm_{u}^{(s)})^\herm}{N_u}+\sum_{s=1}^{t}
	\frac{\Vm^{(s)}(\widehat\Um_u^{(s)})^\herm}{L}\nonumber\\&+
	\Pm_{\Vm^{(1:t)}}^\perp\Sm_u^{(t,t)}\Pm_{\Vm_u^{(1:t)}}^\perp\;.\label{desired}
\end{align}
Using this construction we get
\begin{align}		
	\widehat{\matr\Phi}_u^{(t)}&=
	\sum_{s=1}^{t} \Vm_{u}^{(s)}
	\frac{\langle\widehat\Gm_{u}^{(s)},\Zm^{(t)}\rangle}{\sqrt{\alpha_u}}\nonumber\\&+\sum_{s=1}^{t}\widehat\Um_u^{(s)}\langle\Vm^{(s)},\Zm^{(t)}\rangle+\Fm_u^{(t)}\\
	&=\sum_{s=1}^{t}\widehat\Um_u^{(s)}\langle\Vm^{(s)},\Zm^{(t)}\rangle\nonumber\\&+\sum_{s=1}^{t} \Vm_{u}^{(s)}\left(\frac{\langle\widehat\Gm_{u}^{(s)},\Zm^{(t)}\rangle}{\sqrt{\alpha_u}}+\langle\Vm_u^{(s)},\Fm_u^{(t)}\rangle\right).
\end{align}
Again, as to the latter term above, we note from block Gram-Schmidt orthogonalization that 
\begin{equation}	\Fm_u^{(t)}=\sum_{s=1}^{t}\Vm_u^{(s)}\langle\Vm_u^{(s)},\Fm_u^{(t)}\rangle\;. \label{bgsXm}
\end{equation}
\subsubsection{Iteration Steps \texorpdfstring{$t>2$}{Lg}}
Clearly, for iteration steps $t=3,4,\ldots$, we can repeat the same arguments as for the case $t=2$. Specifically, at each $t\in~[T]$, we utilize equations \eqref{desired0} and \eqref{desired} to compute $\widehat{\matr \Psi}_u^{(t)}$ and $\widehat{\matr\Phi}_u^{(t)}$ in the  residual AMP dynamics \eqref{AMP_algnew}, respectively. By following this process, we arrive at a $\{\Sm_u\}$-free representation of the dynamics \eqref{AMP_algnew}: We commence with the identical initialization of $\Fm_u^{(1)}$ from \eqref{AMP_algnew} for all $u\in [U]$ and proceed for $t=1,2,\ldots, T$ as
\begin{subequations}
	\label{AMP_mf}
	\begin{align}
		\Vm_u^{(t)} &= \GS{\Fm_u^{(t)}}{\Vm_u^{(1:t-1)}} \\
		\widehat{\Gm}_u^{(t)} &= \Pm_{\Zm^{(1:t-1)}}^\perp {\Gm}_u^{(t)} \\
		\matr\Delta_{u}^{(t)}&= \sum_{s=1}^{t-1}{\Vm^{(s)}}(\langle{\widehat{\Um}_u}^{(s)},{\Fm}_u^{(t)}\rangle -\langle\Vm^{(s)},\Zm^{(t-1)}\rangle{\Qm}_u^{(t)}) \label{onsager1} \\
		\widehat{\matr \Psi}_u^{(t)}&=
		{\sqrt{\alpha_u}}\sum_{s=1}^{t}{\widehat{\Gm}_u^{(s)}}\langle\Vm^{(s)}_u, {\Fm}_u^{(t)}\rangle+{\alpha_u}\matr\Delta_{u}^{(t)} \\
		\Zm^{(t)}&=\Nm-\sum_{1\leq u\leq U} \widehat{\matr\Psi}_u^{(t)} \\
		\Vm^{(t)} &= \GS{\Zm^{(t)}}{ \Vm^{(1:t-1)}} \\
		\widehat{\Um}_u^{(t)} &=\Pm_{\Fm_u^{(1:t)}}^\perp  {\Um}_u^{(t)} \\
		\widetilde{\matr 
			\Delta}_{u}^{(t)}&=\sum_{s=1}^ {t}\Vm_u^{(s)}\left(\frac {\langle{\widehat{\Gm}}_u^{(s)},\Zm^{(t)}\rangle} {\sqrt{\alpha_u}}+\langle \Vm_u^{(s)},{\Fm}_u^{(t)}\rangle\right)\label{onsager2}\\		
		\widehat{\matr\Phi}_u^{(t)}&= \sum_{s=1}^{t} \widehat{\Um}_u^{(s)}\langle\Vm^{(s)},\Zm^{(t)}\rangle+\widetilde{\matr\Delta}_{u}^{(t)}\\
		\Fm_u^{(t+1)}&=\eta_{u,t}(\Xm_u+\widehat{\matr\Phi}_u^{(t)})-\Xm_u\;.
	\end{align}
\end{subequations}
Here, by abuse of notation, we set e.g. $\Vm^{(1:0)}=\emptyset$. Moreover, $\{\Gm_u^{(t)}\in\CC^{L\times \bm}, \Um_u^{(t)}\in \CC^{N_u\times \bm}: (u,t)\in [U]\times [T]\}$ is the set of all mutually independent arbitrary random elements with
\[ \Gm_u^{(t)}\sim_{\text{i.i.d.}}\mathcal{CN}(\matr 0;\Id_\bm) ~~ \text{and} ~~\Um_u^{(t)}\sim_{\text{i.i.d.}}\mathcal{CN}(\matr 0;\Id_\bm)\nonumber \;.\] 

Formally, we have verified the following result. 
\begin{lem}\label{lemmaseq} Fix T. Let $\Sm_u\sim_\text{i.i.d.} \mathcal{C N}(0,\Id_{N_u}/L)$ where $\Sm_u$ and $\Sm_{u'}$ are mutually independent for any $u\neq u'$.  Then, the joint probability distribution of the sequence of matrices $\{\widehat{\matr \Psi}_u^{(1:T)},\widehat{\matr \Phi}_u^{(1:T)}\}_{u\in [U]}$ generated by dynamics \eqref{AMP_algnew} is equal to that of the same sequence generated by dynamics~\eqref{AMP_mf}.
\end{lem} 
\subsection{Step-2: The High Dimensional Equivalent}\label{step2}
We next derive the high-dimensional equivalent of the dynamics \eqref{AMP_mf} (and thereby the original residual-AMP dynamics \eqref{AMP_algnew}). As a warm-up, we begin by verifying 
\begin{equation}
	\widehat{\Gm}_u^{(t)}\simeq\Gm_u^{(t)} \text{ and  }\widehat{\Um}_u^{(t)}\simeq\Um_u^{(t)}\;, \quad (u,t)\in[U]\times[T] \label{conprojgaus}
\end{equation}
where we recall the high-dimensional equivalence notation $\widehat\Am\simeq \Am$ in \eqref{simeq}. The concentrations in  \eqref{conprojgaus} are evidently follow from the following result:
\begin{lem}\label{rem2}
	Consider a random matrix $\Gm\in\C^{N\times \bm}$ with $\Gm\sim_\text{i.i.d.} \mathcal{C N}(\matr 0,\Id_\bm)$ and fixed $\bm$ (that does not depend on $N$). For some fixed $t$, let $\Pm \eqdef  \sum_{s \leq t}\Vm^{(s)}(\Vm^{(s)})^\herm$ with $\langle\Vm^{(s)},\Vm^{(t)}\rangle=\delta_{ts}\Id_{\bm}$ for any $t,s$. 
	Let $\Pm$ and $\Gm$ be independent. Then, $\Pm\Gm=\Op{1}$. 
	\begin{proof}
		Since $\Pm^2=\Pm$, we have $\Vert \Pm\Gm\Vert_{\rm F}^2=\sum_{s\leq t}\Vert \widetilde\Gm\Vert_{\rm F}^2$ with $\widetilde\Gm\eqdef(\Vm^{(s)})^\herm\Gm$. Notice also that $\widetilde\Gm\sim \mathcal {CN}(\matr 0,\Id_\bm)$; so  that $\widetilde\Gm=\Op{1}$. Hence, from the sum rule \eqref{eq:op_sum} we have $\Vert \Pm\Gm\Vert_{\rm F}^2=\Op{1}$. This implies from the identity \eqref{eq:op_sqrt_pos} that $ \Pm\Gm=\Op{1}$. 
	\end{proof}
\end{lem}
To derive the high dimensional equivalence of the complete dynamics we essentially need to work on the concentration properties of the $\bm\times \bm$ empirical matrices for $1\leq s\leq t\leq T$:
\begin{align}
	\widehat{{\Bm}}_u^{(t,s)}&\eqdef \langle\Vm_u^{(s)},{\Fm}_u^{(t)}\rangle\\
	\widehat{\Bm}^{(t,s)}&\eqdef \langle\Vm^{(s)},\Zm^{(t)}\rangle\;.
\end{align}
\begin{lem}\label{lem1}
	We have for all $1\leq s\leq t\leq T$ (and $u\in [U])$
	\begin{align}
		\widehat{{\Bm}}_u^{(t,s)}=\Op{1} \text{ and  }
		\widehat{\Bm}^{(t,s)}=\Op{1}\;.\label{BO1}
	\end{align}
	\begin{proof}
		See Appendix~\ref{lem1_proof};
	\end{proof}
\end{lem}
Our goal is to verify that these empirical matrices concentrate around deterministic matrices as
\begin{align}
	\widehat{{\Bm}}_u^{(t,s)}&={\Bm}_u^{(t,s)}+\Op{L^{-\frac 1 2}}\\
	\widehat{{\Bm}}^{(t,s)}&={\Bm}^{(t,s)}+\Op{L^{-\frac 1 2}}\;.
\end{align}
To start we note from block Gram-Schmidt decomposition that for any $t\in [T]$
\begin{align}
	{\Fm}_u^{(t)}
	&=\sum_{1\leq s\le t}\Vm_u^{(s)}\widehat{\Bm}_u^{(t,s)}\label{tildeXm}\\
	\Zm^{(t)}&=\sum_{1\leq s\le t}\Vm^{(s)}\widehat{\Bm}^{(t,s)}\;.\label{fBhat}
\end{align}
Recall that e.g. $\langle\Vm^{(t)},\Vm^{(s)}\rangle=\delta_{ts}\Id_F$ for any $t,s$. Thus, we have from  \eqref{tildeXm} (and resp. \eqref{fBhat}) that  $\widehat{\Bm}^{(t,s)}$ (and resp. $\widehat{\Bm}^{(t,s)}$) satisfy the equations of \emph{block Cholesky decomposition} (see Appendix~\ref{lem3_proof}) for all $1 \leq s\leq t \leq T$:
\begin{align}
	(\widehat{\Bm}_u^{(t,s)})^\herm\widehat {\Bm}_u^{(s,s)}&= \langle\Fm_u^{(t)},\Fm_u^{(s)}\rangle-\sum_{1\leq s'<s}(\widehat{\Bm}_u^{(t,s')})^\herm\widehat{\Bm}_u^{(s,s')}\label{Bhats1}\\
	(\widehat{\Bm}^{(t,s)})^\herm {\Bm}^{(s,s)}&= \langle\Zm^{(t)},\Zm^{(s)}\rangle-\sum_{1\leq s' <s}(\widehat{\Bm}^{(t,s')})^\herm\widehat{\Bm}^{(s,s')}\;.\label{Bhats2}
\end{align}
We now recall the two-time cross-correlation $\Cm^{(t,s)}$ and $\Cm_u^{(t,s)}$ in the state-evolution (i.e., Definition~\ref{SEdef}). 
We then express the deterministic matrices ${\Bm}_u^{(t,s)}$ and ${\Bm}_u^{(t,s)}$
in terms of the \emph{block Cholesky decomposition} equations for all $1 \leq s\leq t \leq T$:
\begin{align}
	({\Bm}_u^{(t,s)})^\herm{\Bm}_u^{(s,s)}&=  \frac {\Cm_u^{(t,s)}}{\alpha_u}-\sum_{1\leq s'<s}({\Bm}_u^{(t,s')})^\herm{\Bm}_u^{(s,s')}\\
	({\Bm}^{(t,s)})^\herm{\Bm}^{(s,s)}&=  \Cm^{(t,s)}-\sum_{1\leq s' <s}({\Bm}^{(t,s')})^\herm{\Bm}^{(s,s')}\;.
\end{align}
Before moving on, we introduce the following notation of matrices:   $\Am^{(1:t')}$ denotes a $t'\bm \times t'\bm$ matrix, and its $(t, s)$ indexed $F \times F$ block matrix is denoted by $\Am^{(t, s)}$, specifically we have
\begin{equation}
	\Am^{(1:t')}\equiv\sum_{1\leq s, t\leq t'}\ev_t^\top \ev_s\otimes\Am^{(t,s)} \label{notation_set_up}
\end{equation}
where $\ev_t\in \RR^{1\times t'}$ denotes the standard basis vector with $(\ev_t)_s=\delta_{ts}$ for all $t,s'\in[t']$.

\begin{lem}\label{lem2}
	Recall the two-time deterministic cross-correlation matrix $\Cm_u^{(t,s)}$ as in \eqref{SE}. Then, in proving Theorem~\ref{Th1} we can assume without loss of generality that 
	\begin{equation}
		\Cm_u^{(1:T)}>\matr 0\;.  \label{singularity}
	\end{equation}
	\begin{proof}
		See Appendix~ \ref{lem2_proof}.
	\end{proof}
\end{lem}
Note that \eqref{singularity} implies $\Cm^{(1:T)}>\matr 0$. Hence,  from Lemma~\ref{lem2} we can assume without loss of generality that the blocks ${\Bm}_u^{(s,s)}$ and ${\Bm}^{(s,s)}$ for $s\in[T]$ are non-singular and ${\Bm}_u^{(t,s)}$ and ${\Bm}^{(t,s)}$ can be uniquely constructed from the block Cholesky decomposition equations \eqref{Bhats1}, and \eqref{Bhats2}, respectively. 

Now let $\mathcal H_{t'}$ stand for the hypothesis that
\begin{subequations}\label{InductiveP}
	\begin{align}
		\widehat{{\Bm}}_u^{(1:t')}&={ {\Bm}}_u^{(1:t')}+\Op{L^{-\frac 1 2}}\\
		\widehat{\Bm}^{(1:t')}&={\Bm}^{(1:t')}+\Op{L^{-\frac 1 2}}\\
		\widehat{\matr \Psi}_u^{(t)}&\simeq{\sqrt{\alpha_u}}\sum_{1\leq s\leq t} \Gm_u^{(s)}{\Bm}_{u}^{(t,s)}~t\in [t'] \label{Gamma1}\\
		\widehat{\matr \Phi}_u^{(t)}&\simeq  \sum_{1\leq s\leq t}\Um_u^{(s)}\Bm^{(t,s)}~~\qquad t\in [t']\;.  \label{Psi1}
	\end{align}
\end{subequations}	
In the following, we verify that $\mathcal H_{t'-1}$ implies $\mathcal H_{t'}$ and that the base case $\mathcal H_1$ holds. Notice that this implies that \eqref{provethis} holds for all $(u,t)\in[u]\times [T]$, and thus Theorem~\ref{Th1} holds.
We first collect some useful results from which the inductive proof follows.

\begin{lem}\label{lem3}
	Let us denote the two-time empirical cross correlation in \eqref{Bhats1} and \eqref{Bhats2} as $\widehat{\Cm}_u^{(t,s)}\eqdef\frac{\langle \Fm_u^{(t)},\Fm_u^{(s)} \rangle}{\alpha_u}$ and $\widehat{\Cm}^{(t,s)}\eqdef\langle \Zm^{(t)},\Zm^{(s)} \rangle$ for all $t,s\in[T]$.  Also, recall the notation setup in \eqref{notation_set_up}.  Let \eqref{singularity} hold. 
	Then for any $t'\in[T]$ we have the implications
	\begin{align}
		&\widehat{\Cm}_u^{(1:t')}-{\Cm}_u^{(1:t')}=\Op{L^{-\frac 1 2}}\nonumber\\&\implies \widehat{\Bm}_u^{(1:t')}-\Bm_u^{(1:t')}=\Op{L^{-\frac 1 2 }}\; \\
		&\widehat{\Cm}^{(1:t')}-{\Cm}^{(1:t')}=\Op{L^{-\frac 1 2}}\nonumber\\&\implies \widehat{\Bm}^{(1:t')}-\Bm^{(1:t')}=\Op{L^{-\frac 1 2}}\;. 
	\end{align}
	\begin{proof}
		See Appendix~\ref{lem3_proof}.
	\end{proof}
\end{lem}

\begin{lem}\label{lem5}
	Let Assumption~\ref{as1} hold.  
	\begin{itemize}
		\item [(i)] If \eqref{Psi1} holds for $t\in[t'-1]$. Then, we have for all $1\leq s\leq t\leq t $ 
		\begin{subequations}
			\label{concen2}
			\begin{align}
				\langle \Fm_u^{(t)},\Fm_u^{(s)} \rangle&=\frac{1}{\alpha_u}\Cm_{u}^{(t,s)}
				+\Op{L^{-\frac 1 2 }}\label{res3}\\
				\langle \widehat{\Um}_u^{(s)}, \Fm_u^{(t)}\rangle&=\Bm^{(t-1,s)}\Qm_u^{(t)}
				+\Op{L^{-\frac 1 2 }}\label{res5}\;.
			\end{align}	
		\end{subequations}
		
		\item [(ii)] If \eqref{Gamma1} holds for $t\in[t']$. Then, we have for all $1\leq s\leq t\leq t'$ 
		\begin{subequations}
			\begin{align}
				\langle\Zm^{(t)},\Zm^{(s)} \rangle&=\Cm^{(t,s)}+\Op{L^{-\frac 1 2}}\\
				\frac{1}{\sqrt{\alpha_u}}\langle\widehat{\Gm }_u^{(s)},\Zm^{(t)} \rangle&=-{\Bm}_u^{(t,s)}+\Op{L^{-\frac 1 2}}\;.
			\end{align}
		\end{subequations}
	\end{itemize}
	\begin{proof}
		See Section~\ref{lem5_proof}.
	\end{proof}
\end{lem}

\subsubsection{Proof of the Induction Step:
	\texorpdfstring{$\mathcal H_{t'-1}\implies \mathcal H_{t'}$}{Lg}}\label{proofHt}
We begin with using  Lemma~\ref{lem3}-(i):
\begin{equation}
	\widehat{{\Cm}}_u^{(1:t')}={{\Cm}}_u^{(1:t')}+\Op{L^{-\frac 1 2}}\;.
\end{equation}
This result implies from Lemma~\ref{lem5}   (together with the assumption~\eqref{singularity}) that
\begin{equation}
	\widehat{ {\Bm}}_u^{(1:t')}={ {\Bm}}_u^{(1:t')}+\Op{L^{-\frac 1 2}}.\label{put2}
\end{equation}
Moreover, from~\eqref{res5} we get for all $1\leq s< t'$
\begin{align}
	&\langle{\widehat{\Um}_u}^{(s)},{\Fm}_u^{(t')}\rangle -\widehat{\Bm}^{(t'-1,s)}{\Qm}_u^{(t')}\nonumber\\
	&=(\Bm^{(t'-1,s)}-\widehat{\Bm}^{(t'-1,s)})\Qm_u^{(t')}+\Op{L^{-\frac 1 2}}\\
	&=\Op{L^{-\frac 1 2}}\;.
\end{align}
This implies (see \eqref{onsager1})
\begin{equation}
	\matr\Delta_u^{(t')}=\Op{1}\;.\label{put3}
\end{equation}
The results \eqref{put2} and \eqref{put3} together imply \eqref{Gamma1} holds for $t=t'$ as well. 

Then, from Lemma~\ref{lem3}-(ii) we get for all $1\leq s\leq t\leq t'$ \begin{align}
	\widehat{{\Cm}}^{(1:t')}&={{\Cm}}^{(1:t')}+\Op{L^{-\frac 1 2}}\\
	\frac 1 {\sqrt{\alpha_u}}\langle \widehat{\Gm}_u^{(s)},\Zm^{(t)} \rangle &=-{{\Bm}}_u^{(t,s)}+\Op{L^{-\frac 1 2}}\;.
\end{align}
These results respectively imply for all $t\in [t']$
\begin{subequations}
	\begin{align}
		\widehat{{\Bm}}^{(1:t')}&\overset{(a)}{=}{{\Bm}}^{(1:t')}+\Op{L^{-\frac 1 2}}\label{final Hd}\\			\widetilde{\matr\Delta}^{(t)}_u&=\Op{1}\;.
	\end{align}   
\end{subequations}
Here, the step (a) uses Lemma~\ref{lem5}. Then, the results in \eqref{final Hd} together imply \eqref{Psi1} holds for $t=t'$ as well. This completes the proof of the induction step $\mathcal H_{t'-1}\implies \mathcal H_t'$. 
\subsubsection{Proof of the Base Case: \texorpdfstring{$\mathcal H_1$}{Lg}}
We recall that $\Fm_u^{(1)}\sim_{\text{i.i.d}}(\xv_u-\xv_u^{(1)})$ such that $\xv_u=\Op{1}$ and $\xv_u^{(1)}=\Op{1}$. Hence, we have by the sum-rule \eqref{eq:op_sum} $(\xv_u-\xv_u^{(1)})=\Op{1}$. Then, following the same steps in the proof of $\mathcal H_{t'-1}\implies \mathcal H_{t'}$ (Section~ \ref{proofHt}) for $t'=1$ yields $\Hc_{1}$.


{ \section{The Proof of Theorem~\ref{the2}}\label{pthe2}
	In the sequel, we verify the convergence of the MD rate, i.e. $\beta_{\rm e}^{\rm md}\overset{(a.s)}{\rightarrow} \beta_{\infty}^{\rm md}$. The proof of the convergence of the FA rate (i.e., $\beta_{\rm e}^{\rm fa}\overset{(a.s)}{\rightarrow} \beta_{\infty}^{\rm fa}$.) follows analogously.

	We introduce the location-based actual set of active messages and the estimated set of active messages, respectively, as
	\begin{align}
		\mathcal A_u&\eqdef  \{(u,n)\in[N_u]: {\rm a}_{u,n} = 1\}\label{la}\\
		\widehat{\mathcal A}_u&\eqdef \{(u,n)\times[N_u]: \widehat {\rm a}_{u,n} = 1\}\label{lad}\;
	\end{align}
	and define the location-based MD rates as
	\begin{equation}
		\beta^{\rm md}_{{\rm e},u}= \frac{\big\vert {\widehat\Ac}_u^{\rm c}\cap \Ac_u \big\vert }{\big\vert  \Ac_u \big\vert}.  
	\end{equation}
	Notice that 
	\begin{equation}
		\beta^{\rm md}_{\rm e}=\sum_{u\in [U]}\frac{\vert\Ac_u\vert}{\vert\Ac\vert}\beta^{\rm md}_{{\rm e},u}\;.
	\end{equation}
	From Lemma~\ref{eq:ip_concentration} we note the concentration
	\begin{equation}
		\frac{1}{N_u}\vert\Ac_u\vert=\frac {1} {N_u} \sum_{n\in [N_u]}{\rm a}_{u,n}= \lambda_u+\Op{N_u^{-\frac 1 2}}.    \label{lamal}
	\end{equation}
	This implies that 
	\begin{equation}
		\frac{\vert\Ac_u\vert}{\vert\Ac\vert}\overset{(a.s)}{\rightarrow}\frac{\alpha_u\lambda_u}{\sum_{u\in [U]}\alpha_u\lambda_u}.\label{invokethis}
	\end{equation}
	So we are left with proving the convergence:
	\begin{equation}
		\beta^{\rm md}_{{\rm e},u} \overset{(a.s)}{\rightarrow}\mathbb P(\Lambda_u(\hv_u+\matr\phi)>\nu_u)\;.
	\end{equation}
	To simplify the notation, in the sequel, we remove the subscript $u$ in the notation (unless it is necessary), e.g. $\Rm\equiv\Rm_u$, ${\rm a}_{n}\equiv {\rm a}_{n,u}$, $\Lambda(\cdot)\equiv\Lambda_u(\cdot)$, 
	etc. Then, we write 
	\begin{align}
		\beta_{{\rm e},u}^{\rm md}\equiv\frac{\frac {1}{N} \sum_{n}{\rm a}_{n}{\rm u}(\Lambda(\rv_{n})-\nu)}{\frac {1} {N} \sum_{n}{\rm a}_{n}}.
	\end{align}
	We have verified the concentration of the denominator in \eqref{lamal}. Hence, we will complete the proof by verifying the (almost sure) convergence of the nominator:
	\begin{align}
		&\lim_{N\to\infty}\frac 1 N \sum_{n}{\rm a}_n{\rm u}(\Lambda(\rv_n)-\nu)\nonumber\\&\overset{?}{=}\mathbb E[{\rm a}{\rm u}(\Lambda({\rm a}\hv+\matr\phi)-\nu)]\label{anun}\\
		&=\lambda \mathbb E[{\rm u}(\Lambda(\hv+\matr\phi)-\nu)]\\
		&=\lambda\mathbb P(\Lambda(\hv+\matr\phi)>\nu).
	\end{align}
	We then complete the proof by verifying \eqref{anun}. 
	Recall that either the decision test function $\Lambda_u(\rv)$ or its reciprocal, $1/\Lambda_u(\rv)$, is assumed to be Lipschitz. We continue the proof by assuming 
	$\Lambda(\rv)$ is Lipschitz. If its reciprocal Lipschitz, the proof follows by applying the trivial transformation ${\rm u}(x-y)={\rm u}(\frac 1 y-\frac 1  x)$ for $x,y>0$. 
	
	By the Lipschitz property of $\Lambda(\rv)$ we have from \eqref{DCP}
	\begin{equation}
		\Lambda(\Rm)-\Lambda(\widetilde\Rm)=\Op{1} ~~\text{with }~~\widetilde\Rm\eqdef \Xm+\matr \Phi\;.
	\end{equation}
	On the other hand, we note that the unit step function ${\rm u}(\cdot)$ is not Lipschitz.  While this creates a hurdle to verify the concentration such a problem can be resolved by using a ``sandwiching'' approach \cite[Eq. (28)-(32)]{kuan22}. In the sequel, we introduce a simple sandwiching approach to bypass the non-Lipschitz aspect of the unit-step function. 
	
	For $\epsilon_N\eqdef N^{-\frac 1 4}$, we introduce the sequence of functions 
	\begin{align}
		{\rm u}^{(N)}_{-}(x)&\eqdef\left\{ \begin{array}{cc}
			1 &   \epsilon_N<x\\
			\frac{x}{\epsilon_N}& 0<x\leq \epsilon_N\\
			0& x\leq 0
		\end{array}\right.\\
		{\rm u}^{(N)}_{+}(x)&\eqdef {\rm u}^{(N)}_{-}(x+\epsilon_N).
	\end{align}
	Notice that we have for any $x\in \RR$
	\begin{equation}
		{\rm u}^{(N)}_{-}(x)\leq {\rm u}(x)\leq {\rm u}^{(N)}_{+}(x).
	\end{equation}
	This implies that 
	\begin{align}
		\sum_n {\rm a}_n{\rm u}_{-}^{(N)}(\Lambda(\rv_n)-\nu)&\leq \sum_{n}{\rm a}_n{\rm u}(\Lambda(\rv_n)-\nu)\\&\leq  \sum_n {\rm a}_n{\rm u}_{+}^{(N)}(\Lambda(\rv_n)-\nu).
	\end{align}
	Furthermore, for any $x,y\in \RR$ we have
	\begin{align}
		\vert {\rm u}^{(N)}_{\mp}(x)-{\rm u}^{(N)}_{\mp}(y)\vert \leq N^{\frac 1 4}{\vert x-y\vert}\;.
	\end{align}
	Then, we have obtained the bounds
	\begin{align}
		&\left\vert \sum_n {\rm a}_n{\rm u}_{\mp}^{(N)}(\Lambda(\rv_n)-\nu)- {\rm a}_n{\rm u}_{\mp}^{(N)}(\Lambda(\widetilde\rv_n)-\nu)\right\vert\\
		&\leq N^{\frac 1 4}\sum_{n}\vert \Lambda(\rv_n)- \Lambda(\widetilde\rv_n)\vert\\
		&\leq N^{\frac 3 4}\Vert\Lambda(\Rm)-\Lambda(\widetilde\Rm) \Vert=\Op{N^{\frac 3 4}}\label{bound1}\;,
	\end{align}
	where the latter inequality follows from a simple application of Cauchy–Schwarz inequality. 
	Furthermore, from Lemma~\ref{eq:ip_concentration} we have the concentration
	\begin{align}
		&\frac 1 N \sum_n {\rm a}_n{\rm u}_{\mp}^{(N)}(\Lambda(\widetilde\rv_n)-\nu)\nonumber\\
		&=\mathbb E[{\rm a}{\rm u}_{\mp}^{(N)}(\Lambda({\rm a}\hv+\matr\phi)-\nu)]+\Op{N^{-\frac 1 2}}\label{bound2}\;.
	\end{align}
	Moreover, by the dominated convergence theorem we have 
	\begin{align}
		&\lim_{N\to\infty}\mathbb E[{\rm a}{\rm u}_{\mp}^{(N)}(\Lambda({\rm a}\hv+\matr\phi)-\nu)]\nonumber\\&\quad=\mathbb E[{\rm a}{\rm u}(\Lambda({\rm a}\hv+\matr\phi)-\nu)]\label{bound3}.
	\end{align}
	Then, by invoking \eqref{bound2} and \eqref{bound3}, we have almost surely
	\begin{align}
		&\lim_{N\to\infty}\frac{1}{N}\sum_n {\rm a}_n{\rm u}_{\mp}^{(N)}(\Lambda(\rv_n)-\nu)\nonumber\\&\quad =\mathbb E[{\rm u}(\Lambda({\rm a}\hv+\matr\phi)-\nu)]\;.
	\end{align}
	This completes the proof. 
	
	\section{The Proof of Theorem~\ref{the3}}\label{pthe3}
	In the sequel, we verify the convergence of the MSE of the channel estimation for the messages in the set $\Ac^{\rm d}$,  i.e. , ${\rm mse}_{\rm e}^{\rm d}\overset{(a.s.)}{\rightarrow}{\rm mse}_{\infty}^{\rm d}$. The convergence of the power of FA, i.e., ${\rm pow}_{\rm e}^{\rm fa}\overset{(a.s.)}{\rightarrow}{\rm pow}_{\infty}^{\rm fa}$, follows analogously. 
	
	We refer to the definitions \eqref{la} and \eqref{lad} for
	the \emph{location-based} actual set of active messages $\mathcal A_u^{\rm d}$
	and the estimated set of active messages $\widehat{\mathcal A}_u $, respectively. Let $\mathcal A_u^{\rm d}\eqdef \widehat{\mathcal A}_u \cap \Ac_u $ stand for the set of correctly detected active messages at a given location $\mathcal L_u$. Then, we define the location-based MSE of channel estimation (for the messages in the set $\Ac_u^{\rm d}$)  as
	\begin{equation}
		{\rm mse}_u^{\rm d}\eqdef \frac{1}{\vert\mathcal A_u^{\rm d} \vert}\sum_{n\in\mathcal A_u^{\rm d}} \Vert \hv_{u,n}-\widehat\hv_{u,n} \Vert^2\;. 
	\end{equation}
	We note that 
	\begin{equation}
		{\rm mse}^{\rm d}=\sum_{u\in [U]}\frac{\vert\Ac_u^{\rm d}\vert}{\vert\Ac^{\rm d}\vert}{\rm mse}^{\rm d}_u.    
	\end{equation}
	In the Appendix~\ref{pthe2}, we have verified the convergences of the location-based detection rates:
	\begin{equation}
		\frac {\vert\mathcal A_u^{\rm d}\vert}{\vert\mathcal A_u\vert}\overset{(a.s)}{\rightarrow}\mathbb P(\mathcal D_u).\label{ldr}
	\end{equation}
	Also, by invoking \eqref{invokethis} together with Theorem~\ref{the2} we have 
	\begin{equation}
		\frac{\vert\mathcal A^{\rm d}\vert}{\vert\mathcal A_u\vert} \overset{(a.s)}{\rightarrow}\frac{Z\beta_\infty^{\rm d}}{\alpha_u\lambda_u}\;.
	\end{equation}
	Hence, we have the convergence  (for each $u\in[U]$)
	\begin{equation}
		\frac{\vert\Ac_u^{\rm d}\vert}{\vert\Ac^{\rm d}\vert}\overset{(a.s)}{\rightarrow}\frac{\alpha_u\lambda_u\mathbb P(\mathcal D_u)}{Z\beta_\infty^{\rm d}}.
	\end{equation}
	So we are left with proving the convergence 
	\begin{equation}
		{\rm mse}_u^{\rm d}\overset{(a.s.)}{\rightarrow}\mathbb E[\Vert \hv_{u}-\eta_{u}(\hv_{u}+\matr \phi) \Vert^2\vert \mathcal D_u]\;.\label{ldmse}
	\end{equation}
	To this end, we write  
	\begin{align}
		&\sum_{n\in\mathcal A_u^{\rm d}}\Vert \hv_{u,n}-\widehat\hv_{u,n} \Vert^2\nonumber\\
		&=\sum_{n\in N_u}{\rm u}(\nu_u-\Lambda_u(\rv_{n,u})){\rm a}_{u,n}\Vert \xv_{u,n}-\eta_{u}(\rv_{u,n})   \Vert^2\\
		&\equiv\Delta+ \sum_{n\in N_u}{\rm u}(\nu_u-\Lambda_u(\rv_{n,u})){\rm a}_{u,n}\Vert \xv_{u,n}-\eta_{u}(\widetilde\rv_{u,n})   \Vert^2\;
	\end{align}
	where we define $\widetilde\Rm_u\eqdef \Xm_u+\matr \Phi_u$ and write $a\equiv b+\Delta$ to imply that $\Delta\equiv a-b$. We have
	\begin{align}
		\vert \Delta\vert&\leq \sum_{n\in [N_u]}\vert\Vert \xv_{u,n}-\eta_{u}(\rv_{u,n}) \Vert^2- \Vert \xv_{u,n}-\eta_{u}(\widetilde\rv_{u,n})   \Vert^2\vert\nonumber  \\
		&\overset{(a)}{\leq}\sum_{n\in [N_u]} (\Vert\xv_{u,n} -\eta_{u}(\rv_{u,n})\Vert+  \Vert\xv_{u,n} -\eta_{u}(\widetilde\rv_{u,n})\Vert)\nonumber \\
		&\qquad\qquad  \times \Vert \eta_{u}(\rv_{u,n})-\eta_{u}(\widetilde\rv_{u,n})\Vert\\
		&\overset{(b)}{\leq} (\Vert \Xm_u -\eta_u(\Rm_u)\Vert_{\rm F} +\Vert \Xm_u -\eta_u(\widetilde\Rm_u)\Vert_{\rm F} ) \nonumber\\
		&\qquad\qquad  \times \Vert\eta_u(\Rm_{u})-\widetilde\eta_u(\Rm_u)\Vert_{\rm F}\\
		&\overset{(c)}{\leq} (2\Vert \Xm_u-\eta_u(\widetilde\Rm_u)\Vert_{\rm F}+\Op{{1}})\Op{1}\label{deltalb}\;.
	\end{align}
	Here, in step (a) we use the reverse triangular inequality (i.e., $\vert\Vert \av\Vert-\Vert \bv\Vert\vert \leq \Vert \av-\bv \Vert$), in step (b) we use the Cauchy–Schwarz inequality (together with the inequality $\sqrt{a+b}\leq \sqrt{a}+\sqrt{b}, a,b\geq 0$) and in step (c) we invoke the result $\Vert\eta_u(\Rm_{u})-\eta_u(\widetilde\Rm_u)\Vert_{\rm F}=\Op{1}$.  We now note from Lemma~\ref{eq:ip_concentration} that $\Vert\Am\Vert_{F}=\Op{\sqrt{N_u}}$ for an $N_u\times F$ matrix $\Am\sim_{\text{i.i.d.}}\av$ for some RV $\av=\Op{1}$. Hence, we have $\Vert \Xm_u-\eta_u(\widetilde\Rm_u)\Vert_{\rm F}=\Op{\sqrt{N_u}}$ and thereby we get the desired bound from \eqref{deltalb} as
	\begin{equation}
		\vert \Delta\vert\leq \Op{\sqrt{N_u}}.
	\end{equation}
	We then follow the steps of the sandwiching approach in Appendix~\ref{pthe2} and verify that 
	\begin{align}
		&\frac {1} {N_u}\sum_{n\in\mathcal A_u^{\rm d}}\Vert \hv_{u,n}-\widehat\hv_{u,n} \Vert^2\nonumber\\\quad &\overset{(a.s)}{\rightarrow}{\lambda_u\mathbb P(\mathcal D_u)}{\mathbb E[\Vert \hv_{u}-\eta_{u}(\hv_{u}+\matr \phi) \Vert^2\vert \mathcal D_u]}. \label{finalth3}  
	\end{align}
	Finally, we note from \eqref{ldr} that $\frac {\vert\mathcal A_u^{\rm d}\vert}{N_u}\overset{(a.s)}{\rightarrow}\lambda_u\mathbb P(\mathcal D_u)$. From \eqref{finalth3} this implies \eqref{ldmse}.}


\section{Proofs of  Lemma~\ref{lem1}, Lemma~\ref{lem2}, Lemma~\ref{lem3}, and Lemma~\ref{lem5} }\label{proof_auxilary}
In this section, we provide the proofs of the Lemmas used in the proof of Theorem~\ref{Th1}, which have relatively lengthy proofs.

\subsection{Proof of Lemma~\ref{lem1}}\label{lem1_proof}
For convenience, we work on the normalized dynamics for all $(u,t)\in[U]\times [T]$  
\begin{align}
	\widetilde \Zm^{(t)}\eqdef \frac{\Zm^{(t)}}{\sqrt L} ~~\text{and }~~\widetilde \Fm_u^{(t)}\eqdef \frac{\Fm_u^{(t)}}{\sqrt L}\;.
\end{align}
Similarly, we  define $\widetilde {\matr\Gamma}_u^{(t)}, \widetilde \Zm^{(t)},\widetilde \Ym$ and etc. 
For example, we note that 
\begin{equation}
\Vert\langle\Vm^{(s)},\Zm^{(t)}\rangle\Vert_{\rm F} \leq \frac{\Vert\Vm^{(s)}\Vert_{\rm F}}{\sqrt L}\Vert \widetilde \Zm^{(t)}\Vert_{\rm F}=\sqrt{F}\Vert \widetilde \Zm^{(t)}\Vert_{\rm F}\;.
\end{equation}
Hence, to prove the Lemma, it is enough to show that $\widetilde \Zm^{(t)}=\Op{1}$ and $\widetilde\Fm_u^{(t)}=\Op{1}$. { Also, by the assumption $\Xm_{u}\sim_{\text{i.i.d.}}\xv_u$ with $\xv_u=\Op{1}$, we have from Lemma~\ref{eq:ip_concentration} that $\widetilde\Xm_u=\Op{1}$. Hence, from the sum-rule \eqref{eq:op_sum}, verifying $ \widetilde\Xm_u^{(t)}=\Op{1}$ implies $\widetilde\Fm_u^{(t)}=\Op{1}$.} 

We note  from the AMP iterations in \eqref{AMP_alg} that
\begin{subequations}\label{tildeAMP}
	\begin{align}
		\Vert \widetilde {\matr\Gamma}_u^{(t)}\Vert_{\rm F}&\leq \Vert\Sm_u\Vert_{2}\Vert \widetilde\Xm_u^{(t)}\Vert_{\rm F}+\alpha_u\Vert \Qm_u^{(t)} \Vert_2\Vert \widetilde\Zm^{(t-1)} \Vert_{\rm F}\\
		\Vert \widetilde {\Zm}^{(t)}\Vert_{\rm F}&\leq \Vert \widetilde\Ym\Vert_{\rm F} +\sum_{u\in U}\Vert \widetilde {\matr\Gamma}_u^{(t)}\Vert_{\rm F}\\
		\Vert \widetilde {\Rm}_u^{(t)}\Vert_{\rm F}&\leq \Vert\Sm_u\Vert_{2}\Vert \widetilde\Zm^{(t)}\Vert_{\rm F}+\Vert \widetilde\Xm_u^{(t)} \Vert_{\rm F}\\
		\Vert \widetilde \Xm_u^{(t+1)}\Vert_{\rm F}&\leq C \sqrt{1+\Vert \widetilde {\Rm}_u^{(t)}\Vert_{\rm F}^2}
	\end{align} 
\end{subequations}
for some constant $C$. Here, the latter step uses the Lipschitz property of the function $\eta_{u,t}(\cdot)$ as
\begin{align}
	\Vert \eta_{u,t}(\Rm)\Vert_{\rm F}&\leq \sqrt{2}\widetilde C\sqrt{\sum_{1\leq i\leq N_u}(1+\Vert \underline\rv_i \Vert)^2 }\\
	&\leq 2\widetilde C\sqrt{N_u+\Vert \matr \Rm\Vert_{\rm F}^2 }\;
\end{align}
for some constant $\widetilde C$.  Latter step above uses the trivial fact $(1+\Vert\underline\rv_i \Vert)^2\leq 2(1+\Vert \underline\rv_i \Vert^2)$. Moreover, $\Vert\Sm_u\Vert_{2}$ stands for the largest singular value of $\Sm_u$ which  is a sub-Gaussian RV \cite{davidson2001local,rudelson2010non} and thereby $\Vert \Sm_u\Vert_2=\Op{1}$. For example, this implies from the arithmetic properties of $\Op{1}$ (see Lemma~\ref{lemma:op_properties}) that 
\begin{equation}
	\Vert \widetilde\Ym\Vert_{\rm F}\leq \sum_{u\in U}\Vert \Sm_u\Vert_2   \Vert \widetilde\Xm_u\Vert_{\rm F}+\Vert \widetilde\Nm\Vert_{\rm F}=\Op{1}\;.
\end{equation}
Here, similar to $\widetilde\Xm_u=\Op{1}$, we have $\widetilde\Nm=\Op{1}$. Then, from \eqref{tildeAMP} it follows inductively (over iteration steps) that all the dynamics in \eqref{tildeAMP} are $\Op{1}$, i.e. 
$\widetilde {\matr\Gamma}_u^{(t)}=\Op{1}$, $\widetilde \Zm^{(t)}=\Op{1}$, $\widetilde {\matr\Gamma}_u^{(t)}=\Op{1}$, $\widetilde\Xm_u^{(t)}=\Op{1}$. This completes the proof. 

\subsection{Proof of Lemma~\ref{lem2}}\label{lem2_proof}
To bypass the singularity issues of the deterministic cross-correlation matrices in Lemma~\ref{lem2},
we consider the perturbation idea of \cite[Section~5.4]{Berthier20}.  
For each time-step $t\in[T]$ , we perturb the residual-AMP dynamics \eqref{AMP_algnew} by arbitrary independent sequence of the $N_u\times \bm$ matrices $\matr \Xi_u^{(t)}\sim_{\text{i.i.d.}}\matr\xi_u^{(t)}\sim\mathcal {CN}(\matr 0,\Id)$  as
\begin{subequations}
	\label{AMP_perturb}
	\begin{align}
		\widehat{\matr \Psi}_{u,\epsilon}^{(t)}&=\Sm_u \Fm_{u,\epsilon}^{(t)}-
		{\alpha_u}\Zm_{\epsilon}^{(t-1)}\Qm_{u,\epsilon}^{(t)}
		\\
		\Zm_{\epsilon}^{(t)}&=\Nm-\sum_{1\leq u\leq U}\widehat{\matr \Psi}_{u,\epsilon}^{(t)}\\
		\widehat{\matr \Phi}_{u,\epsilon}^{(t)}&=\Sm_u^\herm\Zm_\epsilon^{(t)}+{\Xm}_{u,\epsilon}^{(t)}\\
		\Fm_{u,\epsilon}^{(t+1)}&=\eta_{u,t}(\Xm_u+\widehat{\matr \Phi}_{u,\epsilon}^{(t)})-\Xm_u+\epsilon \matr \Xi_u^{(t+1)}\;
	\end{align}	
\end{subequations}
with the initialization $\Fm_{u,\epsilon}^{(1)}=\Fm_u^{(1)}+\epsilon \matr \Xi_u^{(1)}$. For convenience, we will later define the sequence of deterministic matrices $\mathbf{Q}_{u,\epsilon}^{(t)}$ through the state-evolution equations of the perturbed dynamics.
Moreover, we subsequently prove for all $t\in[T]$  
\begin{equation}
	\Qm_{u,\epsilon}^{(t)}-\Qm_u^{(t)}=o(1)\;.
\end{equation}
Here and throughout the section $o(1)$ stands for a \emph{deterministic matrix} of appropriate dimensions such that $\Vert o(1) \Vert_{{\rm F}}\to 0$ as $\epsilon\to 0$.

In the steps of Section~\ref{lem1_proof},
by substituting e.g. $\widetilde {\matr\Gamma}^{(t)}$ with the difference $\widehat{\matr \Psi}_{u}^{(t)}-\widehat{\matr \Psi}_{u,\epsilon}^{(t)}$,  $\widetilde \Rm^{(t)}$ with the difference $\widehat{\matr \Phi}_{u}^{(t)}-\widehat{\matr \Phi}_{u,\epsilon}^{(t)}$, etc. one can verify that for a sufficiently small $\epsilon$ we have 
\begin{align}
	\widehat{\matr \Psi}_{u,\epsilon}^{(t)}\simeq \widehat{\matr \Psi}_{u}^{(t)} \text{~~and~~}
	\widehat{\matr \Phi}_{u,\epsilon}^{(t)}\simeq \widehat{\matr \Phi}_{u}^{(t)}.\label{perpequiv}
\end{align}

Let us accordingly perturb the state-evolution equation Definition~\ref{SEdef}: We introduce the $1\times T\bm$ zero-mean Gaussian vectors $\matr\psi_{u,\epsilon}^{(1:T)}\eqdef (\matr\psi_{u,\epsilon}^{(1)},\matr\psi_{u,\epsilon}^{(2)},\ldots, \matr\phi_{u,\epsilon}^{(T)})$ and $\matr\phi_\epsilon^{(1:T)}\eqdef (\matr\phi_\epsilon^{(1)},\matr\phi_\epsilon^{(2)},\ldots, \matr\phi_\epsilon^{(T)})$ with their two-time covariance matrices $\Cm_{u,\epsilon}^{(t,s)}\eqdef \mathbb E[( \matr\psi_{u,\epsilon}^{(t)})^\herm  \matr\psi_{u,\epsilon}^{(s)}]$  and 
$\Cm_\epsilon^{(t,s)}\eqdef \mathbb E[( \matr\phi_\epsilon^{(t)})^\herm  \matr\phi_\epsilon^{(s)}]$ for all $t,s$ are recursively constructed according to
\begin{subequations}
	\label{statennew}
	\begin{align}
		\Cm_{u,\epsilon}^{(t,s)}&=\alpha_u\mathbb E[(\xv_{u,\epsilon}^{(t)}-\xv_{u})^\herm(\xv_{u,\epsilon}^{(t)}-\xv_{u})]\\
		\Cm_\epsilon^{(t,s)}&=\mathbb E[\nv^\herm\nv]+\sum_{u=1}^{L}\Cm_{u,\epsilon}^{(t,s)}\;, \label{SEnew}
	\end{align}
\end{subequations}
where we have introduced the 
stochastic process for $t\in[T]$
\begin{equation}
	\xv_{u,\epsilon}^{(t+1)}\eqdef \eta_{u,t}(\xv_{u}+\matr\phi_\epsilon^{(t)})+\epsilon\matr\xi_u^{(t+1)}   
\end{equation}
which is independent of $\xv_{u,\epsilon}^{(1)}\eqdef \xv_u^{(1)}+\epsilon\matr\xi_u^{(1)}$. Furthermore, let
\begin{equation}
	\Qm_{u,\epsilon}^{(t+1)}\eqdef\mathbb E[\eta'_{u,t}(\xv_{\epsilon}+\matr\phi_\epsilon^{(t)})].
\end{equation}
In particular, we now have the desired property (for each $u\in[U])$
\begin{equation}
	\Cm_{u,\epsilon}>\zerov. 
\end{equation}
Then, we can follow the same steps in the proof of Theorem~\ref{Th1} with solely substituting $\Fm_{u}^{(t)}$ with $\Fm_{u,\epsilon}^{(t)}$ (and $\xv_{u}^{(t)}$ with $ \xv_{u,\epsilon}^{(t)}$) and verify (under the premises of Theorem~\ref{Th1}) that
\begin{equation}
	\widehat{\matr \Psi}_{u,\epsilon}^{(t)}\simeq {\matr \Psi}_{u,\epsilon}^{(t)} \text{~~and~~}
	\widehat{\matr \Phi}_{u,\epsilon}^{(t)}\simeq {\matr \Phi}_{u,\epsilon }^{(t)}
\end{equation}
where $\matr\Psi_{u,\epsilon}^{(t)}\sim_{\text{i.i.d.}} \matr\psi_{u,\epsilon}^{(t)}$ and  
$\matr\Phi_{u,\epsilon}^{(t)}\sim_{\text{i.i.d.}} \matr\phi_\epsilon^{(t)}$, etc. 
In the sequel, we verify that for a small enough $\epsilon$ we have
\begin{equation}
	{\matr \Psi}_{u,\epsilon}^{(t)}\simeq {\matr \Psi}_{u}^{(t)} \text{~~and~~}
	{\matr \Phi}_{u,\epsilon}^{(t)}\simeq {\matr \Phi}_{u}^{(t)}\;.
\end{equation}

Let $\mathcal H_{t'}$ stand for the hypothesis that for all $t,s\leq t'$ we have
\begin{align}
	\Cm_{u,\epsilon}^{(t,s)}&=\Cm_u^{(t,s)}+o(1)\label{pertub1}\\
	\Qm_{u,\epsilon}^{(t)}&=\Qm_u^{(t)}+o(1)\;
\end{align}
with recalling that $o(1)$ stands for a \emph{deterministic matrix} of appropriate dimensions such that $\Vert o(1) \Vert_{{\rm F}}\to 0$ as $\epsilon\to 0$. Note also that \eqref{pertub1} implies that $\Cm_{\epsilon}^{(t,s)}=\Cm^{(t,s)}+o(1)$. We will complete the proof by verifying that $\mathcal H_{t'}$ implies $\mathcal H_{t'+1}$ in the sequel. The proof of the base case $\mathcal H_1$ is analogous (and rather simpler).

By the inductive hypothesis $\mathcal H_{t'}$ we have for all $t,s\leq t'$
\begin{align}
	&(\matr\phi_{\epsilon}^{(t)},\matr\phi_{\epsilon}^{(s)})\eqdef (\gv_1,\gv_2)\left(
	\begin{array}{cc}
		\Cm_\epsilon^{(t,t)}&\Cm_\epsilon^{(t,s)}\\
		\Cm_\epsilon^{(s,t)}& \Cm_\epsilon^{(s,s)}
	\end{array}\right)^{\frac 1 2}\\
	&= \underbrace{(\gv_1,\gv_2)\left(
		\begin{array}{cc}
			\Cm^{(t,t)}&\Cm^{(t,s)}\\
			\Cm^{(s,t)}& \Cm^{(s,s)}
		\end{array}\right)^{\frac 1 2}}_{\eqdef(\matr\phi^{(t)}, \matr\phi^{(s)})}+(\gv_1,\gv_2)o(1)
\end{align}
where $\gv_1$ and $\gv_2$ are auxiliary independent $1\times \bm$ vectors with i.i.d standard complex normal Gaussian entries.  Furthermore, for each $t\leq t'$ we define
\begin{align}
	\rv_{u,\epsilon}^{(t)}&\eqdef \xv_{u}+\matr \phi_\epsilon^{(t)}\\
	\rv_u^{(t)}&\eqdef \xv_{u}+\matr\phi^{(t)}\;.
\end{align}
Note that 
\begin{align}
	\Vert\rv_{u,\epsilon}^{(t)}-\rv_u^{(t)}\Vert&\leq \Vert (\gv_1,\gv_2)\Vert o(1).
\end{align}
Now, for an arbitrarily chosen pair of indices $i, j' \in [\bm]$ and iteration indexes $t, s \leq t'$, we introduce the function
\begin{align}
	f_1(\xv_1,\xv_2)&\eqdef (\eta_{u,t}(\xv_1))^*_i(\eta_{u,s}(\xv_2))_{j}\\ f_{2}(\xv_1,\xv_2)&\eqdef (\xv_1)^*_{i}(\eta_{u,t}(\xv_2))_{j}\\
	f_{3}(\xv_1,\xv_2)&\eqdef (\eta_{u,s}(\xv_2))_{i}^* (\xv_1)_j
\end{align}
where for short we neglect the dependencies of the indices $i,j$ and $t,s$ (and $u$) in the notations. In particular, by Lipschitz continuity of the function $\eta_{u,t}(\cdot)$ the functions $f_{1,2,3}(\xv)$ and for $\xv\eqdef (\xv_1,\xv_2)$ are pseudo-Lipschitz, i.e. 
\begin{equation}
	\Vert f_{1,2,3}(\xv)-f_{1,2,3}(\xv') \Vert\leq C(1+\Vert \xv \Vert+ \Vert \xv' \Vert)\Vert \Vert \xv-\xv'\Vert \label{usefulbound} 
\end{equation}
for some absolute constant $C$. We then write
\begin{align}
	&\mathbb E[(\xv_{u,\epsilon}^{(t+1)}-\xv_u)^\herm (\xv_{u,\epsilon}^{(s+1)}-\xv_{u})]_{ij}\nonumber \\
	&=(\mathbb E[\xv_u^\herm \xv_u])_{ij}+\mathbb E[f_1(\rv_{u,\epsilon}^{(t)},\rv_{u,\epsilon}^{(s)})]\nonumber\\&+\mathbb E[f_{2}(\xv_{u},\rv_{u,\epsilon}^{(t)})]+\mathbb E[f_{3}(\xv_{u},\rv_{u,\epsilon}^{(s)})]\nonumber \;.
\end{align}
For short, let $\rv_{u,\epsilon}^{(t,s)}\eqdef(\rv_{u,\epsilon}^{(t)},\rv_{u,\epsilon}^{(s)})$ and $\rv_u^{(t,s)}\eqdef(\rv_u^{(t)},\rv_u^{(s)})$. From \eqref{usefulbound} we write for any $t,s\leq t'$
\begin{align}
	&\vert \mathbb E[f_{1}(\rv_{u,\epsilon}^{(t,s)})]-\mathbb E[f_{1}(\rv_u^{(t,s)})]\vert\nonumber\\&\leq C\mathbb E\left[(1+\Vert\rv_{u,\epsilon}^{(t,s)}\Vert+\Vert\rv_u^{(t,s)}\Vert)\Vert \rv_{u,\epsilon}^{(t,s)}-\rv_u^{(t,s)}\Vert \right]\\
	&\leq \mathbb E\left[(1+\Vert \rv_{u,\epsilon}^{(t,s)}\Vert+\Vert\rv_u^{(t,s)}\Vert)\Vert (\gv_1,\gv_2)\Vert\right]o(1)\\
	&=o(1).  
\end{align}
Similarly, one can show for $i=2,3$ 
\begin{equation}
	\vert \mathbb E[f_{i}(\xv_{u},\rv_{u,\epsilon}^{(t)})]-\mathbb E[f_{i}(\xv_{u},\rv_{u}^{(t)})]\vert=o(1).
\end{equation}
These results together imply that for all $s\leq t'+1$
\begin{equation}
	\Cm_{u,\epsilon}^{(t'+1,s+1)}-\Cm_u^{(t+1,s+1)}=o(1).\label{res1}
\end{equation}

For $t=t'$, we define 
\begin{align}
	\widetilde{\Qm}_u^{(t+1)}&=\mathbb E[(\matr\phi^{(t)})^\herm \eta_{u,t}(\rv_{u}^{(t)})]\\
	\widetilde{\Qm}_{u,\epsilon}^{(t+1)}&=\mathbb E[(\matr\phi_{\epsilon}^{(t)})^\herm \eta_{u,t}(\rv_{u,\epsilon}^{(t)})]\;.
\end{align}
In particular, by Stein' lemma \cite[Lemma~2.3]{campese2015fourth} we have 
\begin{align}
	\widetilde{\Qm}_u^{(t+1)}&=\Cm^{(t,t)}{\Qm}_u^{(t+1)}\\
	\widetilde{\Qm}_{u,\epsilon}^{(t+1)}&=\Cm_\epsilon^{(t,t)}{\Qm}_{u,\epsilon}^{(t+1)}.
\end{align}
Since $\Cm_\epsilon^{(t,t)}>\zerov$ for all $t$, we have the implication
\begin{equation}
	\widetilde{\Qm}_{u,\epsilon}^{(t+1)}-\widetilde{\Qm}_u^{(t+1)}=o(1)\implies{\Qm}_{u,\epsilon}^{(t+1)}-{\Qm}_u^{(t+1)}=o(1).
\end{equation}
For short let $\matr\epsilon_\phi^{(t)}\eqdef \matr\phi_\epsilon^{(t)}-\matr\phi^{(t)}$ and $\matr\epsilon_r^{(t)}\eqdef \rv_\epsilon^{(t)}-\rv^{(t)}$. Then, we complete the proof by
\begin{align}
	&\vert \mathbb E[f_{2}(\matr\phi_{\epsilon}^{(t)},\rv_ {u,\epsilon}^{(t)})]-\mathbb E[f_{2}(\matr\phi^{(t)},\rv_u^{(t)})]\vert  \nonumber \\
	&\leq C\mathbb E\left[\left(1+\Vert(\matr\phi_{\epsilon}^{(t)},\rv_{u,\epsilon}^{(t)})\Vert+\Vert(\matr\phi^{(t)},\rv_u^{(t)})\Vert\right)\right.\nonumber\\&\left.\qquad \qquad   \times\Vert(\matr\epsilon_\phi^{(t)},\matr\epsilon_r^{(t)})\Vert\right]\label{blublu}\\
	&\leq \mathbb E[(1+\Vert(\matr\phi_{\epsilon}^{(t)},\rv_{u,\epsilon}^{(t)})\Vert+\Vert(\matr\phi^{(t)},\rv_{u}^{(t)})\Vert)\Vert (\gv_1,\gv_2)\Vert]o(1)\label{tired}\\
	&=o(1)\nonumber, 
\end{align}
where from \eqref{blublu} to \eqref{tired} we have used the result \eqref{res1}. 

\subsection{Proof of Lemma~\ref{lem3}}\label{lem3_proof}
We begin with defining the block-Cholesky decomposition. To this end, we recall the notation in \eqref{notation_set_up}
such that $\Am^{(1:t')}$ denote a $t' \bm \times t'\bm $ matrix  with its the $(t,s)$ indexed $\bm\times \bm$ block matrix denoted by $\Am^{(t,s)}$. Let $\Am^{(1:t')}\geq \matr 0$. Then, from an appropriate application of the block Gram-Schmidt process (to the columns of $(\Am^{(1:t')})^{\frac 1 2}$), we can always write the decomposition 
$\Am^{(1:t')} =(\Bm^{(1:t')})^\herm \Bm^{(1:t')}$
where $\Bm^{(1:t')}$ is a $t'F \times t'F$ { upper} triangular matrix with its { upper} triangular blocks $\Bm^{(t,s)}$ satisfying for all $1\leq s\leq t\leq t'$
\begin{subequations}\label{BCD}
	\noeqref{bcd1,bcd2}
	\begin{align}
		\Bm^{(s,s)}=\text{chol}\left(\Am^{(s,s)}- \sum_{s'=1}^{s-1}(\Bm^{(s,s')})^\herm\Bm^{(s,s')}\right)\label{bcd1}\\
		(\Bm^{(t,s)})^\herm\Bm^{(s,s)}=\Am^{(t,s)}- \sum_{s'=1}^{s-1}(\Bm^{(t,s')})^\herm\Bm^{({s,s'})}\;,\label{bcd2}
	\end{align}
\end{subequations}
with $\Bm=\text{chol}(\Am)$ for $\Am\geq 0$ standing for the { upper}-triangular matrix such that $\Am=\Bm^\herm\Bm$. For short, let us also denote the block-Cholesky decomposition by
\begin{equation}
	\Bm^{(1:t')}=\text{chol}_\bm(\Am^{(1:t')})\;.
\end{equation}
If $\Am^{(1:t')}>\matr 0$, $\Bm^{(s,s)}$ for all $s\in[t']$ are non-singular and then $\Bm^{(1:t')}$ can be uniquely constructed from the equations \eqref{BCD}.

From Lemma~\ref{lem1} we note that $\widehat{\Cm}_u^{(1:T)}=\Op{1}$ and $\widehat{\Cm}^{(1:T)}=\Op{1}$. Then, the proof of Lemma~\ref{lem3} follows evidently from the following result. 
\begin{lem}\label{lemmaBCD}
	Consider the $t'F\times t'F$ matrices $\widehat{\Cm}^{(1:t')}\geq \matr 0$ and $\Cm^{(1:t')}>\matr 0$
	where $\widehat{\Cm}^{(1:t')}=\Op{1}$ and ${\Cm}^{(1:t')}$ is deterministic. Suppose  
	$\widehat{\Cm}^{(1:t')}-{\Cm}^{(1:t')}=\Op{L^{-c}}$ for some constant $c>0$. Let $\widehat{\Bm}^{(1:t')}\eqdef\text{\rm chol}_{F}(\widehat{\Cm}^{(1:t')})$ and $\Bm^{(1:t')}\eqdef\text{\rm chol}_{F}({\Cm}^{(1:t')})$\;
	Then, we have \[\widehat{\Bm}^{(1:t')}-\Bm^{(1:t')}=\Op{L^{-c}}\;.\] 
	\begin{proof}
		Firstly, by using Lemma~\ref{lemmaCD} below we have
		\begin{align}
			\widehat{\Bm}^{(1,1)}&=\text{chol}(\widehat{\Cm}^{(1,1)})=\underbrace{\text{chol}({\Cm}^{(1,1)
				})}_{={\Bm}^{(1,1)}}+\Op{L^{-c}}\label{uselccd1}
		\end{align}
		Secondly, with an appropriate application of Lemma~\ref{rem3} below we write 
		\begin{align}
			(\widehat{\Bm}^{(t,1)})^\herm&=\underbrace{\Cm^{(t,1)}(\Bm^{(1,1)})^{-1}}_{=(\Bm^{(t,1)}})^\herm+\Op{L^{-c}} \quad \forall 1<t\leq t'. \label{giargument}
		\end{align}
		Let $H_{t-1}$ denote the hypothesis that
		\begin{align}
			\widehat{\Bm}^{(s,s')}={\Bm}^{(s,s')}+\Op{L^{-c}}\quad  ~~\forall 1\leq s'\leq s<t\;. 
		\end{align}
		Suppose $H_{t-1}$ hold. We then start from \eqref{giargument} and proceed for $s'=2,3,\ldots,t-1$ with appropriately using  Lemma~\ref{rem3}
		\begin{align}		
			&(\widehat{\Bm}^{(t,s')})^\herm\nonumber=\Op{L^{-c}}\\&+\underbrace{\left(\Cm^{(t,s')}-\sum_{1\leq s< s'}(\Bm^{(t,s)})^\herm\Bm^{(s',s)}\right)({\Bm}^{(s's')})^{-1}}_{=({\Bm}^{(t,s')})^\herm}\;.
		\end{align}
		Also, similar to \eqref{uselccd1}, from Lemma~\ref{lemmaCD}, one can verify 
		$\widehat{\Bm}^{(t,t)}={\Bm}^{(t,t)}+\Op{L^{-c}}.$
		Thus, $H_{t-1}$ implies $H_{t}$. This completes the proof. 
	\end{proof}
\end{lem} 
\begin{lem}\label{lemmaCD}
	Consider the matrices $\widehat{\Cm}\geq \matr 0$ and $\Cm>\matr 0$
	where the latter is deterministic. Suppose  
	$\widehat{\Cm}-{\Cm}=\Op{L^{-c}}$ for some constant $c>0$.
	Then, we have $\text{\rm chol}(\widehat{\Cm})-\text{\rm chol}({\Cm})=\Op{L^{-c}}$\;.
	\begin{proof}
		Firstly, for a real RV ${a}=\Op{L^{-c}}$ with $\vert a\vert \leq C$ for some constant C, we note the identity 
		\begin{equation}
			\vert\sqrt{a+C}-\sqrt{C} \vert=\frac{\vert a\vert}{\sqrt{a+C}+\sqrt{C}}\leq \frac{\vert a\vert}{\sqrt{C}}=\Op{L^{-c}}.\label{eq:op_sqrt}
		\end{equation}
		Then, the proof follows from the steps in the proof of Lemma~\ref{lemmaBCD} where we set $F=1$ and in the steps where Lemma~\ref{lemmaCD} was used we use the property \eqref{eq:op_sqrt} instead. 
	\end{proof}
\end{lem}
\begin{lem}\label{rem3}
	Suppose $
	\widehat{\Bm}_1\widehat{\Bm}_2=\Cm+\Op{L^{-c}}$ where $\widehat{\Bm}_1=\Op{1}$ and $\widehat{\Bm}_2={\Bm}_2+\Op{L^{-c}}$ with $\Bm_2$ being determinisitic and non-singular. Then, we have
	\begin{equation}
		\widehat{\Bm}_1=\Cm\Bm_2^{-1}+\Op{L^{-c}}\;.
	\end{equation}
	\begin{proof}
		For short let $\Dm\eqdef \widehat{\Bm}_2-{\Bm}_2$. Note from $\widehat{\Bm}_1\Dm=\Op{L^{-c}}$. Hence, we have 
		\begin{equation}
			\widehat{\Bm}_1{\Bm}_2=\Cm+\Op{L^{-c}}+\widehat{\Bm}_1\Dm=\Cm+\Op{L^{-c}}\;.
		\end{equation}
	\end{proof}
\end{lem}


\subsection{Proof of Lemma~\ref{lem5}}\label{lem5_proof}
By the premises of Lemma~\ref{lem5}-(i) and the Lipschitz-continuity of $\eta_{u,t}$ we have 
for any $t\in[t']$
\begin{align}	\Fm_u^{(t)}&\simeq\eta_{u,t-1}\left(\Xm_{u}+ \sum_{1\leq s<t}\Um_u^{(s)}\Bm^{(t-1,s)}\right)-\Xm_u\\
	&\sim_{\text{i.i.d.}}\eta_{u,t-1}(\xv_u+\matr\phi^{(t-1)})-\xv_u\;.
\end{align}
Furthermore, recall also from \eqref{conprojgaus} that 
$\widehat\Um_u^{(t)}\simeq \Um_u^{(t)}$.
Moreover,  as to the result \eqref{res5}  we note from Stein's lemma \cite[Lemma~2.3]{campese2015fourth} that 
\begin{align}
	\mathbb E [(\uv_u^{(s)})^\herm \eta_{u,t}(\rv_u^{(t-1)})]&=\Bm^{(t-1,s)}\mathbb E[\eta_{u,t-1}'(\rv_u^{(t-1)})]\\
	&=\Bm^{(t-1,s)}\Qm_u^{(t)}\;,
\end{align}
where $\rv_u^{(t-1)}\eqdef \xv_{u}+ \sum_{1\leq s<t}\uv_u^{(s)}\Bm^{(t-1,s)}$ and we assume that $\Um_u^{(t)}\sim_\text{i.i.d.}\uv_u^{(s)}$ for all $s<t$.
Hence, the concentrations in (i) follow evidently from  Lemma~\ref{eq:ip_concentration}. 

By the premises of Lemma~\ref{lem5}-(ii) we have
\begin{align}
	\Zm^{(t)}&\simeq \Nm-\sum_{u\in U}\sqrt{\alpha_u}\sum_{1\leq s\leq t}\Gm_u^{(s)}\Bm_u^{(t,s)}\\
	&\sim_{\text {i.i.d.}}\nv+\sum_{u\in [U]}\matr\psi_u^{(t)}\;.
\end{align}
Furthermore, recall also from \eqref{conprojgaus} that 
$\widehat\Gm_u^{(t)}\simeq \Gm_u^{(t)}$. Then, the concentrations in (ii) evidently follow from  Lemma~\ref{eq:ip_concentration}.

\section{The RS Prediction of the Mutual Information}\label{free_energy}
We denote the negative log-likelihood by 
\begin{align}
	&\mathcal E_\Sm(\Ym,\Xm)\eqdef {L}\ln\pi\vert \Cm_{\rm n} \vert\nonumber \\&+\sum_{i=1}^{L}[{\yv_i}-\sum_{u=1}^{U}(\Sm_u\Xm_u)_i]\Cm_{\rm n}^{-1}[{\yv_i}-\sum_{u=1}^{U}({\Sm}_u\Xm_u)]^\herm\;.
\end{align}
Moreover, we define the partition function, i.e., the probability of data matrix $\Ym$, as 
\begin{equation}
	Z_\Sm(\Ym)\eqdef \int {\rm dP}(\Xm)\; {\rm e}^{-\mathcal E_\Sm(\Ym,\Xm)}\;.
\end{equation}
We make use of the mild assumption that the normalized mutual information is asymptotically (as $L\to \infty$) self-averaging (w.r.t. $\Sm$) and study the limit
\begin{align}
	&\lim_{L\to \infty}\frac{1}{L}\mathbb E_\Sm[\mathcal I(\Xm;\Ym)]=-(\ln \vert \pi\Cm_{\rm n}\vert +\bm)\nonumber\\
	&\qquad\underbrace{-\lim_{L\to \infty}\frac{1}{L}\int {\rm d}\Ym\; \mathbb E_{\Sm}[Z_\Sm(\Ym)\ln Z_\Sm(\Ym)]}_{\eqdef \mathcal F} \label{fenergy}\;.
\end{align}
The term $\mathcal F$ is often called to the (thermodynamic) ``free energy'' of the input-output system. Remark that
\begin{align}
	&\int {\rm d}\Ym\; \mathbb E_{\Sm}[Z_\Sm(\Ym)\ln Z_\Sm(\Ym)]\nonumber\\\qquad&=\lim_{n\to 1}\frac{\partial}{\partial n}\ln\int {\rm d}\Ym\; \mathbb E_{\Sm}[Z^{n}_\Sm(\Ym)]\;.
\end{align}
Then, assuming we can exchange the order of the limits $n\to 1$ and $L\to \infty$, we write
\begin{equation}
	\mathcal F=-\lim_{n\to 1}\frac{\partial}{\partial n}\underbrace{\lim_{L\to \infty}\frac 1 L\ln\int {\rm d}\Ym\; \mathbb E_{\Sm}[Z^{n}_\Sm(\Ym)]}_{\eqdef \mathcal M^{(n)}}\;.
\end{equation}

By using the following elementary result
\begin{equation}
	\mathbb E_{\sv}[\delta(\matr\gamma-{\sv}\Xm)]=\textswab{g}(\matr\gamma \vert \matr 0,\tau \Xm^\herm\Xm)
\end{equation}
where ${\sv} \sim \mathcal{CN}(\matr 0;\tau\Id_N)$ and $\Xm\in \CC^{N\times \bm}$,  one can verify 
\begin{align}
	&\int {\rm d}\Ym\;\mathbb E_{\Sm}[ Z^{n}_\Sm(\Ym)]\nonumber\\
	&=\int\{{\rm d}\mathcal Q_u^{(1:n)}\}_{u\in[U]}\; {\rm e}^{L (\mathcal G(\{\mathcal Q_u^{(1:n)}\})+\sum _{u=1}^{U}{\alpha_u}{\mathcal I}_{u}(\mathcal Q_u^{(1:n)}))}\;.
\end{align}
Here, we have defined
\begin{align}
	&\mathcal G(\{\mathcal Q_u^{(1:n)}\})\eqdef
	\int {\rm d}\yv {\rm d}\mathcal{CN}(\matr\gamma^{(1:n)}\big\vert \matr 0; \sum_{u\leq U}\mathcal Q_{u}^{(1:n)})\nonumber\\&\qquad \times{\rm e}^{-\sum_{a=1}^{n}\Vert({\yv}-{\matr \gamma^{(a)}})\sqrt{\Cm_{\rm n}^{-1}}\Vert^2}-n\ln\vert\pi \Cm_{\rm n}\vert\\
	&= -n\ln\vert\pi \Cm_{\rm n}\vert -F\ln n +\ln \vert\Cm_{\rm n}\vert \nonumber\\\quad&-\ln\left\vert\Id +(\sum_{u\leq U}\mathcal Q_u^{(1:n)})(\Id_n- \frac 1 n\Em_n)\otimes \Cm_{\rm n}^{-1}\right\vert\label{GQ}\;,
\end{align}
where the entries of the $n\times n$ matrix $\Em_n$ are all one. 
Moreover, we have defined
\begin{align}
	&{\mathcal I}_u(\mathcal Q_u^{(1:n)})\eqdef\frac{1}{N_u}\ln \int \prod_{a=1}^{n}{\rm dP}_u(\Xm_u^{(a)})\; \nonumber\\&\quad\times\delta \left({(\Xm_u^{(1:n)})}^\herm{\Xm_u}^{(1:n)}-L\mathcal Q_u^{(1:n)}\right)\;\label{latdef}
\end{align}
where e.g. $\Xm^{(1:n)}\eqdef[\Xm^{(1)},\Xm^{(2)},\ldots,\Xm^{(n)}]$. 

By the saddle-point method, we can express \eqref{latdef} as
\begin{align}
	&\mathcal I_u(\mathcal Q_u^{(1:n)})=o(1)\nonumber\\
	&+\inf_{\widehat{\mathcal Q}_u^{(1:n)}}\left(\mathcal G_u(\widehat{\mathcal Q}_u^{(1:n)})-\frac 1 {\alpha_u}{\rm tr}(\widehat{\mathcal Q}_u^{(1:n)}\mathcal Q_u^{(1:n_)})\right)
\end{align}
where we have introduced the moment-generating function
\begin{equation}
	\mathcal G_u(\widehat{\mathcal Q})\eqdef\ln\int \prod_{a=1}^{n}{\rm dP}_u(\xv^{(a)})\; {\rm e}^{\xv^{(1:n)}\widehat{\mathcal Q}(\xv^{(1:n)})^\herm}\;\label{eqc}
\end{equation}
and $o(1)\to 0$ as $L\to \infty$. 
Again by invoking the saddle-point method, we get 
an expression of $\mathcal M(n)$ given in \eqref{extrm} 
\begin{figure*}[t!]
	\normalsize
	\setcounter{equation}{247}
	\begin{align}
		\mathcal M(n)&=\lim_{L\to \infty}\frac{1}{L}\ln \int \{{\rm d}\mathcal Q_u^{(1:n)}\} e^{L\left(\mathcal G(\{\mathcal Q_u^{(1:n)}\})+\sum_{u=1}^{U}{\alpha_u}\inf_{\widehat{\mathcal Q}_u^{(1:n)}}\left(\mathcal G_u(\widehat{\mathcal Q}_u^{(1:n)})-\frac 1 {\alpha_u}{\rm tr}(\widehat{\mathcal Q}_u^{(1:n)}\mathcal Q_u^{(1:n_)})\right)+o(1)\right)}\label{pextrm}\\
		&=\sup_{\mathcal Q_u^{(1:n)}}\inf_{\widehat{\mathcal Q}_u^{(1:n)}}\left(\mathcal G(\{\mathcal Q_u^{(1:n)}\})+\sum_{u=1}^{U}{\alpha _u}\mathcal G_u(\widehat{\mathcal Q}_u^{(1:n)})-{\rm tr}(\widehat{\mathcal Q}_u^{(1:n)}\mathcal Q_u^{(1:n_)})\right)\;.\label{extrm}
	\end{align}
	\setcounter{equation}{249}
	\hrulefill
	\vspace*{4pt}
\end{figure*}
at the top of the next page. There, from \eqref{pextrm} to \eqref{extrm} we neglect the contribution of the vanishing term $o(1)$ to the saddle point. 

\begin{rem}\label{remrs1}
	Let the saddle-point $\mathcal Q^{\star(1:n)}$ and $\widehat{\mathcal Q}_u^{\star(1:n)}$ achieve the extremal condition in \eqref{extrm}. Then, for all $ a,b\in[n]$ (and for all $u\in[U]$) we have
	\begin{align}
		&\mathcal Q_u^{\star(a,b)}(n)=\frac{1}{\alpha_u}\mathbb E[(\xv_u^{(a)})^\herm\xv_u^{(b)}]_{\mathcal G_u}\quad \label{rem11}\\
		&~\quad \widehat{\mathcal Q}_u^{\star(1:n)}=-(\Id_n-\frac{1}{n}\Em_n)\otimes {\Id_\bm}\nonumber\\&\times\left(\Id_n\otimes\Cm_{\rm n}+(\sum_{u\leq U}\mathcal Q_u^{(1:n)})(\Id_n- \frac 1 n\Em_n)\otimes \Id_{\bm}\right)^{-1}\label{rem12}
	\end{align} 
	where $\mathbb E[(\cdot)]_{\mathcal G_u}$ denotes the expectation w.r.t. the moment generating function $\mathcal G_u(\widehat{\mathcal Q}_u^{\star(1:n)})$.  \hfill $\lozenge$
\end{rem}
Recall that we are interested in the limit $\lim_{n\to 1}\frac{\partial \mathcal M(n)}{\partial n}$. Thus, from the property of stationary points of multivariate functions \cite[Appendix~G]{Tulino13}, we can assume without loss of generality
\begin{align}
	{\mathcal Q}_u^{\star(a,b)}&={\mathcal Q}_u^{\star(a,b)}(1)\\
	\widehat{\mathcal Q}_u^{\star(a,b)}&=\widehat{\mathcal Q}_u^{\star(a,b)}(1)\;.
\end{align}
\begin{asmp}[RS Assumption]
	For all $u\in[U]$  we assume the replica-symmetry 
	\begin{equation}
		\mathcal Q_u^{\star(a,b)}=\left\{\begin{array}{cc}
			\Qm_{u0} & a=b \\
			\Qm_u  & \text{elsewhere}
		\end{array}
		\right.\;.\label{RSass}
	\end{equation}
\end{asmp}
The remaining part involves simplifications using the RS assumption \eqref{RSass}. First, by utilizing \eqref{GQ}, we obtain
\begin{align}
	&\mathcal G^{(n)}(\{\Qm_{u0};\Qm_u\})\eqdef \mathcal G(\{\mathcal Q_{u}^{\star(1:n)}\})\nonumber\\
	&=-n\ln\vert\pi \Cm_{\rm n}\vert -F\ln n +\ln \vert\Cm_{\rm n}\vert \nonumber\\&\quad- (n-1)\ln\left\vert\Id +\Cm_{\rm n}^{-1}\sum_{u=1}^{U}\Qm_{u0}-\Qm_u\right\vert\;.
\end{align}
Hence, we have
\begin{equation}
	\lim_{n\to 1}\frac{\partial \mathcal G^{(n)}}{\partial n}=-\ln\vert\pi{\rm e} \Cm_{\rm n}\vert  -\ln\left\vert\Id +\Cm_{\rm n}^{-1}\sum_{u=1}^{U}\Qm_{u0}-\Qm_u\right\vert.
\end{equation}
Second, from \eqref{rem12} it yields for all $u\in [U]$
\begin{align}
	\widehat{\mathcal Q}^{(a,b)}&=\widehat{\mathcal Q}_u^{(a,b)}(1)\nonumber\\&=
	(1-\delta_{ab})\underbrace{\left(\Cm_{\rm n}+\sum_{u'=1}^{U}\Qm_{0u'}-\Qm_{u'} 
		\right)^{-1}}_{\eqdef\widehat \Qm}\label{Qhatsymery}\;.
\end{align}
Then, from the symmetry \eqref{Qhatsymery} we easily obtain 
\begin{equation}
	\mathcal G_u^{(n)}(\widehat\Qm)\eqdef \mathcal G_u(\tilde{\mathcal Q}_u^{(1:n)})=\ln \int {\rm d}\mathcal {CN}(\yv\vert\matr 0,\widehat\Qm)(\mathcal L_u(\yv))^{n}\label{qhatsym}
\end{equation}
where we have defined 
\begin{equation}
	\mathcal L_u(\yv)\eqdef
	\int {\rm dP}_u(\xv)\;{\rm e}^{-\xv \widehat\Qm \xv^\herm+2{\rm Re}(\yv\xv^\herm)}\;.
\end{equation}
Hence, we have 
\begin{align}
	\lim_{n\to 1}\frac{\mathcal G_u^{(n)}}{\partial n}&=\int {\rm d}\mathcal {CN}(\yv\vert\matr 0,\widehat\Qm)\;\mathcal L_u(\yv)\ln \mathcal L_u(\yv)\;.
\end{align}
Consider now the auxiliary input-output observation model
\begin{equation}
	\yv=\xv_u\widehat\Qm+\nv ~~\text{with}~~ \nv\sim {\mathcal{CN}}(\matr 0;\widehat\Qm)
\end{equation}
Notice that
\begin{equation}
	\mathcal L_u(\yv)=\frac{p_{\yv}(\yv)}{p_n(\yv)}\;.
\end{equation}
Then, it is easy to verify that
\begin{align}
	&\int {\rm d}\mathcal {CN}(\yv\vert\matr 0,\widehat\Qm)\;\mathcal L_u(\yv)\ln \mathcal L_u(\yv)
	\nonumber\\&={\rm tr}(\widehat\Qm\mathbb E[\xv_u^\herm\xv_u])-\mathcal I\left(\xv_u;\widehat\Qm\xv_u+\zv\sqrt{\widehat \Qm}\right)\;\\
	&={\rm tr}(\widehat\Qm\mathbb E[\xv_u^\herm\xv_u])-\mathcal I\left(\xv_u;\xv_u+\zv\sqrt{\widehat \Qm^{-1}}\right)\;.
\end{align}
We plug \eqref{qhatsym} into the saddle-point condition \eqref{rem11} and obtain 
\begin{align}
	&\mathcal Q_u^{\star(a,b)}(n)=\frac{1}{\alpha_u}
	\left(\int {\rm d}\mathcal {CN}(\yv;\matr 0,\widehat\Qm)\mathcal L^n_u(\yv)\right)^{-1}\nonumber\\
	&~\times \left\{\begin{array}{cc}
		\int {\rm d}\mathcal {CN}(\yv\vert\matr 0,\widehat\Qm)\;\mathbb E[\xv^\herm\xv\vert \yv]\mathcal L_u^{n-1}(\yv) & a=b\\
		\int {\rm d}\mathcal {CN}(\yv\vert\matr 0,\widehat\Qm)\;\mathbb E[\xv^\herm\vert \yv]\mathbb E[\xv\vert \yv]\mathcal L_u^{n}(\yv) & a\neq b
	\end{array}\right..
\end{align}
Hence, we get for $a\neq b$
\begin{align}
	\Qm_{u0}-\Qm_u&=\mathcal Q_u^{\star(a,a)}(1)-\mathcal Q_u^{\star(a,b)}(1)\\
	&={\alpha_u}{\rm mmse}(\xv_u\mid\widehat\Qm\xv_u+\nv)\\
	&={\alpha_u}{\rm mmse}(\xv_u
	\mid\xv_u+\zv\sqrt{\widehat \Qm^{-1}})\;. 
\end{align}
This completes the derivation.

\section{Proof of Corollar~\ref{Cormse}}\label{Proof_Cormse}
To simplify notation, we omit the subscript $(\cdot)_{u}$ throughout the proof, e.g. $N_u$ instead of $N$ and $\Xm$ instead of $\Xm_u$, etc.  From \eqref{etause} we get
\begin{align}
	&\left \Vert \frac{(\Xm-\Xm^{(t)})^\herm (\Xm-\Xm^{(t)})}{N} -\mathbb E[(\xv-\xv^{(t)})^\herm (\xv-\xv^{(t)})]\right\Vert_{\rm F}\nonumber\\&\leq \left\Vert \matr \Delta^{(t)}\right \Vert_{\rm F}+\Op{L^{-\frac 1 2}}
\end{align}
where for short we define 
\begin{align}
	&\matr\Delta^{(t)}\eqdef  \mathbb E[(\xv-\xv^{(t)})^\herm (\xv-\xv^{(t)})]\nonumber\\&-\frac{(\Xm-\eta_{t-1}(\Xm+\matr \Phi^{(t-1)})^\herm (\Xm-\eta_{t-1}(\Xm+\matr \Phi^{(t-1)})}{N}\;.
\end{align}
Note also that 
\begin{equation}
	(\Xm-\eta_{t-1}(\Xm+\matr \Phi^{(t-1)}))\sim_{\text{ i.i.d.}}(\xv-\xv^{(t)})\;.
\end{equation}
Since we assume $\xv=\Op{1}$, from the sum rule \eqref{eq:op_sum} we have $\xv+\matr\phi^{(t-1)}=\Op{1}$. This implies from the Lipschitz property of $\eta_{t-1}(\cdot)$ that $\xv^{(t)}=\Op{1}$. Thus, from the sum rule \eqref{eq:op_sum} we have $\xv-\xv^{(t)}=\Op{1}$. Hence, from Lemma~\ref{keycon} we get $	\matr \Delta^{(t)}=\Op{L^{-1/2}}\;. $ This completes the proof. 
\section{Proof of Equation~(\ref{dPT})}\label{proof_cordecop}
Conditioned on  a random permutation matrix $\Pm\in\RR^{N_u\times N_u}$ independent of $\Sm_u$, we have that $\Sm_u\Pm\sim\Sm_u$ and thereby $\Sm_u\Pm$ is independent of $\Pm$. Thus, by construction, the rows of $\Rm_{u}^{(t)}$ are statistically exchangeable, $\rv_{u,n}^{(t)}\sim \rv_{u,n'}^{(t)}$ for any $n\neq n'$. For a given pair $(u,t)\in[U]\times [T]$ let
\begin{equation}
	\matr\Delta\equiv \Rm_u^{(t)}-(\Xm_{u}+\matr \Phi_u^{(t)})\;.
\end{equation}
Since the rows of $\matr\Delta$, say $\{ \matr \delta_n\}$ are statistical exchangeable, we have
\begin{align}
	N_u\mathbb E[\Vert \matr\delta_n\Vert^{2p}]&=\mathbb E\sum_{n\leq N_u}\Vert \matr\delta_n\Vert^{2p} \leq \mathbb E(\sum_{n\leq N_u}\Vert \matr\delta_n\Vert^{2})^{p}\\
	&=\mathbb E\Vert\matr \Delta \Vert_{\rm F}^{2p}\leq C_p\;. 
\end{align}
This completes the proof. 
\section{Proof of Example~\ref{conmmse}}\label{Prop1_proof}

Given $\xv_u^{(1)}=\matr 0$ and $\xv_u$ having a Bernoulli-Gaussian distribution \eqref{xu_distrib} with a diagonal covariance, then following the steps of \cite[Appendix~B]{bai2022activity} one can verify (inductively over iteration steps) that $\Cm^{(t,t)}$ is diagonal for each $t\in[T]$.  The proof then follows from the monotonicity property of the scalar MMSE \cite{wu2011functional}.
Specifically, we introduce the mapping for the SNR
$\rho\in[0,\infty)$
\begin{equation}
	f(\rho)\eqdef(\sigma^2+\sum_{u\in [U]}\alpha_u {\rm mmse}(x_u\vert \sqrt{\rho}x_u+z))^{-1}
\end{equation}
where $x_u\eqdef a_uh_u$ and the random variables $a_u\sim{\rm Bern}(\lambda_u)$, $h_u\sim\mathcal{ CN}(0,\sigma^2_u)$ and $z\sim\mathcal{CN}(0,1)$ are all independent. In particular, from the decreasing property of ${\rm mmse}$ with respect to $\rho$ (see \cite[Corollary 5]{wu2011functional}) we have
\begin{equation}
	f(\rho)\leq f(\rho')\;,\quad \rho\leq \rho'\;. \label{monoticity}
\end{equation}
Now, for any arbitrarily chosen index $f\in[F]$, let us write $\tau^{(t)}\equiv(\Cm^{(t,t)})_{ff}$ and $\sigma^2_u\equiv(\Sigmam_u)_{ff}$. Notice that for any $t\in[T]$ we have
\begin{equation}
	\tau^{(t+1)}=\frac{1}{f(\frac 1 {\tau^{(t)}})}\;.
\end{equation}
Also, from $\xv_u^{(1)}=\matr 0$, the initial value $\tau^{(1)}$ reads as
\begin{equation}
	\tau^{(1)}=\frac 1{f(0)}\;. \label{initialtau}
\end{equation}
Given $\tau^{(t-1)}\geq \tau^{(t)}$, it follows from the property
\eqref{monoticity} that $\tau^{(t)}\geq \tau^{(t+1)}$. Also from \eqref{initialtau} we have $\tau^{(2)}\leq \tau^{(1)}$. So it follows inductively over iteration steps that 
\begin{equation}
	\tau^{(t)}\geq \tau^{(t+1)} \quad t\in[T]. 
\end{equation}
This implies that as $t\to \infty$ we have the convergence $\tau^{(t)}\to \tau^\star$ for some $\tau^\star=1/f(1/\tau^\star)$.

{ \section{The Proof of Proposition~\ref{geniemmse}}\label{pthe4}
	We first write the location-based genie MMSEs:
	\begin{equation}
		g_u\eqdef\frac{1}{\vert \mathcal A_u \vert  }{\left\Vert \Hm_u-\mathbb E[\Hm_u\vert \Ym,\{\widetilde\Sm_u\}] \right\Vert_{\rm F}^2 }.
	\end{equation}
	One can verify that
	\begin{align}
		g_u&=\sum_{f\in [F]}\tau_{uf}\frac{1}{\vert \mathcal A_u \vert }{\rm tr}\left((\Id+\tau_{uf}\widetilde\Sm_u\Am_{fu}^{-1}\widetilde\Sm_u^\herm)^{-1}\right)\\
		&=\sum_{f\in [F]}\tau_{uf}\frac{L}{\vert \mathcal A_u \vert }\left[\frac{1}{L}{\rm tr}\left((\Id+\Am_{fu}^{-1}\Bm_{fu})^{-1}\right)\nonumber\right.\\&\qquad \qquad \qquad \qquad\left.+1-\frac{\vert \mathcal A_u \vert }{L}\right]
	\end{align}
	where for short we define the random matrices
	\begin{align}
		\Bm_{fu}\eqdef \tau_{uf}\widetilde\Sm_u\widetilde\Sm_u^\herm\;~~\text{and}~~
		\Am_{fu}\eqdef \sigma^2\Id +\sum_{u'\neq u} \Bm_{fu'}\;.
	\end{align}
	Then, we write 
	\begin{align}
		{\rm mmse}_{\rm e}^{\rm genie}&=\sum_{u\in [U]}\frac{\vert\Ac_u\vert}{\vert\Ac\vert}g_u\\&=\frac{L}{\vert\mathcal A\vert}\sum_{(u,f)\in [U]\times [F]}\tau_{uf}(g_{uf}+\frac{\vert \mathcal A_u \vert}{L}-1)
	\end{align}
	where for short we define $g_{uf}\eqdef\frac{1}{L}{\rm tr}\left((\Id+\Am_{uf}^{-1}\Bm_{uf})^{-1}\right)$.
	We note from \eqref{lamal} that $\frac{L}{\vert\mathcal A\vert}$ and $\frac{\vert \mathcal A_u \vert}{L}$ converge (as $L\to \infty$) almost surely to $1/Z$ and $\alpha_u\lambda_u$, respectively. Then we complete the proof by deriving the large-system limit expression of $g_{uf}$ using standard random matrix theory results. 
	
	The so-called R-transform of the limiting spectral distribution (LSD) of $\Bm_{fu}$ reads as \cite{tulino2004random} 
	\begin{equation}
		{\rm R}_{\Bm_{fu}}(\omega)=\frac{\tau_{uf}\alpha_u\lambda_u}{1-\tau_{uf}\omega}.
	\end{equation}
	Moreover, we have by the additive property of the R-transform that
	\begin{equation}
		{\rm R}_{\Am_{fu}}(\omega)=\sigma^2+\sum_{u\in [U]}{\rm R}_{\Bm_{fu'}}(\omega).
	\end{equation}
	Then, the proof evidently follows from the following free probability result. 
	\begin{lem}
		Let the $N\times N$ matrices $\Am>\matr 0$ and $\Bm\geq \matr 0$ be independent and unitarily invariant and have a compactly supported LSD each. Then, we have almost surely
		\begin{align}
			g^\star&\eqdef\lim_{N\to \infty}\frac{1}{N}{\rm tr}((\Id+\Am^{-1}\Bm)^{-1})\\
			&=1-\frac {1}{c^\star}{\rm R}_{\Bm}\left(-\frac{1} {c^\star}\right)    
		\end{align}
		where $c^\star$ is the unique solution of 
		$c^\star={\rm R}_{\Am}\left(-1/ {c^\star}\right)+{\rm R}_{\Bm}\left(-1/ {c^\star}\right)\;$.  
		\begin{proof}
			
			Let $\frac{1}{c^\star}\equiv \lim_{N\to \infty}\frac{1}{N}{\rm tr}((\Am+\Bm)^{-1}).$ From \cite[Lemma 1]{ccakmak2016self} and the additivity property of R-transform of random matrices we write respectively 
			\begin{equation}
				c^\star={\rm R}_{\Am+\Bm}\left(-\frac{1} {c^\star}\right)={\rm R}_{\Am}\left(-\frac{1} {c^\star}\right)+{\rm R}_{\Bm}\left(-\frac{1} {c^\star}\right)\;. 
			\end{equation}
			Without loss of generality, we can assume $\Am$ is diagonal. Then, from the local-law of addition of random matrices \cite{bao2017local,cakmak2018expectation} we have for each $n\in[N]$
			\begin{equation}
				\lim_{N\to \infty}\left\vert [(\Am+\Bm)^{-1}]_{nn}-[(\Am+{\rm R}_{\Bm}(-\frac{1} {c^\star})\Id)^{-1}]_{nn}\right\vert=0. 
			\end{equation} 
		\end{proof}
	\end{lem}
}

\section{Computing the FA and MD probabilities}\label{laplace-inversion}

{ Considering the motivated fading model \eqref{channel_u}, i.e. $\hv_u\in \Nc(\matr 0,\Sigmam_u)$,
	both the asymptotic of the MD rate (i.e., $\beta_\infty^{\rm md}$) and the FA rate (i.e., $\beta_\infty^{\rm fa}$) can be computed in closed form, as long as we have the closed-form expression of}
\begin{equation}
	{\rm P}(\gamma)\eqdef\PP\left( \zv \Dm \zv^\herm\leq\gamma\right)
\end{equation}
$\zv\sim\mathcal {CN}(\matr 0,\Id_\bm)$ for the Hermitian symmetric PSD deterministic $\Dm$. 
In fact, since $\zv\Dm \zv^\herm\sim \zv \tilde\Dm \zv^\herm$ where $\tilde \Dm$ stands for the diagonal matrix with the diagonal entries being the eigenvalues of $\Dm$, we can assume without loss of generality that $\Dm={\rm diag}(d_1,d_2,\cdots,d_{\bm})\;$.

Let ${\rm p}$ denote the density function of the RV $\zv\Dm \zv^\herm$. In particular, the Laplace transform of the density function reads as
\begin{align}
	\Lc_{\rm p}(s)\eqdef \mathbb E[ {\rm e}^{-s\zv \Dm \zv^\herm}]=\prod_{f\in[F]}\frac{1}{1+d_fs}\;.
\end{align}
Hence, the Laplace transform of the corresponding cumulative distribution function
${\rm P}$ is given by 
\[\Lc_{\rm P}(s) = \frac{1}{s} \Lc_{\rm p}(s) = 
\frac{1}{s} \prod_{f\in[F]}\left(\frac{1}{1+d_f{s}}\right). \]
Using the Laplace inversion we can write
\begin{subequations}
	\begin{align}
		{\rm P}(\gamma) 
		&=\frac{1}{2\pi{j}}\int_{c-j\infty}^{c+j\infty}
		\frac{1}{s}
		\prod_{f\in[F]}\left(\frac{1}{1 + d_{f} s} \right)e^{s \gamma}\, ds
	\end{align}
\end{subequations}
where $c$ is chosen so that the integration path is contained in the 
region of convergence of the Laplace transform. Since the Laplace transform has a pole of order 1 at the origin, and 
poles of order $F$ at the points $-1/d_{f}$  on the real negative line, the integration path must be chosen with $c > 0$.

Using the Cauchy residue rule, we obtain the above Laplace inversion in analytic closed form as the sum of the residues of all poles in the left complex half-plane $\Re\{s\} \leq 0$. This yields
\begin{align} 
	{\rm P}(\gamma) = & 1 + 
	\sum_{f\in[F]}{\rm Res}\left \{\prod_{f\in[F]}\left(\frac{1}{1 + d_{f} s} \right) {\rm e}^{s \gamma}, s =\frac{-1}{d_{f}} \right \}
\end{align}
where the term 1 is the residue at $s = 0$.  
A problem with this closed-form approach is that the poles at 
$-1/d_{f}$ may have multiplicity larger than 1. Hence, the residue computation requires 
high-order differentiation which may quickly become cumbersome. For example, in our specific application case, the explicit form of the (log-)likelihood ratio test in (\ref{LLR-Test}) is given by 
\begin{align} 
	\rv \left [  \diag \left ( \frac{g_{u,1}}{\tau(\tau + g_{u,1})}, \ldots,  \frac{g_{u,B}}{\tau(\tau + g_{u,B})}\right ) \otimes \Id_M \right ] \rv^\herm 
	\nonumber\\\underset{\Hc_0}{\overset{\Hc_1}{\gtrless}} M \sum_{b=1}^B \log \left (1 + \frac{g_{u,b}}{\tau} \right )  - \log \nu_u \label{LLR-Test-true}
\end{align}
where $\rv \sim \zv \Cm^{1/2}$ under hypothesis $\Hc_0$ (i.e., $a = 0$), and 
$\rv \sim \zv  ( \Sigmam_u + \Cm)^{1/2}$ under hypothesis $\Hc_1$ (i.e., $a = 1$), 
with $\zv \sim_{\rm i.i.d.}  \Cc\Nc(0,1)$, $\Cm = \tau \Id_F$ and $\Sigmam_u$ given in \eqref{bdiag}. 
We see that in this case the matrix $\Dm$ is given by 
\[ \Dm = \diag \left ( \frac{g_{u,1}}{(\tau + g_{u,1})}, \ldots,  \frac{g_{u,B}}{(\tau + g_{u,B})}\right ) \otimes \Id_M \; \mbox{under $\Hc_0$} \]
and by 
\[ \Dm = \diag \left ( \frac{g_{u,1}}{\tau}, \ldots,  \frac{g_{u,B}}{\tau}\right ) \otimes \Id_M \;\;\; \mbox{under $\Hc_1$}. \]
Hence, in both cases we have poles of multiplicity $M$ (provided that the LSFCs $\{g_{u,b}\}$ are distinct) or 
higher if some of the LSFCs are non-distinct, e.g., due to some symmetry in the network geometry. 

In order to avoid the case-by-case complication of higher order differentiation, 
a more direct and yet numerically  accurate approach consists of computing the Laplace inversion using a Gauss-Chebyshev quadrature, as introduced in \cite{ventura1997impact}. The result is applicable once the Laplace transform $\Lc_{\rm p}(s)$ is known. Since probabilities are real numbers, we can write
\begin{subequations}
	\begin{align}
		{\rm P}(\gamma)  &= \frac{1}{2\pi} \int_{-\infty}^{+\infty}  \frac{\mathcal L_{\rm p}(c + j\omega) e^{\gamma(c + j\omega)}}{c + j\omega}  \;d\omega\\
		& = \frac{1}{2\pi}\int_{-\infty}^{+\infty}\frac{c\Re\{\mathcal L_{\rm p}(c+j\omega) e^{\gamma(c + j\omega)} \} }{c^2+\omega^{2}}\; d\omega\nonumber\\
		&+\frac{1}{2\pi}\int \frac{\omega\Im\{\mathcal L_{\rm p}(c+j\omega)e^{\gamma(c + j\omega)}\}}{c^2+\omega^{2}}\;d\omega\;.
	\end{align}
\end{subequations}
From the change of variables $\omega = c\sqrt{1-x^2}/x$ we get \eqref{finint} {(at the top of next page)}
\begin{figure*}[t!]
	\normalsize
	\setcounter{equation}{297}
	\begin{align}
		{\rm P}(\gamma)  =  &\frac{1}{2\pi}\int_{-1}^{1} \left \{ \Re\left[\mathcal L_{\rm p}\left(c+jc\frac{\sqrt{1-x^2}}{x}\right) 
		\exp\left (\gamma\left(c+jc\frac{\sqrt{1-x^2}}{x}\right) \right ) \right] \right . \nonumber \\
		& + \left . \frac{\sqrt{1-x^2}}{x}\Im\left[\mathcal L_{\rm p}\left(c+jc\frac{\sqrt{1-x^2}}{x}\right)
		\exp\left ( \gamma \left(c+jc\frac{\sqrt{1-x^2}}{x}\right) \right ) \right] \right \} \frac{dx}{\sqrt{1-x^2}} \label{finint}\;.
	\end{align}
	\setcounter{equation}{298}
	\hrulefill
	\vspace*{4pt}
\end{figure*}
Finally, using a Gauss-Chebyshev quadrature with $v$ nodes, we have
\begin{align}
	&{\rm P}(\gamma)  =  \frac{1}{v} \sum_{n=1}^{v/2} \Big \{ \Re\left [\mathcal L_{\rm p}(c+jc\tau_n) \exp( \gamma_u(c+jc\tau_n)) \right ] \nonumber \\
	& +  \tau_{n} \Im[\mathcal L_{\rm p}(c+jc\tau_{n}) \exp ( \gamma_u (c+jc\tau_n)) ] \Big \} + E_v, 
\end{align}
where $\tau_{n} = \tan((2n-1)\pi/(2v))$ and the error term $E_v$ vanishes as $v \rightarrow \infty$. 

Numerical evidence shows that the numerical computation of the integral is most stable by choosing the value of $c$ that minimizes the function $\mathcal L_{\rm p}(s) e^{\gamma_u s}$ on the real line. This value of $c$ corresponds to the Chernoff bound, defined as 
\begin{equation}
	{\rm P}(\gamma) \leq \min_{c \geq 0} \EE[ e^{-c(\zv\Dm \zv^\herm - \gamma)}] = \min_{c \geq 0} \mathcal L_{\rm p}(c) e^{c \gamma}. 
\end{equation}
The Chernoff bound is easier to compute (numerically) than the Laplace inversion and serves as a sanity check  to verify that the numerical computation of the Laplace inversion gives an accurate and numerically stable result.

\section{Block Gram-Schmidt Process}\label{BGS}
We begin with the classical Gram-Schmidt  process: Let 
$\underline\vv^{(1:t-1)}= \{\underline\vv^{(1)},\underline\vv^{(2)},\ldots, \underline\vv^{(t-1)}\}$ 
be a collection of  vectors in $\CC^{N}$ with $\langle \underline\vv^{(i)},\underline\vv^{(j)} \rangle = \delta_{ij}$ for all $i,j$. Also, let us denote the projection matrix to the orthogonal complement of $\text{span}(\underline\vv^{(1:t-1)})$ by
\begin{equation}
	\Pm^\perp_{\underline\vv^{(1:t-1)}}\eqdef \Id_N- \frac 1 N\sum_{1\leq s<t}\underline\vv^{(s)}(\underline\vv^{(s)})^\herm 
\end{equation}
Then, for any vector $\underline\bv\in \CC^{N}$, we construct a new orthogonal vector $\underline\vv^{(t)}=\GS{\bv}{\vv^{(1:t-1)}}$ where
\begin{equation}
	\GS{\bv}{\underline\vv^{(1:t-1)}} \eqdef \sqrt{N}\frac{\Pm^\perp_{\underline\vv^{(1:t-1)}}\bv}{\Vert \Pm^\perp_{\underline\vv^{(1:t-1)}}\bv \Vert }
\end{equation}
unless $\bv\in{\rm span}(\underline\vv^{(1:t-1)})$. 
In case $\bv\in{\rm span}(\underline\vv^{(1:t-1)})$ we generate an arbitrary vector $\underline\vv^{(t)}$ such that $\langle \underline\vv^{(t)},\underline\vv^{(i)}\rangle=\delta_{ti}$. For example,  we can set  $\underline \vv^{(t)}=\GS{\hv}{\vv^{(1:t-1)}}$ for an arbitrary Gaussian random vector $\underline\hv\sim\mathcal{CN}(\matr 0,\Id_N)$ with noting that as long as $N>t$ we have $\underline\hv\notin{\rm span}(\underline\vv^{(1:t-1)})$ a.s. \cite{Berthier20}.  

Similarly, let $\Vm^{(1:t)}=\{\Vm^{(1)},\ldots ,\Vm^{(t-1)}\}$ be a collection of matrices in $\CC^{N\times \bm}$ with $\langle \Vm^{(i)}, \Vm^{(j)}\rangle=\delta_{ij}\Id_{\bm}$ for all $i,j$. Then, for any matrix $\Bm\in \RR^{N\times \bm}$, we construct the new  matrix  $\Vm^{(t)}\eqdef \GS{\Bm}{\Vm^{(1:t-1)}} $
such that its $f$th column sequentially constructed as ${\underline\vv}^{(s)}_{f}=\GS{\underline\bv_{k}}{{\underline\vv_{1:\bm}^{(1:t-1)},{\underline \vv}^{(s)}_{1:f-1}}}\;$ with ${\underline\vv}_{1:f}=\{\underline\vv_{1} ,\underline\vv_{2},\ldots, \underline\vv_{f}\}$ and $\underline\vv_{1:0}= \emptyset$.  Notice that we have by construction $\langle\Vm^{(t)}, \Vm^{(s)}\rangle=\delta_{ts}\Id_\bm$ and 
\begin{equation}
	\Bm=\sum_{1\leq s\leq t}\Vm^{(s)}\langle\Vm^{(s)},\Bm\rangle\;.
\end{equation}

\bibliographystyle{IEEEtran}
\bibliography{report,massive-MIMO-references}

\end{document}